\documentclass[preprint,12pt,numbers]{elsarticle}

\usepackage{subcaption}
\usepackage[title]{appendix}
\usepackage{amssymb}
\usepackage{amsmath}
\usepackage{amsbsy}
\usepackage{bm} 
\usepackage{calligra}
\usepackage{tikz}%
\usepackage{booktabs} 
\usepackage{multirow}
\usepackage{multicol} 
\usepackage{hyperref}

\usepackage{graphicx}%
\usepackage{multirow}%
\usepackage{amsmath,amssymb,amsfonts}%
\usepackage{amsthm}%
\usepackage{mathrsfs}%
\usepackage[title]{appendix}%
\usepackage{xcolor}%
\usepackage{float}
\usepackage{ulem}
\usepackage{ragged2e}
\usepackage{graphicx}
\usepackage{stackengine}
\usepackage{textcomp}%
\usepackage{manyfoot}%
\usepackage{booktabs}%
\usepackage{tikz}%
\usepackage{adjustbox}
\usepackage[utf8]{inputenc}
\usepackage{setspace}
\usepackage{bm}
\usepackage{psfrag} 
\usepackage{url}
\usepackage{stmaryrd}
\usepackage{algorithm}%
\usepackage{algorithmicx}%
\usepackage{algpseudocode}%
\usepackage{listings}%
\usepackage{gensymb}
\usepackage{comment}
\usepackage{svg}
\usepackage[colorinlistoftodos]{todonotes}
\usepackage{siunitx}
\usepackage[T1]{fontenc}
\usepackage{textcomp}
\usepackage{lmodern}

\journal{Nuclear Physics B}

\begin{document}
\biboptions{numbers,sort&compress}

\begin{frontmatter}



\title{Phase-field analysis of fracture in casing-cement-rock wellbore systems: effects of casing eccentricity and cement-rock interface strength} 

 \author[label1]{Tharunsarathy}
 \author[label1]{Wasim Niyaz Munshi}
 \author[label1]{Chandrasekhar Annavarapu\texorpdfstring{\cormark[cor1]}{}}
 \author[label2]{Birendra Jha}

 \cortext[cor1]{Corresponding author: annavarapuc@civil.iitm.ac.in}
 
 \affiliation[label1]{organization={Department of Civil Engineering},
             addressline={Indian Institute of Technology Madras},
             city={Chennai},
             postcode={600036},
             state={Tamil Nadu},
             country={India}}

 \affiliation[label2]{organization={Mork Family Department of Chemical Engineering and Materials Science},
             addressline={University of Southern California},
             city={Los Angeles},
             postcode={90007},
             state={CA},
             country={USA}}

\begin{abstract}

Predicting the initiation and propagation of cracks in heterogeneous wellbore systems under complex in-situ conditions remains challenging. We present a hybrid phase-field fracture framework to model crack growth in heterogeneous wellbore systems with weak interfaces. The framework is first validated against benchmark problems with available analytical and numerical solutions. Subsequently, numerical experiments are conducted to isolate the effects of interface strength and casing eccentricity on crack growth. The results show that casing eccentricity strongly influences both the pressure at crack initiation and the resulting crack paths, reducing the crack initiation pressure by up to 30\% relative to the concentric configuration. For the examples considered here, beyond a critical eccentricity threshold of 50\%, localized shear stresses drive the nucleation of inclined cracks in the rock formation in addition to radial cracking -- a failure mode absent in concentric configurations. For sufficiently weak interfaces (i.e., interfaces with 30\% of the toughness of the surrounding bulk material), radially propagating cracks in the cement sheath are deflected along the interface rather than penetrating into the rock formation. This deflection delays stress relaxation within the sheath, promotes the nucleation of additional radial cracks, and increases the risk of sustained casing pressure and wellbore failure. Finally, a three-dimensional simulation reveals depth-dependent crack nucleation, stress-shadow effects that suppress full-depth crack growth, and crack coalescence along the cement–rock interface -- phenomena that are fundamentally inaccessible under plane-strain assumptions -- demonstrating the applicability of the framework to realistic heterogeneous wellbore systems.
\end{abstract}

\begin{highlights}
\item Effects of casing eccentricity and interface weakening in the wellbore are studied using a phase-field fracture model.
\item Applicability of the phase-field fracture model to large-scale problems is demonstrated using a 3D simulation of a wellbore.
\end{highlights}


\end{frontmatter}

\section{Introduction}\label{sec:introduction}

 Wellbores provide engineered pathways from the surface to subsurface reservoirs while ensuring zonal isolation (keeping different rock layers separated) and containment of fluids for safe and efficient operations in many  applications~\cite{driscoll1986,Bourgoyne1986}. Examples include oil and gas production, geothermal
energy extraction, borehole mining and leaching, underground waste disposal,
natural gas and hydrogen storage,  carbon sequestration, and groundwater pumping. In practice, these systems comprise heterogeneous material assemblies, typically consisting of a cylindrical steel casing surrounded by a cement sheath that bonds the casing to the rock formation to ensure structural integrity of the wellbore over the life of the well, which is typically expected to be several decades. The casing-cement and cement-rock interface regions are often occupied by fluids (drilling mud, sludge from mud-cement interaction, formation fluids such as hydrocarbons and brine, and cement filtrate) and fine particles (mud cake, fly ash, and silica from cement slurry), which can cause interface weakening and poor bonding between casing and cement or cement and rock. Furthermore, casing position inside the borehole can be eccentric or off-centered due to borehole inclination and rock surface roughness (rugosity), introducing azimuthal variation in the cement sheath thickness, which can impact cement bond integrity~\citep{Bourgoyne1986}. Both interface weakening and casing eccentricity can negatively impact the well log data quality and interpretation. 

\begin{figure}[h!]
\centering
    \begin{subfigure}[b]{0.3\textwidth}
        \centering
        \includegraphics[trim=0 90 0 150, clip,
        width=0.75\linewidth]{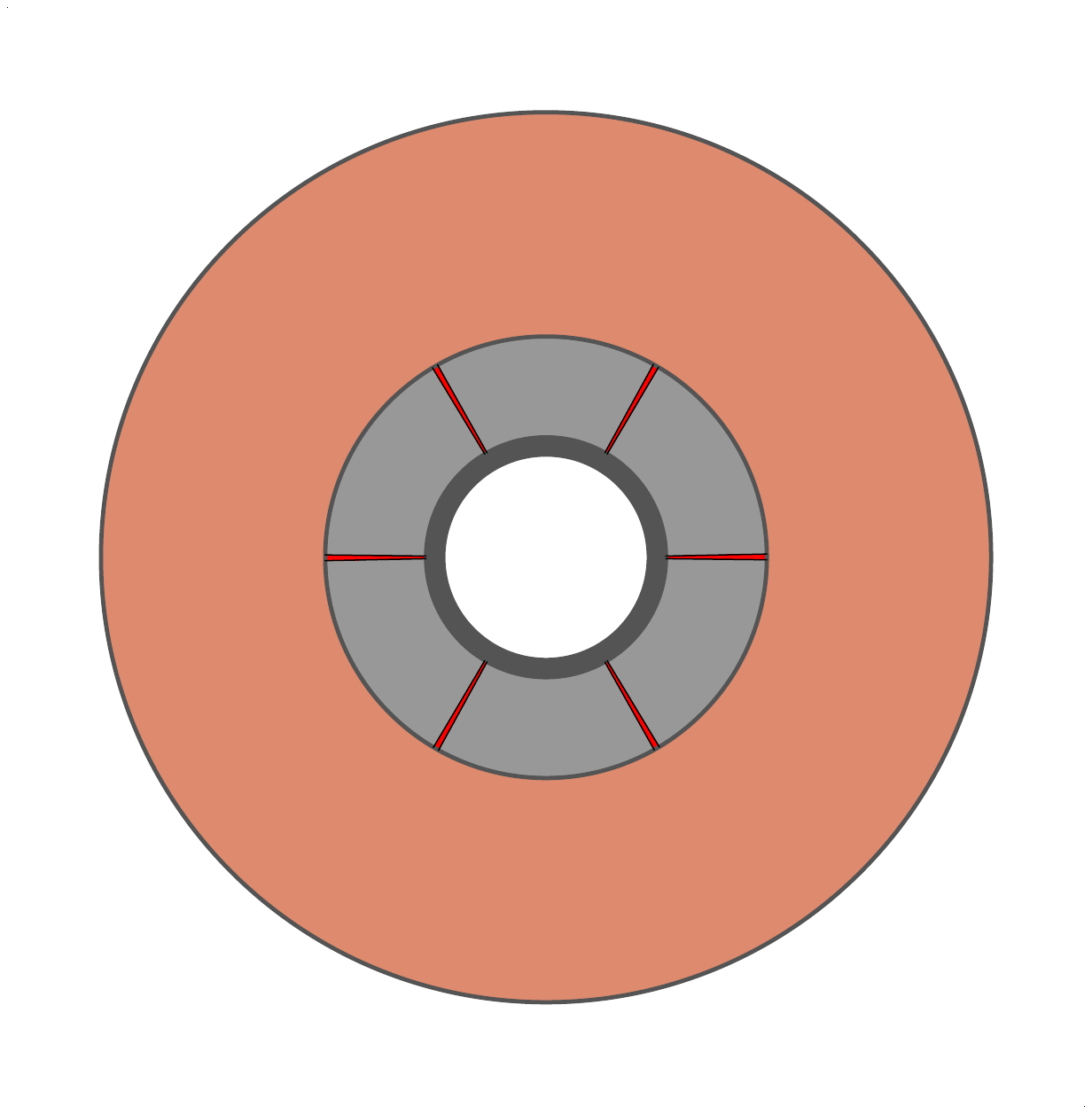}
        \caption{Radial cracks\\ }
        \label{fig:radial_cracks}
    \end{subfigure}
    \hfill
    \begin{subfigure}[b]{0.3\textwidth}
        \centering
        \includegraphics[trim=0 90 0 150, clip,
        width=0.75\linewidth]{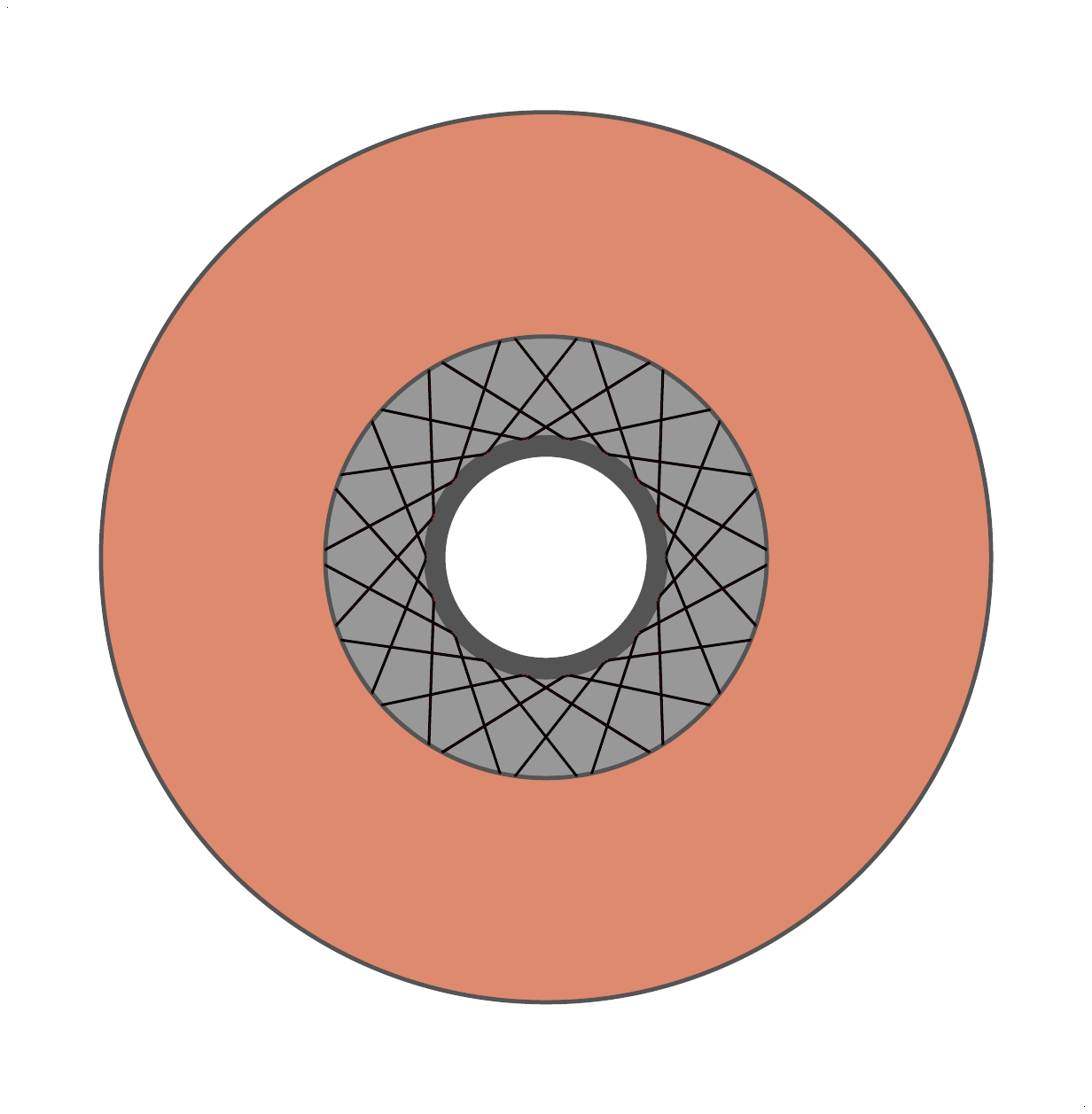}
        \caption{Shear cracks\\ }
        \label{fig:shear_cracks}
    \end{subfigure}
    \hfill
    \begin{subfigure}[b]{0.3\textwidth}
        \centering
        \includegraphics[trim=0 150 0 150, clip,
        width=0.75\linewidth]{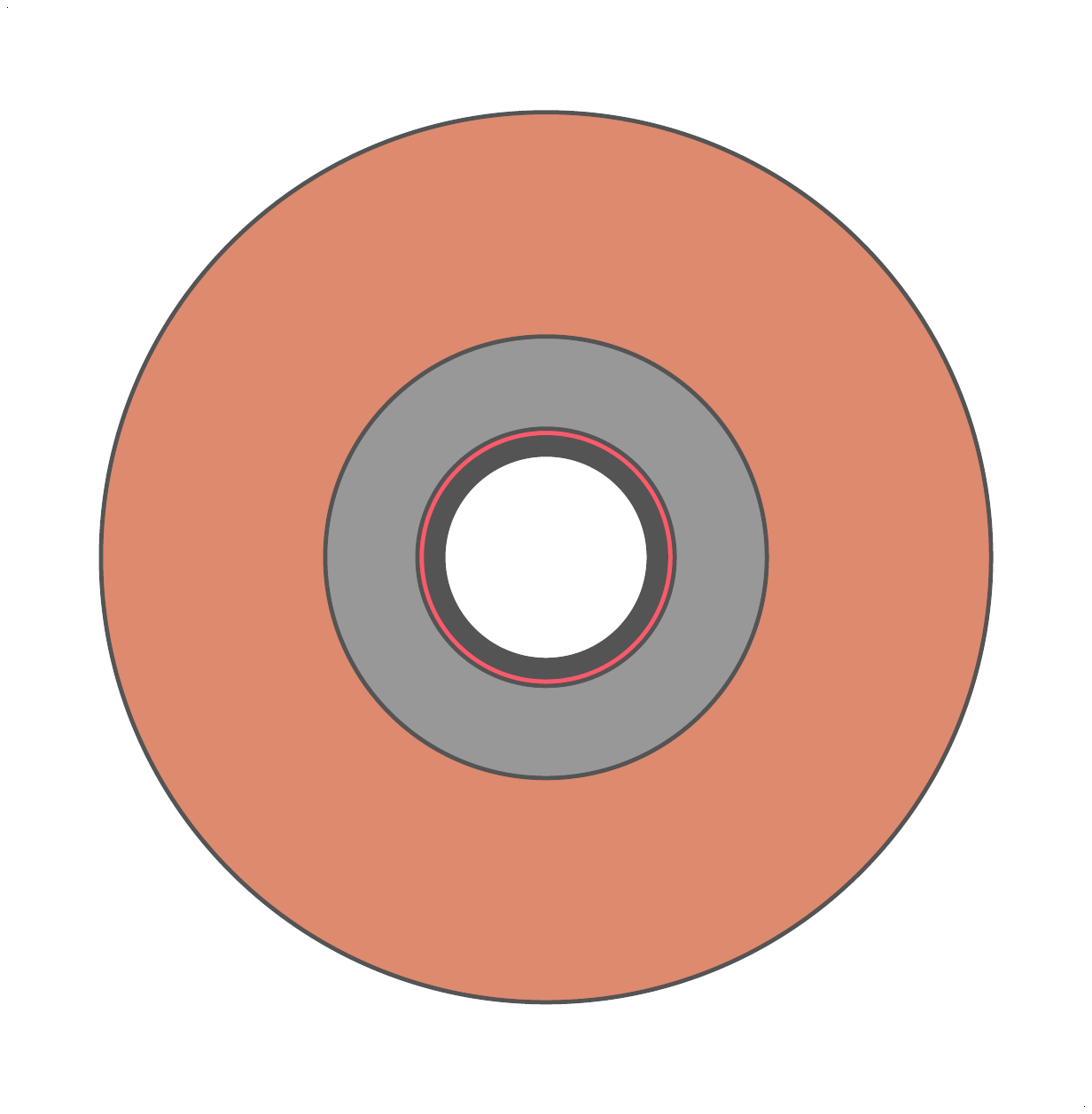}
        \caption{Debonding of casing-cement sheath interface}
        \label{fig:ca_cs_debonding}
    \end{subfigure}
    \hfill
    \begin{subfigure}[b]{0.3\textwidth}
        \centering
        \includegraphics[trim=0 150 0 150, clip,
        width=0.75\linewidth]{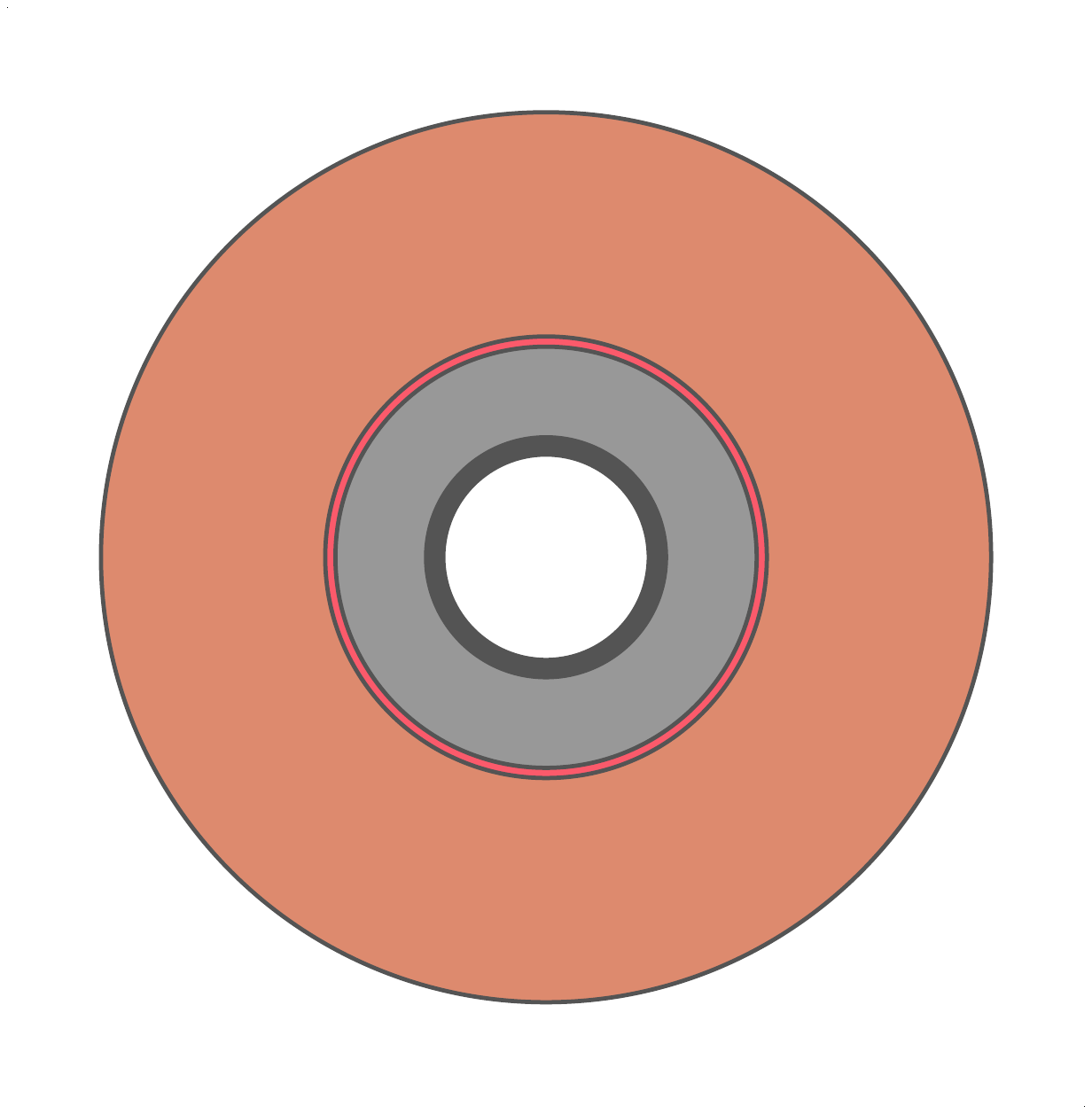}
        \caption{Debonding of cement sheath-rock interface}
        \label{fig:cs_f_debonding}
    \end{subfigure}
    \begin{subfigure}[b]{0.3\textwidth}
        \centering
        \includegraphics[width=0.75\linewidth]{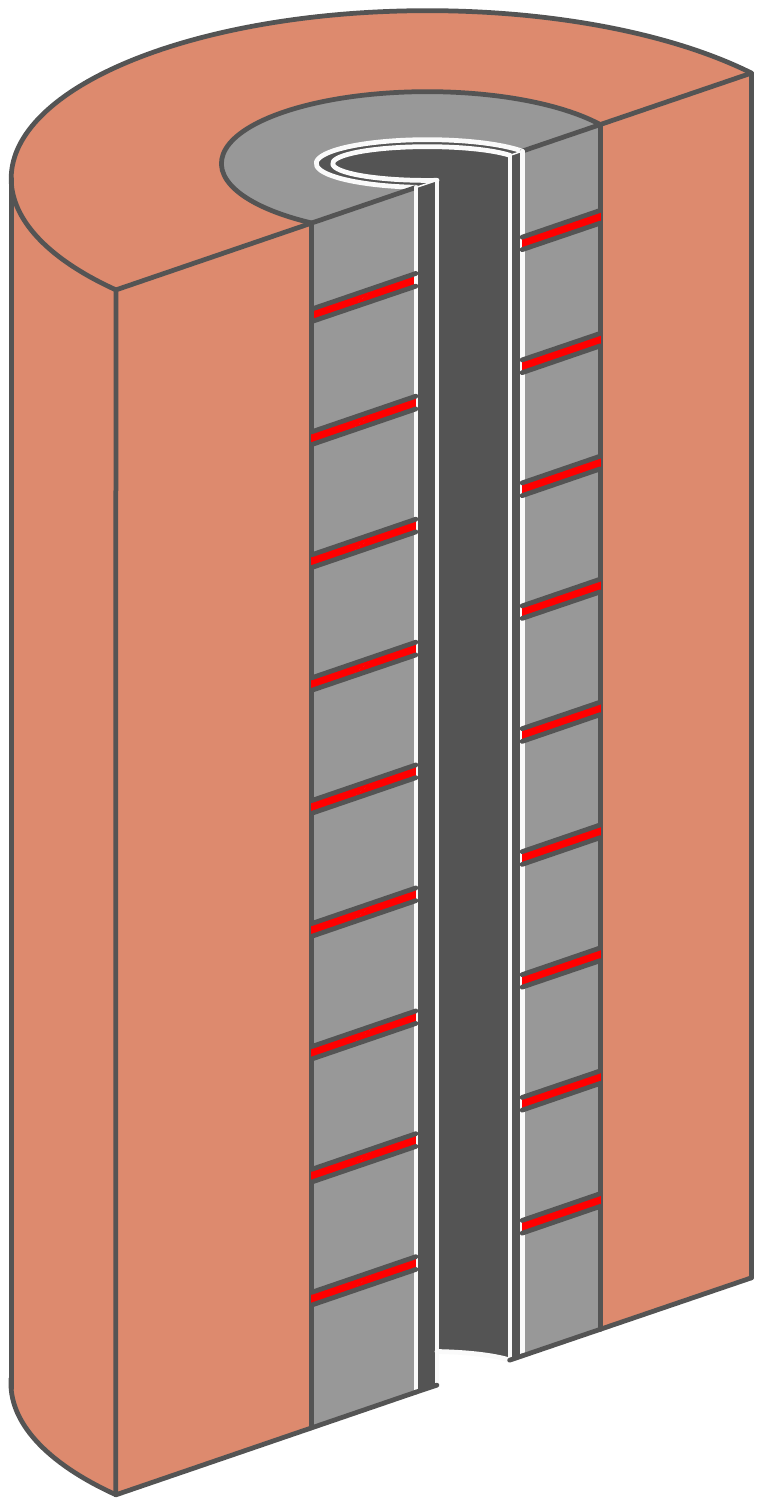}
        \caption{Disking\\ } 
        \label{fig:disking}
    \end{subfigure}
    \caption{Different types of wellbore failures. (a-d) show a cross-sectional view on a plane perpendicular to the wellbore axis, and (e) shows a cut-away 3D view parallel to the axis.}
    \label{fig:integrity_failures}
\end{figure}

Under deep subsurface conditions, cemented wellbore systems experience complex in-situ stresses present in the rock and pressure variations due to fluid flow and cement shrinkage that can compromise the structural integrity of the well and its hydraulic isolation property. A range of cement failure modes have been observed, including radial cracking, annular debonding, disking, and shear failure (Figure~\ref{fig:integrity_failures}), which result in microannuli (channels filled with fluid) and a loss of hydraulic isolation~\cite{Bourgoyne1986}. Despite extensive research, predicting the initiation and subsequent evolution of such failures under complex in-situ loading conditions remains a significant challenge. A first-principles understanding of these mechanisms is therefore essential for the reliable design and long-term integrity of wellbore systems.

 A wide range of experimental and numerical studies have been conducted to investigate wellbore integrity. Among numerical approaches, cohesive zone-based finite element methods have been widely employed to model interfacial failure and fracture processes in the cement sheath. For instance, \citet{wang2014three} examined the influence of cement stiffness and interface strength on fracture arrest, while \citet{gu2022numerical} and \citet{xi2024failure} analyzed the effects of wellbore trajectory, casing eccentricity, and plastic deformation on microannulus creation. Similarly, \citet{zhao2020cohesive} performed parametric studies under high-temperature and high-pressure conditions using cohesive zone modeling. Collectively, these studies underscore the critical role of interface properties and geometric imperfections in governing wellbore integrity.

In parallel, several studies have focused on evaluating the stress state in wellbore systems using analytical and numerical techniques. The influence of casing eccentricity has been investigated using the boundary perturbation method \citep{zheng2023influence}, boundary element methods \citep{cheng2017numerical}, and fracture mechanics-based approaches such as stress intensity factor evaluation via contour integrals \citep{dou2020numerical}. Analytical formulations combined with yield criteria, such as the Drucker-Prager model, have also been employed to characterize failure in the cement sheath \citep{liu2018impact}. Additionally, finite element-based analyses have been used to assess stress distributions under specific applications, including coalbed methane fracturing \citep{zhao2023integrity}. These approaches primarily provide insight into stress evolution and failure initiation, but offer limited capability in capturing post-initiation fracture propagation and interaction.

Experimental investigations have primarily focused on characterizing wellbore failure through indirect measurements, such as fluid leakage tests \citep{goodwin1992cement, xi2020experimental, nassan2024experimental, vraalstad2020digital}. While these studies provide valuable insights into the loss of hydraulic integrity, they offer limited understanding of the underlying fracture mechanisms. More recent efforts have explored advanced imaging techniques, such as X-ray computed tomography (CT), to directly observe crack initiation and propagation at the laboratory scale \citep{taghipour2022novel, anya2023novel, vraalstad2020digital}. However, such experiments remain constrained in replicating field-scale conditions and the inherent heterogeneity of wellbore materials.

Despite these advances, existing approaches face significant limitations in capturing the full fracture process in heterogeneous wellbore systems. Laboratory-scale investigations are often insufficient to replicate field-scale conditions, while analytical methods are generally limited to pre-failure response. Furthermore, state-of-the-art numerical approaches, including cohesive zone-based finite element models, typically require \textit{a priori} assumptions on crack paths or predefined interfaces, thereby restricting their ability to simulate arbitrary crack initiation and complex propagation. As a result, critical fracture phenomena such as branching, coalescence, and interaction across material interfaces remain inadequately captured. In addition, key factors such as cement-rock interface strength and casing eccentricity, which strongly influence stress distribution and fracture evolution, have not been systematically studied in a coupled manner. These limitations motivate the need for modeling frameworks capable of naturally capturing complex crack evolution in heterogeneous wellbore systems without restrictive assumptions on crack geometry.

Cohesive-zone based finite element approaches that rely on conforming grids consider cracks as sharp surfaces of discontinuity~\cite{Ortiz1999,Camacho1996,moes1999finite,Ghosh2019,Ghosh2021,Annavarapu2016}. In contrast, the last decade has seen the emergence of regularized fracture models such as phase-field methods~\citep{francfort1998revisiting,bourdin2000numerical,amor2009regularized,miehe2010thermodynamically,Karma2001}, which treat cracks as diffuse regions of damaged material. Phase-field fracture models are rooted in Griffith's theory, where crack evolution is governed by energy minimization principles. Within this framework, crack initiation, branching, coalescence, and complex crack-path evolution emerge naturally, without the need for explicit crack tracking or ad-hoc criteria. These features make phase-field methods particularly attractive for heterogeneous geomaterials and multi-material systems. While such models have been widely applied in diverse contexts~\citep{Ding2024,seiler2020efficient,zhang2019phase,Jain2023a,Jain2023b,Khan2023a, Khan2023b, KHAN2025110672,denli2020phase,messaoudi2025fracture}, their application to wellbore systems remains relatively limited. The most closely related work applies phase-field methods to hydraulic fracture nucleation and propagation in the near-wellbore rock formation~\citep{Cusini_2287725}, focusing on fracture driven by fluid injection into the borehole. This is a fundamentally different problem class from the structural integrity of the annular multi-material system -- comprising the casing, cement sheath, and surrounding rock formation -- which is the focus of the present work. A key challenge limiting their applicability to heterogeneous material systems such as wellbores lies in representing the competition between bulk fracture and interface debonding. Recent advances address this by incorporating interface-specific fracture energy contributions within the phase-field framework, enabling the modeling of coupled bulk and interfacial failure~\citep{KHAN2025110672, munshi2025modeling, munshi2025phase}.

Building on these advances and to address the limitations of existing approaches in modeling complex fracture processes in heterogeneous wellbore systems, this study applies the phase-field fracture formulation for layered media with weak interfaces described in Khan et al.~\citep{KHAN2025110672} and Munshi et al.~\citep{munshi2025modeling, munshi2025phase} to study wellbore integrity. This framework is capable of naturally representing intricate crack patterns, incorporating interface effects, and maintaining computational efficiency for simulating complex fracture processes in wellbore systems. To our knowledge, this is the first study to simulate critical fracture phenomena such as branching, coalescence, and interaction across material interfaces in a wellbore system, going beyond the pre-failure response without being limited to a predefined crack path. In this work, we:
\begin{itemize}
    \item Apply a hybrid phase-field framework to model crack growth in heterogeneous wellbore systems
    \item Quantify the influence of the cement-rock interface properties on wellbore failure
    \item Investigate the role of casing eccentricity on stress distribution and fracture evolution
\end{itemize}

 The rest of the paper is structured as follows. Section~\ref{sec:methodology} presents the governing equations of the hybrid phase-field fracture model with weak interfaces, with particular emphasis on the interface energy contribution and its role in controlling the competition between bulk fracture and interfacial debonding. Section~\ref{sec:nv} validates the framework through three benchmark problems: elastic stress recovery in a pipe with an eccentric bore, crack deflection and interfacial debonding in a bi-material plate, and crack propagation in a realistic multi-material wellbore geometry. Section~\ref{sec:ne} presents targeted numerical experiments to isolate the effects of casing eccentricity and cement–rock interface strength on wellbore fracture behavior, followed by a three-dimensional simulation that reveals depth-dependent crack nucleation, stress-shadow effects, and crack coalescence phenomena that are inaccessible under plane-strain assumptions. Section~\ref{sec:limitations} briefly discusses the limitations of the present study and outlines potential future directions, including fluid coupling, frictional sliding, and field-scale applications. Section~\ref{sec:conclusion} summarizes the key findings of the present work.

\section{Methodology}
\label{sec:methodology}
\begin{figure} [h!]
    \centering
    \includegraphics[trim=0 115 0 110, clip, 
    width=0.4\linewidth]{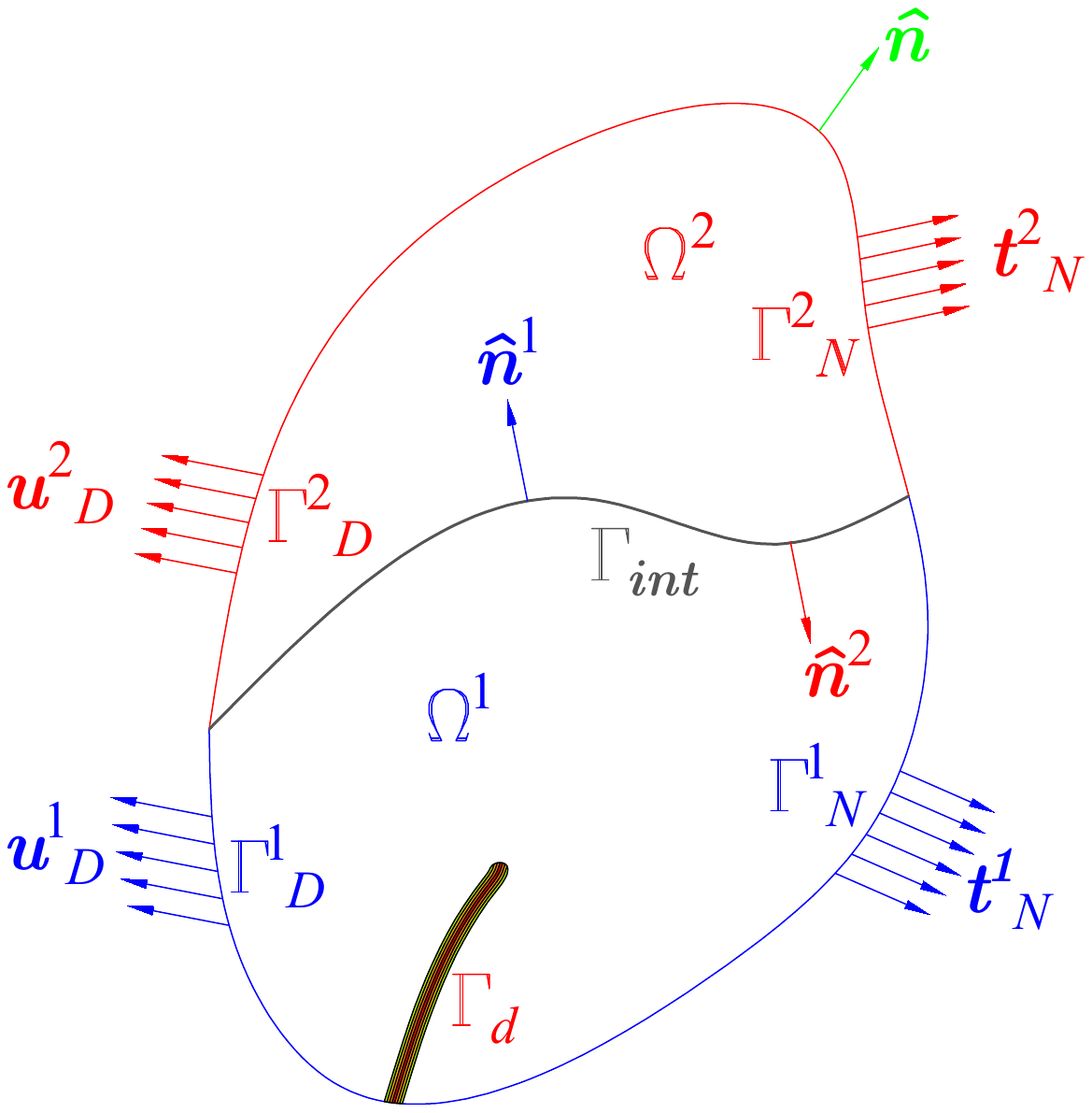}
    \caption{Schematic showing a bi-material elastic body with a crack $\Gamma_d$. $\Gamma_\text{int}$ represents the interface between the two subdomains $\Omega^1$ and $\Omega^2$. The internal discontinuity $\Gamma_d$ is modeled in a smeared manner via a scalar phase-field variable $d$, defined throughout the domain.}
    \label{fig:pf}
\end{figure}

The formulation summarized in this section follows the phase-field fracture framework for weakly bonded interfaces developed in Khan et al.~\cite{KHAN2025110672} and Munshi et al.~\cite{munshi2025modeling,munshi2025phase}, and is reproduced here for completeness. No modifications to the formulation are introduced in the present work. Consider a domain $\Omega$, with a boundary $\partial\Omega$, partitioned into two non-overlapping subdomains $\Omega^1$ and $\Omega^2$ by a sharp interface $\Gamma_\text{int}$, as shown in Figure~\ref{fig:pf}. The subdomains comprise distinct materials, and the interface $\Gamma_\text{int}$ represents the sharp material boundary separating them. The domain boundary $\partial\Omega$ is partitioned into disjoint sets $\partial\Omega^1$ and $\partial\Omega^2$ such that $\partial\Omega = \overline{\partial\Omega^1\cup\partial\Omega^2}$, and is associated with an outward-pointing unit normal denoted by $\boldsymbol{\hat{n}}$. The boundary $\partial\Omega^\mathrm{m}$ of the subdomain $\Omega^\mathrm{m}$ is decomposed as $\partial\Omega^\mathrm{m}
  = \overline{\Gamma_D^\mathrm{m} \cup \Gamma_N^\mathrm{m} \cup \Gamma_{N_0}^\mathrm{m} \cup \Gamma_\text{int}}$, where $\mathrm{m}$ denotes the domain (or material) index. Here, $\Gamma_D^\mathrm{m}$ and $\Gamma_N^\mathrm{m}$ denote the Dirichlet and Neumann boundaries, with prescribed displacements $\boldsymbol{u}_D^\mathrm{m}$ and tractions $\boldsymbol{t}_N^\mathrm{m}$, respectively. $\Gamma_{N_0}^\mathrm{m}$ denotes a traction-free Neumann boundary, while $\Gamma_\text{int}$ is the material interface shared by both subdomains. The unit normals on $\Gamma_\text{int}$ pointing outward from 
$\Omega^1$ and $\Omega^2$ are denoted by $\boldsymbol{\hat{n}}^{\,1}$ and $\boldsymbol{\hat{n}}^{\,2}$, respectively. The total potential energy 
of the bi-material system, including both bulk and interface contributions, is given by

\begin{equation}
\begin{split}
 \text{\calligra{U}}\;\left(\boldsymbol{u}^1, \boldsymbol{u}^2, d^1, d^2\right) &= \sum_{\mathrm{m}=1}^2 \int_{\Omega^\mathrm{m}} g\left(d^\mathrm{m}\right) \psi^\mathrm{m} \left(\boldsymbol{\varepsilon}\left(\boldsymbol{u}^\mathrm{m}\right)\right) \, \mathrm{d}\Omega \\
&\quad + \sum_{\mathrm{m}=1}^2 \int_{\Omega^\mathrm{m}} G_{c}^\mathrm{m} \left\{\frac{1}{2}\left[\frac{1}{\ell}\, (d^\mathrm{m})^2 + \ell\, \left|\boldsymbol{\nabla} d^\mathrm{m}\right|^2\right]\right\} \, \mathrm{d}\Omega \\
&\quad + \int_{\Gamma_\text{int}} \frac{1}{2} \kappa\, G_{c}^1\, (d^1)^2 \, \mathrm{d}\Gamma + \int_{\Gamma_\text{int}} \frac{1}{2} \kappa\, G_{c}^2\, (d^2)^2 \, \mathrm{d}\Gamma.
\label{eq:potential_energy_1}
\end{split}
\end{equation}

For each material $\mathrm{m}$, $g(d^\mathrm{m}) = (1-d^\mathrm{m})^2$ denotes the stiffness degradation function, and $\psi^\mathrm{m} \left(\boldsymbol{\varepsilon} \left( \boldsymbol{u}^\mathrm{m} \right) \right)$ 
is the elastic strain energy density functional. Here, $\boldsymbol{\varepsilon}$ denotes the small-strain tensor defined as the symmetric gradient of the displacement field $\boldsymbol{u}^\mathrm{m}$. $G_c^\mathrm{m}$ is the critical strain energy release rate of the material, and $\ell$ is the phase-field length scale. The interface contributions penalize damage at the material interface, allowing the modeling of interface debonding in conjunction with bulk fracture. The scalar parameter \(\kappa\) controls the relative toughness of the interface with respect to a neutrally bonded interface. Enforcing damage continuity at the interface, i.e., setting $d^1 = d^2 = \widetilde{d}$ in Equation~\eqref{eq:potential_energy_1}, yields

\begin{equation}
\begin{split}
    \text{\calligra{U}}\;\left(\boldsymbol{u}^{1}, \boldsymbol{u}^{2}, d^{1}, d^{2}\right) &= \sum_{\mathrm{m}=1}^{2} \int_{\Omega^\mathrm{m}} g(d^\mathrm{m}) \psi^\mathrm{m} \left(\boldsymbol{\varepsilon} \left( \boldsymbol{u}^\mathrm{m} \right) \right) \mathrm{d}\Omega \\
    &\quad + \sum_{\mathrm{m}=1}^{2} \int_{\Omega^\mathrm{m}} G_{c}^\mathrm{m} \left\{ \frac{1}{2} \left[ \frac{1}{\ell} \left( d^\mathrm{m} \right)^{2} + \ell \left| \boldsymbol{\nabla} d^\mathrm{m} \right|^{2} \right] \right\} \mathrm{d}\Omega \\
    &\quad + \int_{\Gamma_\text{int}}  G_{c}^\text{ib} (\tilde{d})^{2} \mathrm{d}\Gamma
    \label{eq:potential_energy_2}
\end{split}
\end{equation}

where $G_c^\text{ib} = \left(\dfrac{G_c^1+G_c^2}{2}\right)\kappa$ denotes the critical energy release rate of the interface, distinct from those of the bulk materials. The first and second terms on the right-hand side of Equation~\eqref{eq:potential_energy_2} represent the total elastic strain energy of the two materials and the sum of their regularized crack surface energies, which are standard in phase-field fracture models. The third term introduces an interface-specific crack surface energy contribution, enabling the modeling of interfaces that are inherently weaker or stronger than the bulk material. To ensure the convexity of the total energy functional, the interface term is expressed in a quadratic form (see, also, \cite{kuhn2019phase,KHAN2025110672,munshi2025modeling}). This term is evaluated directly as a surface integral without additional regularization, since the interface location is known a priori. 

Equation~\eqref{eq:potential_energy_2} shows that the interface-specific contribution vanishes in the energy functional when $G_c^\text{ib} = 0$. A positive value, $G_c^\text{ib} > 0$, represents interface strengthening with respect to a reference state, whereas $G_c^\text{ib} < 0$ represents interface weakening. The reference state corresponds to a neutrally bonded interface, denoted by $G_c^\text{nb}$, which is neither stronger nor weaker than the surrounding bulk material. In this work, the strength of such a neutrally bonded interface is considered to be the arithmetic average of the critical energy release rates of the surrounding bulk material, i.e., $G_c^\text{nb}=\left(G_{c}^1+G_{c}^2\right) / 2$.
 
The effective critical energy release rate of the interface $G_c^\mathrm{eff}$ is then given in terms of the critical energy release rate of the neutrally bonded interface and the critical energy release rate associated with the interface bond strength as 

\begin{equation}
\begin{split}
 & G_c^{\mathrm{eff}} = G_c^\text{nb} + G_c^{ib} = G_c^\text{nb} + \left(\dfrac{G_c^1+G_c^2}{2}\right)\kappa,
 \\
 & G_c^{\mathrm{eff}} = G_c^\text{nb}+\kappa G_c^\text{nb} = G_c^\text{nb}(1+\kappa).
\end{split}
\label{eq:toughness}
\end{equation}

The parameter $\kappa$ controls the interface bond strength. When $\kappa=0$, the interface is neutrally bonded. $\kappa>0$ represents interface strengthening, i.e., its fracture toughness is higher than that of the neutrally bonded interface (e.g., in strongly adhesively bonded interfaces or welded metals). When $\kappa<0$ -- a likely scenario in wellbore systems -- the interface is weakened, i.e., its fracture toughness is lower than that of the neutrally bonded interface. It is worth emphasizing that the effective value of the critical energy release rate at the interface $G_c^\mathrm{eff}$ is the sum of two contributions: the bulk regularized crack surface energy, which by construction supplies $G_c^\mathrm{nb}$ per unit area at the interface and the interface correction $\kappa G_c^\mathrm{nb}$. The net contribution is therefore non-negative for all $\kappa \geq -1$, ensuring that the total energy functional remains coercive over the admissible range of $\kappa$. For more details and a physical interpretation of the scalar parameter $\kappa$, we refer the reader to the Appendix of \citet{munshi2025modeling}. It should be noted that the present model only considers crack opening at the interface and does not account for sliding.

\sloppy{The stationarity of the potential energy functional with respect to the displacement and damage fields yields the equations governing the evolution of these fields. The displacement field satisfies the following equilibrium equation together with the associated boundary and interface conditions.}

\begin{subequations}\label{eq:hybrid_r}
\begin{align}
    & \boldsymbol{\nabla} \cdot\boldsymbol{\sigma}^\mathrm{m} \left( \boldsymbol{\varepsilon}(\boldsymbol{u}^\mathrm{m}),d^\mathrm{m} \right)=\boldsymbol{0} && \text{ in } \Omega^\mathrm{m} \label{eq:lm_1}, \\
    & \boldsymbol{u}^\mathrm{m} = \boldsymbol{u}_D^\mathrm{m} && \text{ on } \Gamma_D^\mathrm{m} \label{eq:lm_2}, \\  
    & \boldsymbol{\sigma}^\mathrm{m}\boldsymbol{\hat{n}}=\boldsymbol{t}_N^\mathrm{m} && \text{ on } \Gamma_N^\mathrm{m} \label{eq:lm_3}, \\
    & \boldsymbol{u}^1 = \boldsymbol{u}^2 && \text{ on } \Gamma_\text{int} \label{eq:lm_4}, \\
    & \boldsymbol{\sigma}^1\boldsymbol{\hat{n}}^1 = -\boldsymbol{\sigma}^2\boldsymbol{\hat{n}}^2 && \text{ on } \Gamma_\text{int} \label{eq:lm_5} .
\end{align}
\end{subequations}

Here, $\boldsymbol{\sigma}^\mathrm{m}$ denotes the Cauchy stress tensor in material $m$, $\boldsymbol{\nabla}$ is the nabla operator, and $\boldsymbol{\hat{n}}$ is the outward-pointing normal to the boundary $\partial\Omega_N^\mathrm{m}$. Equations \eqref{eq:lm_2} and \eqref{eq:lm_3} represent the standard Dirichlet and Neumann boundary conditions, with prescribed displacements $\boldsymbol{u}_D^\mathrm{m}$ and tractions $\boldsymbol{t}_N^\mathrm{m}$, respectively. Equations \eqref{eq:lm_4} and \eqref{eq:lm_5} enforce displacement continuity and traction equilibrium across the interface. 

Under the assumption of isotropic linear elasticity, the bulk stress–strain constitutive relation, including the degradation function $g(d)$, is given by
\begin{subequations}
\begin{align}
     \boldsymbol{\sigma}^\mathrm{m} &= g(d^\mathrm{m}) \frac{\partial\psi^\mathrm{m}(\boldsymbol{\varepsilon})}{\partial\boldsymbol{\varepsilon}} 
                       = g(d^\mathrm{m}) \mathbb{C}^\mathrm{m}\colon\boldsymbol{\varepsilon} \label{eq:constitutive} \\
                       & g(d) = \begin{cases} (1-d)^2 \;\;\; \text{if} \;\;\;\psi^+ > \psi^-\\
1 \;\;\;\;\;\; \text{otherwise.} \end{cases} \label{eq:degradation}
\end{align}
\end{subequations}

Here, $\mathbb{C}^\mathrm{m}$ is the fourth-order elasticity tensor for the material $\mathrm{m}$, while $\psi$, $\psi^+$, and $\psi^-$ represent the total, tensile, and compressive strain energy densities respectively. These quantities are computed in terms of the strain tensor $\boldsymbol{\varepsilon}$ and the Lam\'e parameters $\lambda$ and $\mu$ as follows
\begin{subequations}
\begin{align}
& 
\psi(\boldsymbol{\varepsilon})=\frac{1}{2} \lambda \operatorname{tr}(\boldsymbol{\varepsilon})^2+\mu \operatorname{tr}\left(\boldsymbol{\varepsilon}^2\right),
\label{eq:str_energy} \\
& 
\psi^{ \pm}(\boldsymbol{\varepsilon})=\frac{1}{2} \lambda\langle\operatorname{tr}(\boldsymbol{\varepsilon})\rangle_{ \pm}^2+\mu \operatorname{tr}\left(\boldsymbol{\varepsilon}_{ \pm}^2\right).
\label{eq:str_energy_Split}
\end{align}
\end{subequations}
A tension–compression split based on the spectral decomposition of the strain tensor, following Miehe~\cite{miehe2010thermodynamically}, is employed to suppress nonphysical crack growth under compression, such that
\begin{equation}
\boldsymbol{\varepsilon}_{ \pm}=\sum_{j=1}^{n_{s d}}\left\langle\varepsilon_j\right\rangle_{ \pm} \boldsymbol{e}_{\boldsymbol{j}} \otimes \boldsymbol{e}_{\boldsymbol{j}}.
\end{equation}
Here, $\varepsilon_{j}$ are the principal strains, $\boldsymbol{e}_{\boldsymbol{j}}$ are the principal directions, and $n_{sd}$ is the number of spatial dimensions. $\left\langle \odot \right\rangle_{\pm}$ are the Macaulay brackets defined as $\left\langle \odot \right\rangle_{\pm} = (\odot \pm |\odot|)/2$. Equation~\eqref{eq:degradation} serves as the constraint equation, which uses the standard quadratic degradation function $g(d) = (1-d)^2$ if the tensile strain energy, $\psi^+$, exceeds the compressive strain energy, $\psi^-$ and sets $g(d) = 1$ otherwise, consistent with the hybrid phase-field model of \citet{ambati2015review}. Equation~\eqref{eq:degradation} ensures that stiffness is degraded only in regions where tension dominates over compression, thereby preventing the interpenetration of crack faces under compression.

The damage evolution equation is given by,
\begin{subequations}\label{eq:damage_growth}
\begin{align}
&G_c^\mathrm{m} \frac{d^\mathrm{m}}{\ell} - G_c^\mathrm{m} \ell \Delta d^\mathrm{m} = 2 \left(1 - d^\mathrm{m} \right) \mathcal{H}^{+} && \text{in } \Omega^\mathrm{m}, \label{eq:damage_1}\\
&\left(G_c^\mathrm{m} \ell \right) \boldsymbol{\nabla} d^\mathrm{m} \cdot \boldsymbol{\hat{n}} = 0 
&& \text{on } \partial \Omega^\mathrm{m} \setminus \Gamma_\text{int}, \label{eq:damage_2}\\
&d^1 = d^2 = \tilde{d} && \text{on } \Gamma_\text{int}, \label{eq:damage_3}\\
&G_c^1 \ell \boldsymbol{\nabla} d^1 \cdot \boldsymbol{\hat{n}}^1 
+ G_c^2 \ell \boldsymbol{\nabla} d^2 \cdot \boldsymbol{\hat{n}}^2 
= -\kappa \tilde{d} \left(G_c^1 + G_c^2\right) 
&& \text{on } \Gamma_\text{int}. \label{eq:damage_4}
\end{align}
\end{subequations}

Here, $\ell$ denotes the phase-field length-scale parameter. Although different length scales could, in principle, be assigned to the two bulk materials, a single value is used throughout this study to keep the parameter set minimal. Since the material interface is modeled as a sharp boundary, no additional interfacial length scale is required. The quantity $G_c$ represents the critical energy release rate, which may be defined as a piecewise continuous function taking distinct values $G_{c}^1$ and $G_{c}^2$ in the subdomains $\Omega^1$ and $\Omega^2$, respectively.  Equation~\eqref{eq:damage_2} defines flux-free boundary conditions over the external boundary. For heterogeneous materials, Equation~\eqref{eq:damage_3} enforces damage continuity at the interface. Equation~\eqref{eq:damage_4} is a mixed boundary condition at the interface, where $\kappa$ is a scalar parameter introduced earlier to control the interface strength. When $\kappa=0$, the right-hand side of Equation~\eqref{eq:damage_4} is zero, resulting in damage flux compatibility across the interface. In this case, the interface has neither damage sources nor sinks. In contrast, when $\kappa<0$, the right-hand side of Equation~\eqref{eq:damage_4} is positive, indicating the presence of a damage source at the interface. The interface, therefore, acts as a damage attractor (i.e., promotes localization of damage at the interface), consistent with the interpretation that $\kappa < 0$ corresponds to weakly bonded interfaces.

$\mathcal{H}^{+}$ is a history field variable introduced to ensure the irreversibility of damage growth (see \cite{miehe2010thermodynamically}) and is defined as 
\[ \mathcal{H}^{+} = \max_{\tau \in[0, t]} \psi^{+}(\boldsymbol{\varepsilon}(\boldsymbol{x},\tau)). \]

Note that in Equations~\eqref{eq:hybrid_r} and \eqref{eq:damage_growth}, the field variables $\boldsymbol{u}^\mathrm{m}, d^\mathrm{m},$ and $\tilde{d}$ refer to the displacements and damage fields evaluated in $\Omega^\mathrm{m}$ and at $\Gamma_\mathrm{int}$ respectively. The formulation does not explicitly solve for separate field variables in each of the subdomains and at the interface. Therefore, the history field variable as defined above ensures irreversibility of damage everywhere, i.e., both in the bulk domains and at the interface.

The equations are discretized using the standard Bubnov-Galerkin finite element approximations in their weak forms. The solution follows a staggered iterative approach until convergence, i.e., Equations~\eqref{eq:hybrid_r} are solved assuming that the damage field is frozen and subsequently, the damage evolution problem (Equations~\eqref{eq:damage_growth}) is solved for frozen displacements. The convergence criterion mandates that the $L_2$ norm of the difference between the nodal damage vectors in two consecutive iterations is less than a specified tolerance. It is worth recalling that when solved using a staggered approach, the hybrid phase-field model maintains linearity in the two sub-problems of linear momentum balance and damage evolution while simultaneously enforcing damage irreversibility, suppressing crack growth, and preventing crack face interpenetration in compression. The above framework is implemented in the deal.II finite-element library~\citep{bangerth2007deal}. Its data structures facilitate adaptive mesh-refinement and parallel computing, both of which are necessary features for large-scale three-dimensional simulations at the field scale. Although the present manuscript does not provide implementation details of the hybrid phase-field framework in deal.II for brevity, the interested reader is referred to Munshi et al.~\citep{MUNSHI2026109901} for a detailed discussion of the implementation.

\section{Numerical validation}
\label{sec:nv}
In this section, we validate the numerical framework introduced in Section~\ref{sec:methodology} using benchmark problems for which either closed-form analytical solutions or published numerical results are available. Each of the benchmark problems serves a distinct purpose. The first example validates the purely elastic framework for a pipe with an eccentric bore by comparing our numerical results with the analytical solution provided in \citet{jeffery1921ix}. Next, we assess the accuracy of the phase-field fracture formulation for weakly bonded interfaces by comparing our results with those reported in \citet{gustafsson2022phase} for a bi-material setup with a circular interface, which resembles the multimaterial wellbore geometries. Finally, we compare our results with the findings of \citet{taghipour2022novel} on a real wellbore geometry. All the simulations in this section adhere to the following assumptions. 

All two-dimensional simulations are performed under plane strain conditions. The horizontal and vertical directions in the two-dimensional simulations correspond to the \(x\)- and \(y\)-axes, respectively. Spatial discretization employs bilinear quadrilateral elements for two-dimensional analyses and trilinear hexahedral elements for three-dimensional simulations. The convergence tolerance for the staggered solution scheme is set to \(10^{-2}\). For visualization purposes, the damage surface is displayed by plotting regions where the damage field exceeds 0.9.

\subsection{Validation example 1: Elastic stress recovery in a pipe with an eccentric bore}
\label{nv:subsec1}
\begin{figure}[hbtp]
    \centering
    \begin{subfigure}[b]{0.35\textwidth}
    \centering
    \includegraphics[
    width=1\linewidth]{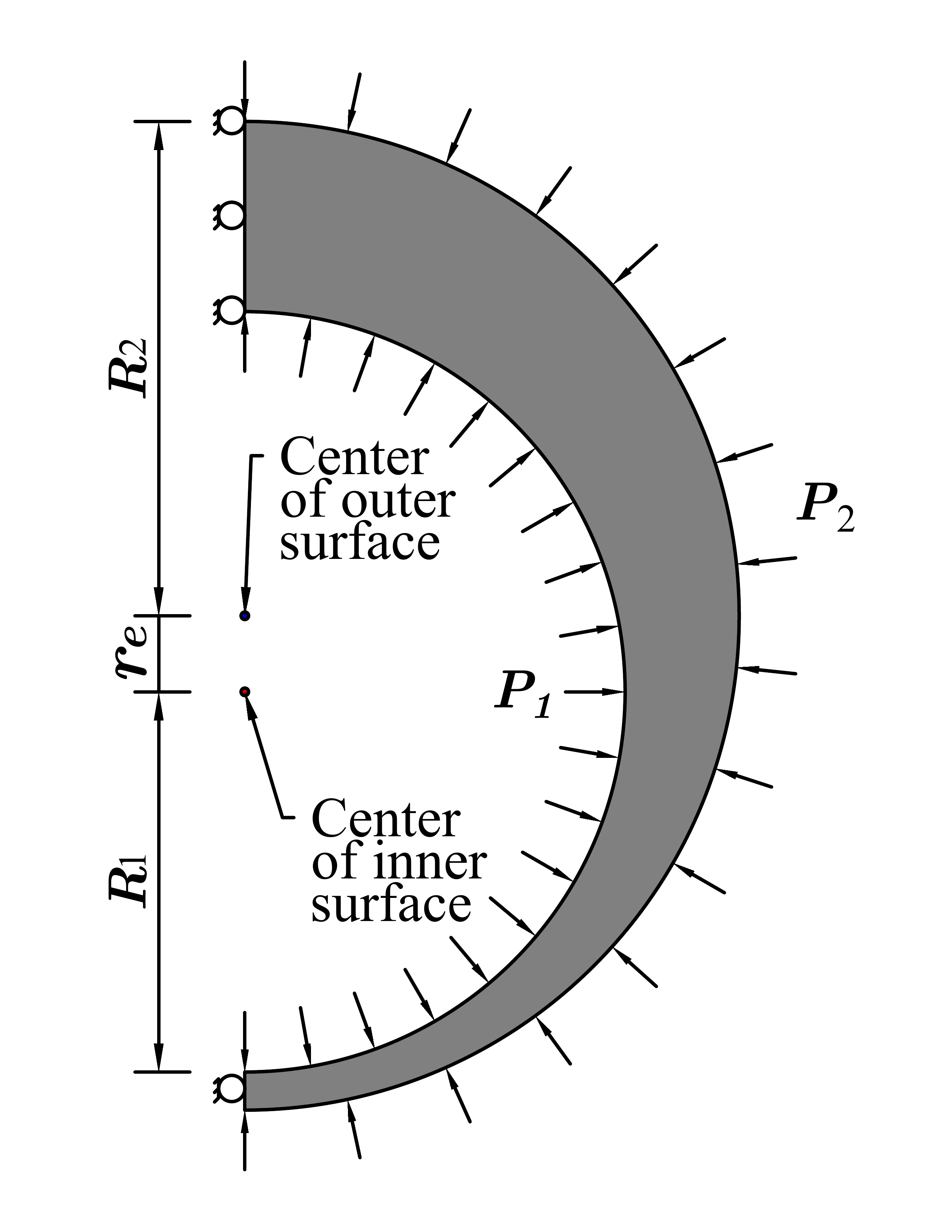}
    \caption{}
    \label{fig:jeff_model}
    \end{subfigure}
    \hfill
    \begin{subfigure}[b]{0.6\textwidth}
    \centering
    \includegraphics[
    width=1\linewidth]{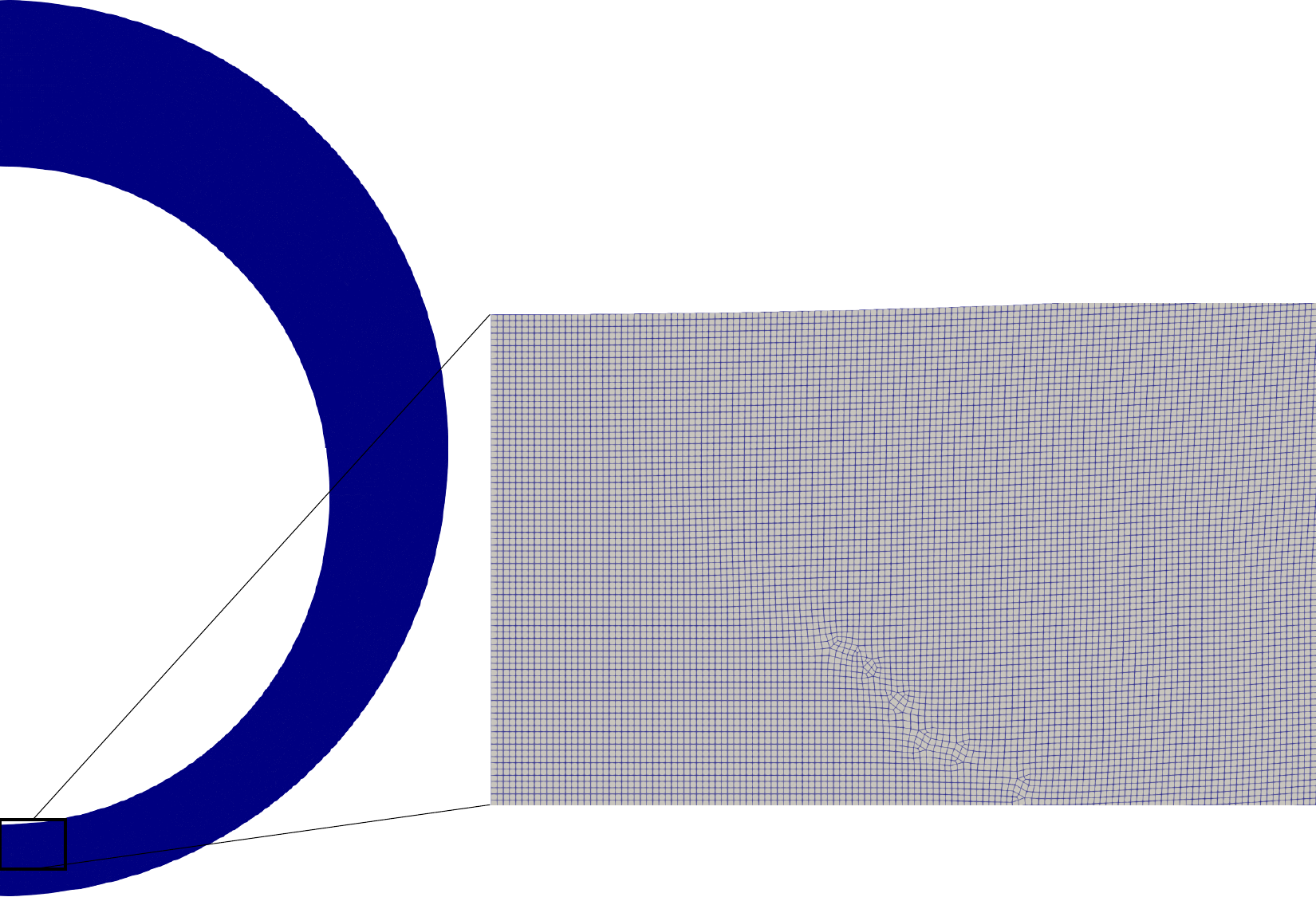}
    \caption{}
    \label{fig:jeff_mesh}
    \end{subfigure}
    \hfill
     \begin{subfigure}[b]{0.49\textwidth}
        \centering
        \includegraphics[
                    width=0.6\linewidth]{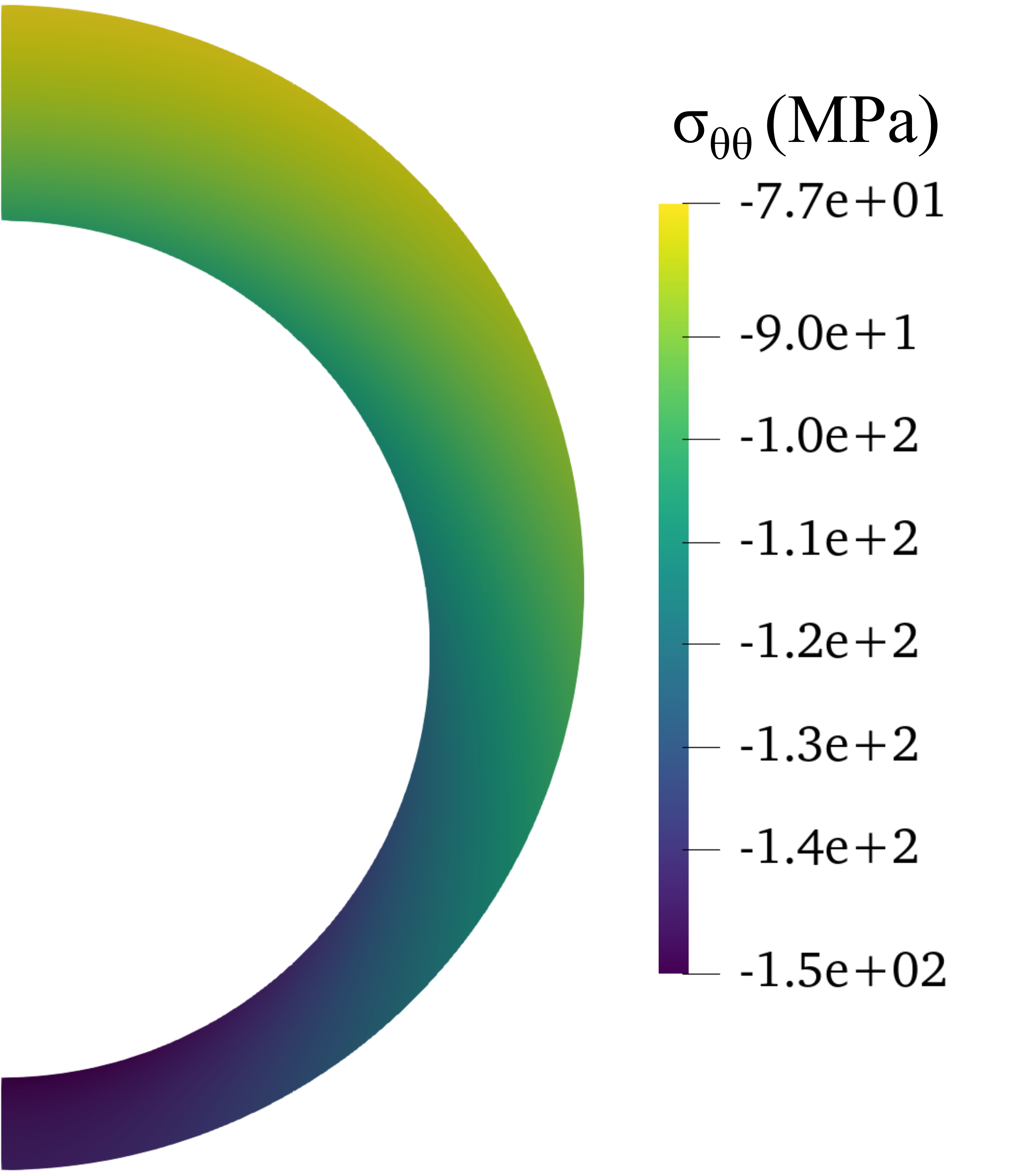}
        \caption{}
        \label{fig:jeff_circum_stress_contour}
        \end{subfigure}
      \hfill
    \begin{subfigure}[b]{0.5\textwidth}
        \centering
        \includegraphics[
    width=1\linewidth]{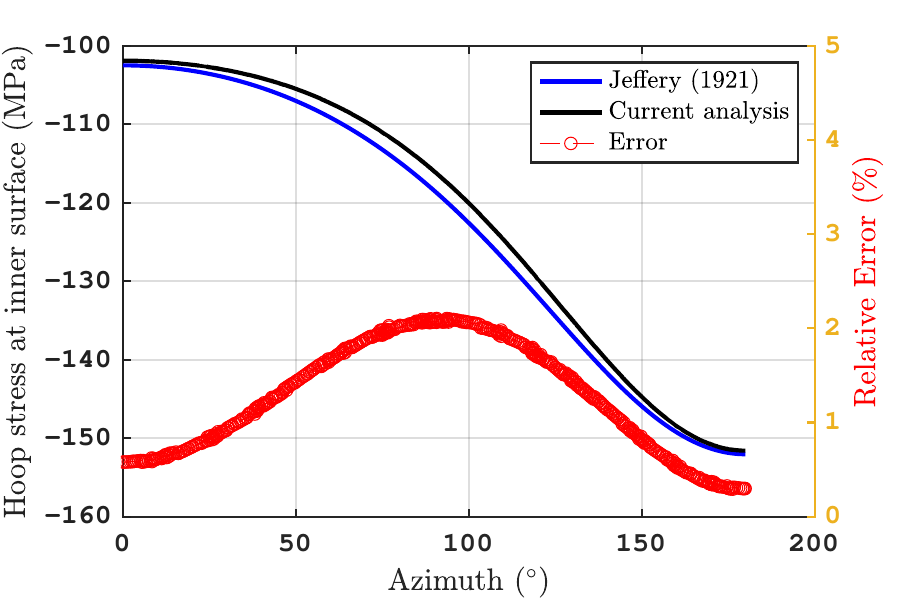}
      \caption{}
       \label{fig:jeff_circum_stress_validation_plot}
    \end{subfigure}
    \caption{(a) Geometry and boundary conditions of the pipe with eccentric bore as considered by \citet{jeffery1921ix}. For the current analysis, the dimensions of the pipe are as follows: $R_1=69.5$~mm, $R_2=94.5$~mm, $r_e=10$~mm, $e=0.4$~(40\%). (b) A mesh of unstructured bilinear quadrilateral elements with an average size of 0.06 mm was used for this simulation. (c) Distribution of the tangential stress in the cemented pipe with an eccentric bore (right half shown). Due to higher confining pressure on the outer surface, the tangential stress is compressive in nature. The stress concentration is higher at the thinner region of the pipe, as expected from the analytical solution. (d) Comparison of the tangential stress along the azimuth of the inner surface of the pipe with eccentric bore from our FEM analysis with those obtained from the analytical solution of \citet{jeffery1921ix}. The maximum error was observed to be $2.1\%$ at an azimuth of $94^\circ$.}
    \label{fig:jeff}
\end{figure}

As a first validation example, we consider the elastic deformation of a pipe with an eccentric bore pressurized at its internal and external surfaces. This example is used to verify the accuracy of the elastic formulation and stress recovery in eccentric wellbore geometries. The closed-form analytical solutions for the tangential stresses developed along the inner and outer surfaces, expressed in bipolar coordinates, are provided in \citet{jeffery1921ix}. Here, the numerical tangential stresses obtained from the finite-element model are compared with the analytical solution along the inner and outer boundaries for a specified eccentricity. For completeness, the closed-form solutions are provided in \ref{sec:app_closed_form}.

We consider a cemented pipe with an eccentric bore subjected to an internal pressure of 30~MPa and an external pressure of 50~MPa, as shown in Figure~\ref{fig:jeff_model}. The eccentricity of the inner bore is quantified using the dimensionless eccentricity parameter $e$, defined as
\begin{equation}
e = \frac{r_e}{R_{2}-R_{1}}
\label{eq:eccentricity}
\end{equation}
where $r_e$ denotes the eccentric distance between the centers of the inner and outer boundaries of the pipe. For the present case, $R_1 = 69.5$~mm and $R_2 = 94.5$~mm, with $r_e = 10$~mm, yielding $e = 0.4$ (40\%).

The elastic modulus $E$ of the cement pipe is specified as 25~GPa and a Poisson's ratio $\nu$ of 0.3 is used. The domain is discretized using bilinear quadrilateral finite elements, with mesh refinement near the inner and outer boundaries to accurately capture stress gradients. The mesh used in the simulation is shown in Figure~\ref{fig:jeff_mesh} which consists of approximately 7.6 million unstructured elements with an average size of 0.06~mm.

The variation of the tangential stress in the cement pipe is shown in Figure~\ref{fig:jeff_circum_stress_contour}. Since the outer confining pressure is higher than that of the inner pressure, the tangential stresses are compressive in nature. As expected, the stress concentration occurs at the thinner section of the pipe with an eccentric bore. Figure~\ref{fig:jeff_circum_stress_validation_plot} shows a comparison between the values of the tangential stress obtained from the closed-form expressions of~\citet{jeffery1921ix} with the FEM model used in this study. The numerical results show good agreement with the analytical solution, with a maximum relative error of 2.1\% at an azimuth of approximately $94^\circ$. A mesh convergence study is also performed to ensure that the results are insensitive to discretization -- the results of this mesh convergence study are presented in \ref{app:mesh_convergence}.

\subsection{Validation example 2: Crack propagation in a bi-material plate with round interface}
\label{nv:subsec2}

\begin{figure} [hbtp]
    \centering
    \begin{subfigure}[b]{0.46\textwidth}
    \centering
    \includegraphics[trim=45 300 50 290, clip, width=0.8\linewidth]{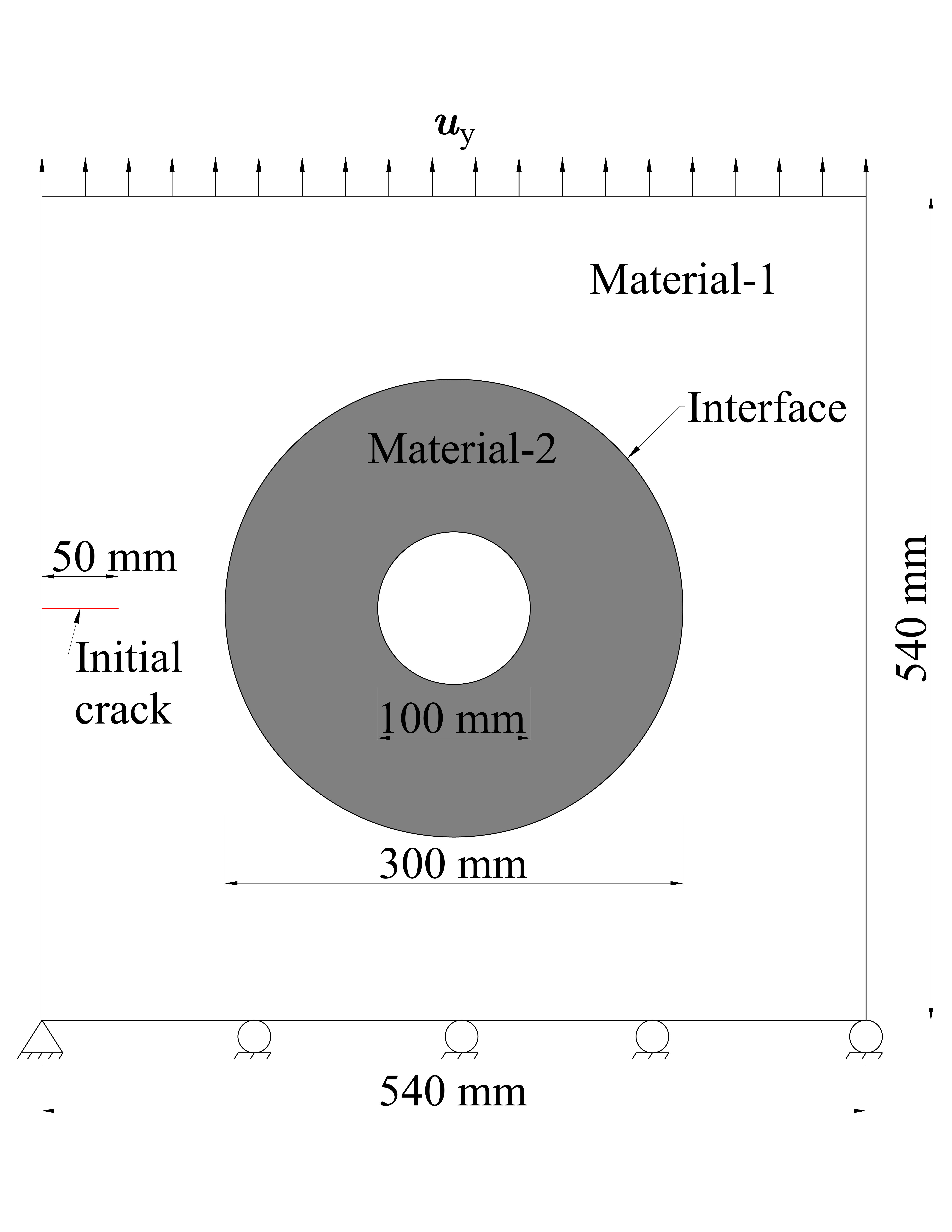}
    \caption{}
    \label{fig:geo_bmp}
    \end{subfigure}
    \hfill
    \begin{subfigure}[b]{0.49\textwidth}
    \centering
    \includegraphics[
    width=1\linewidth]{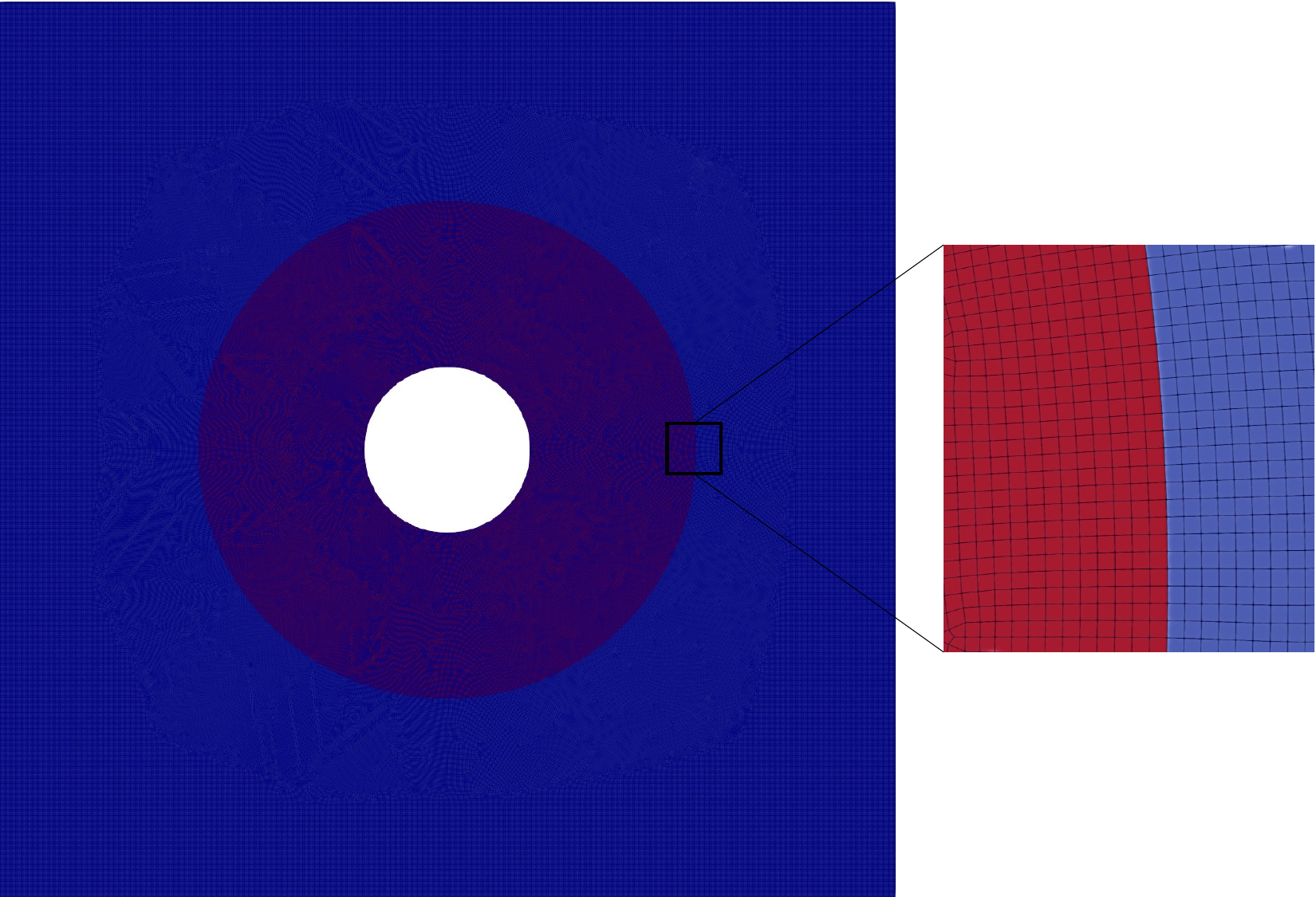}
    \hfill
    \caption{}
    \label{fig:bmp_mesh}
    \end{subfigure}
    \hfill
    \begin{subfigure}[b]{0.45\textwidth}
        \centering
        \includegraphics[width=1\linewidth]{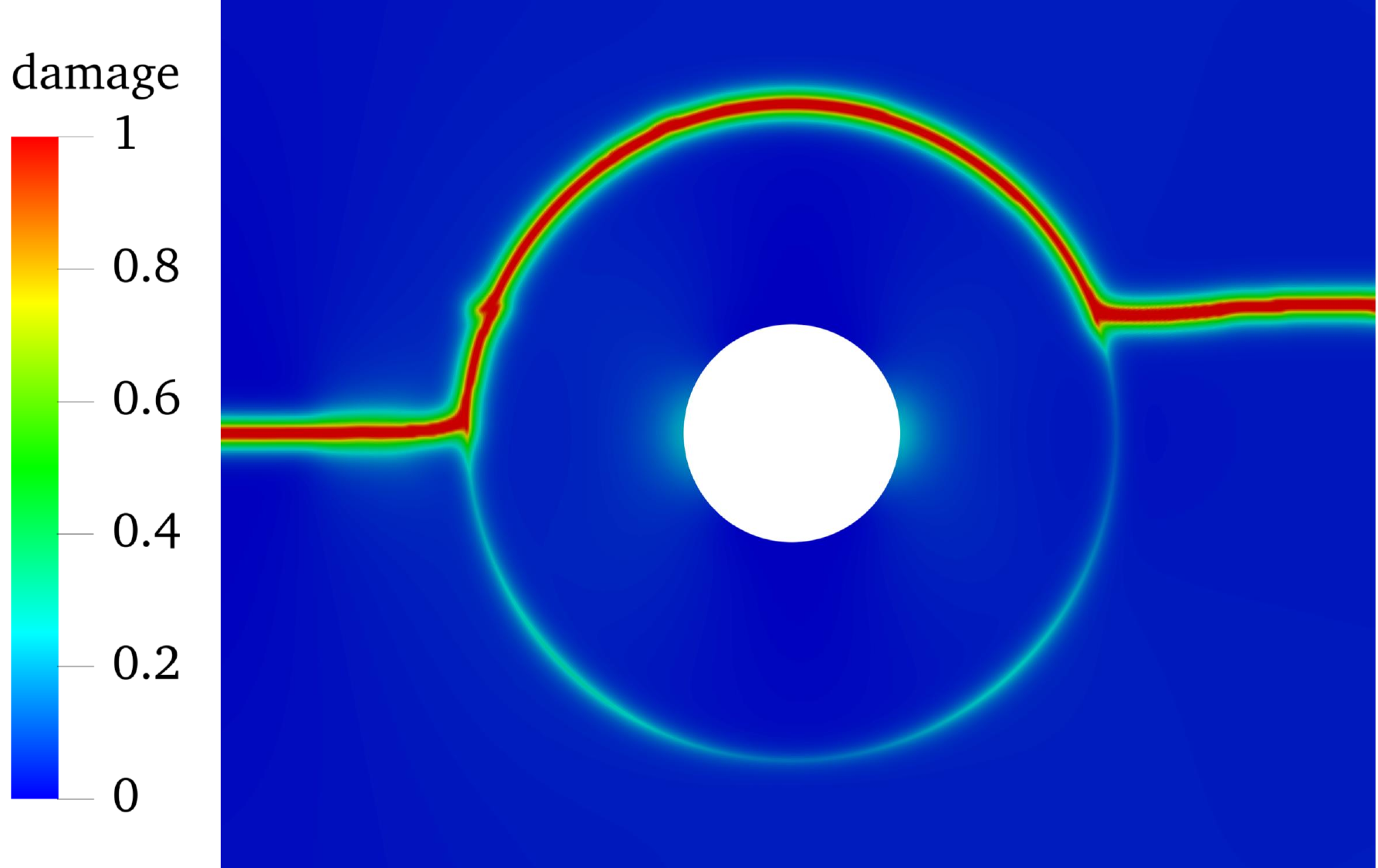}
        \caption{Present work.}
        \label{fig:subfig2}
    \end{subfigure}
    \hfill
    \begin{subfigure}[b]{0.45\textwidth}
        \centering
        \includegraphics[width=1\linewidth]{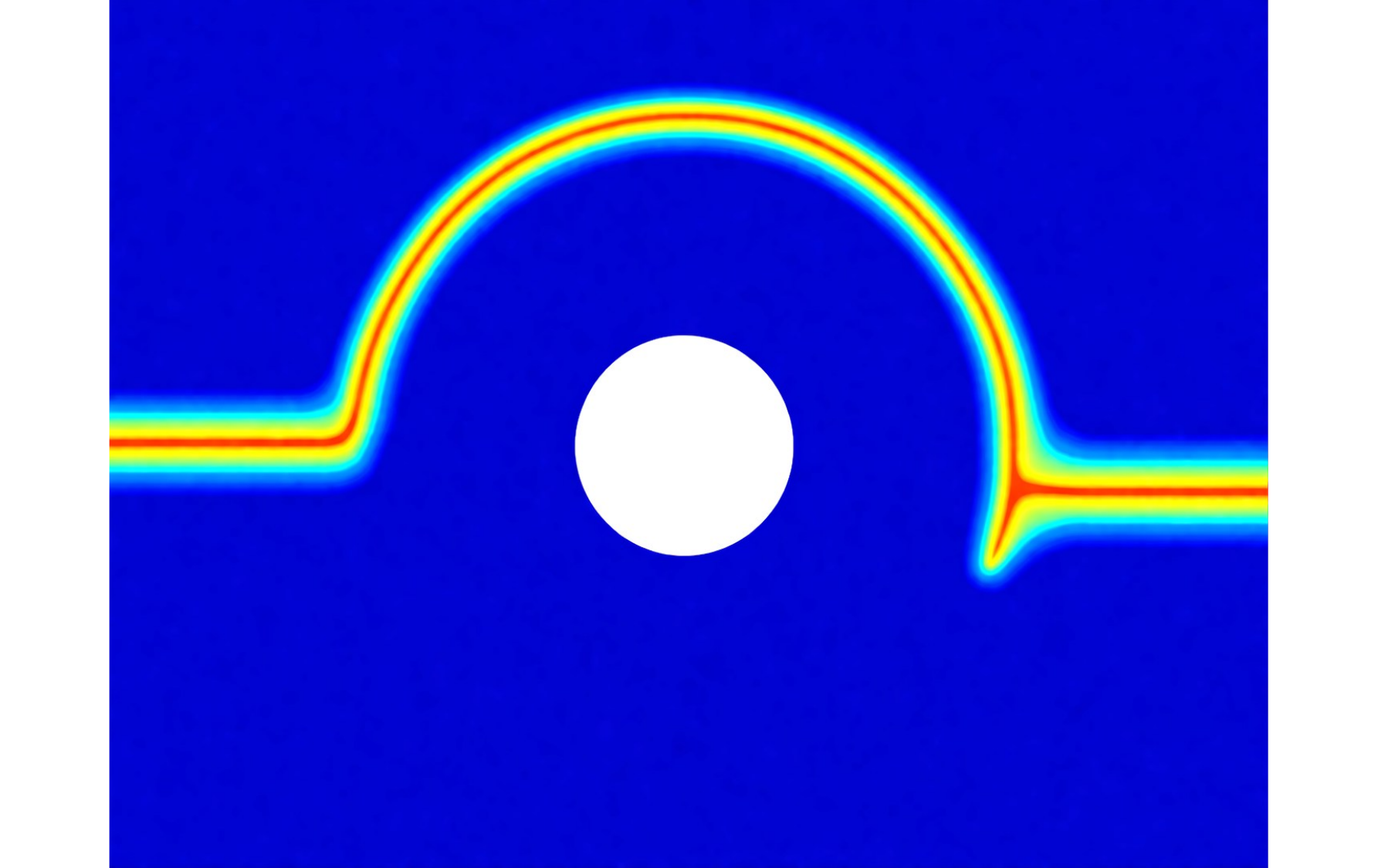}
        \caption{\citet{gustafsson2022phase}.}
        \label{fig:subfig1}
    \end{subfigure}
    \caption{(a) Geometry and boundary conditions for the bi-material plate with a circular interface and an initial crack of length 50~mm located at the mid-height of the vertical axis. The plate is pinned at the bottom-left corner, supported by rollers along the bottom edge, and subjected to a prescribed tensile displacement on the top edge. (b) A mesh of unstructured bilinear quadrilateral elements with an average size of 1.2 mm was used for this simulation. (c) and (d): Comparison of damage profiles for a bi-material square plate with a circular interface: (c) results from the present work and (d) results from \citet{gustafsson2022phase}. The damage patterns show strong qualitative agreement with the reference results. The damage profiles from~\citet{gustafsson2022phase} are reproduced with appropriate permissions from the publisher.}
    \label{fig:bimaterial_plate}
\end{figure}

\renewcommand{\arraystretch}{1.2}
\begin{table}[h!]
\small
\centering
\caption{Material and interface properties for the bi-material plate with a single edge crack. The two materials have different elastic moduli but identical bulk critical energy release rates, while the interface is significantly weaker.}
\label{table:prop_bmp}
\begin{tabular}{lccc}
\hline
\textbf{Property} & \textbf{Material 1} & \textbf{Material 2} & \textbf{Source}\\
\hline
Young's modulus, $E$ (N/m$^2$) & $15\times10^{9}$ & $25\times10^{9}$ & \citep{gustafsson2022phase}\\
Poisson's ratio, $\nu$         & 0.25             & 0.30             & \citep{gustafsson2022phase}\\
\hline
Fracture toughness, $G_c$ (N/m)          & \multicolumn{2}{c}{400} & \citep{gustafsson2022phase}\\
Effective interface toughness, $G_c^{\text{eff}}$ (N/m) 
                                         & \multicolumn{2}{c}{50} & \citep{gustafsson2022phase} \\
Weakening parameter, $\kappa$             & \multicolumn{2}{c}{-0.875} \\
Length-scale parameter, $\ell$ (m)       & \multicolumn{2}{c}{$3.75\times10^{-3}$} & \citep{gustafsson2022phase}\\
\hline
\end{tabular}
\end{table}

As a second validation example, we consider a bi-material square plate with an edge crack and a weakly bonded circular interface, first presented in \citet{gustafsson2022phase}, to validate the proposed phase-field formulation. The geometry closely resembles wellbores in the field, where a weak circular material interface exists between the cement sheath and the rock formation. The objective of this example is to validate the ability of the proposed phase-field formulation described in Section~\ref{sec:methodology} to accurately capture crack deflection and interfacial debonding in the presence of weak material interfaces.

The material properties are summarized in Table~\ref{table:prop_bmp} and are specified in accordance with the reference simulation of~\citet{gustafsson2022phase}. They report an interface fracture toughness of $G_c^\text{eff} = 50$~N/m which, using Equation~\eqref{eq:toughness}, yields $\kappa = -0.875$, indicating a strongly weakened interface relative to the surrounding bulk materials. The geometry, boundary conditions, and initial notch are illustrated in Figure~\ref{fig:geo_bmp}. The mesh used in the simulation is shown in Figure~\ref{fig:bmp_mesh} and comprises 295~thousand unstructured bilinear quadrilateral elements of average size 1.2~mm. The plate is pinned at the bottom-left corner, supported by rollers along the bottom edge, and subjected to an incrementally applied tensile displacement of $\Delta u = 10^{-6}$~mm on the top edge.

Since the applied loading is aligned with the normal to the initial notch, the crack initially propagates in Mode~I without deviation. Upon reaching the material interface, the crack may either penetrate into Material~2 or deflect along the interface, depending on the relative fracture toughness of the bulk material and the interface. For the present parameters, interfacial delamination is energetically more favorable due to the significantly lower interface toughness, and the crack is therefore expected to follow the curved interface.

Figure~\ref{fig:subfig2} shows the final damage profile obtained in the present work. The crack follows the expected trajectory, exhibiting Mode~I failure in Material~1, followed by interfacial delamination, and subsequently re-entering Material~2 under Mode~I conditions. Figure~\ref{fig:subfig1} presents the corresponding damage profile reported in \citet{gustafsson2022phase}, which shows a similar crack propagation pattern. The only noticeable difference is that, in the present study, the crack exits the interface earlier than in the reference work. This discrepancy can be attributed to differences in how the interface is modeled in the two approaches. In \citet{gustafsson2022phase}, the interface is represented as a third material with a finite thickness and a lower fracture toughness than the surrounding materials. Once the crack enters this weaker interfacial layer, it prefers to remain within it due to its lower toughness. As a result, it follows a curved path along the interface until its eventual exit at a location where the loading direction becomes normal to it. In contrast, the present analysis assumes a sharp interface with no finite thickness. Consequently, the crack leaves the weak interface earlier, as there is no extended low-toughness region to guide or delay its propagation along it. A detailed comparison of the modeling approaches is omitted for brevity. However, the interested reader is directed to the papers by~\citet{KHAN2025110672} and~\citet{munshi2025modeling} for a more detailed description of the underlying methodology. Despite these differences, the close agreement in crack trajectory and failure modes demonstrates that the proposed formulation accurately captures interfacial fracture behavior. The minor differences in crack exit location are consistent with the differing interface representations and do not affect the overall physical interpretation, thereby validating the robustness of the present approach.

\subsection{Validation example 3: Crack propagation in a wellbore}
\label{nv:subsec3}
\begin{figure} [hbtp]
    \centering
    \begin{subfigure}{0.46\textwidth}
    \centering
        \includegraphics[trim=0 50 0 50, clip, width=1\linewidth]{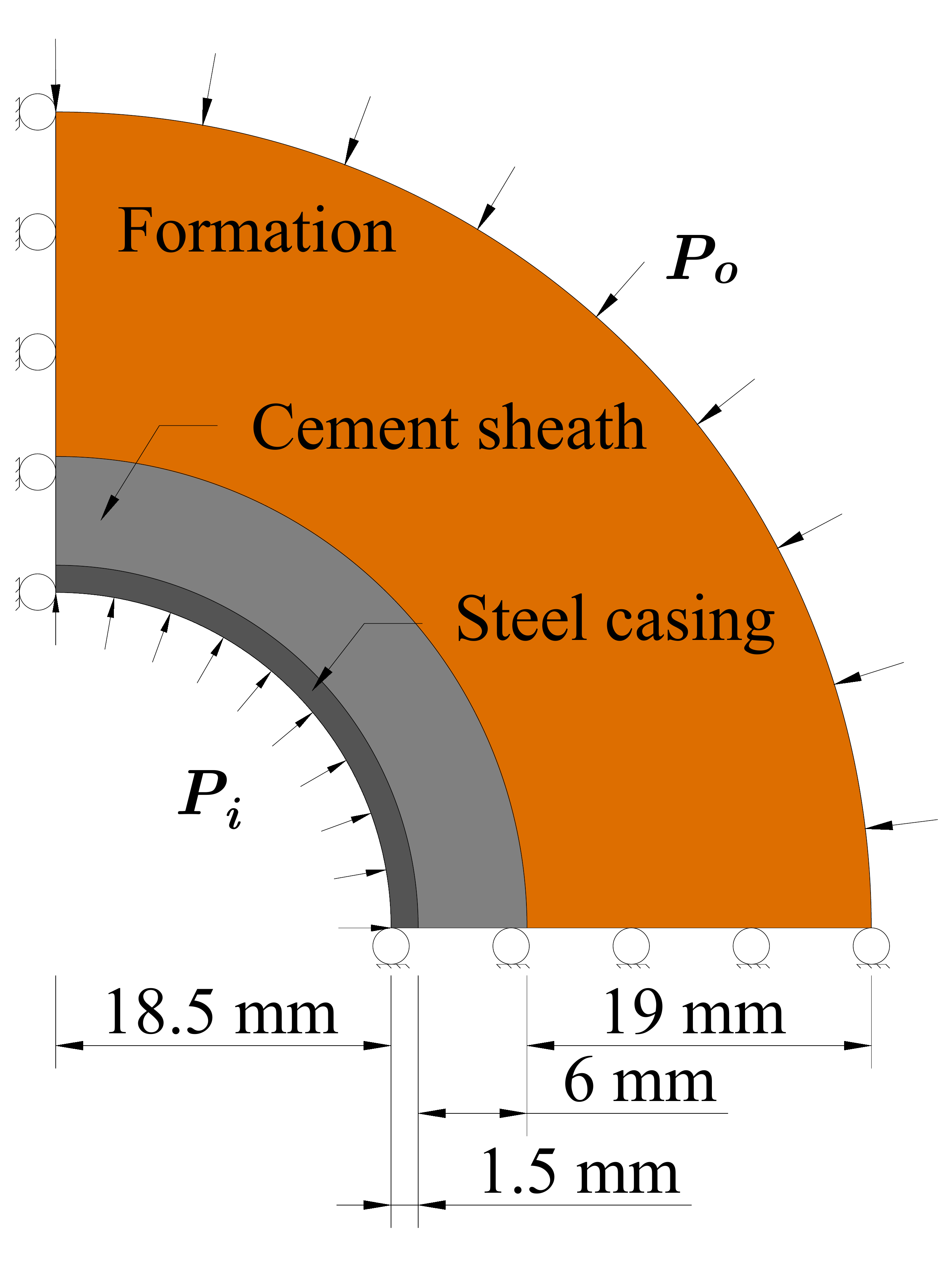}
        \caption{}
        \label{fig:wellbore_tag}        
    \end{subfigure}
    \hfill
    \begin{subfigure}{0.49\textwidth}
    \centering
        \includegraphics[
        width=1\linewidth]{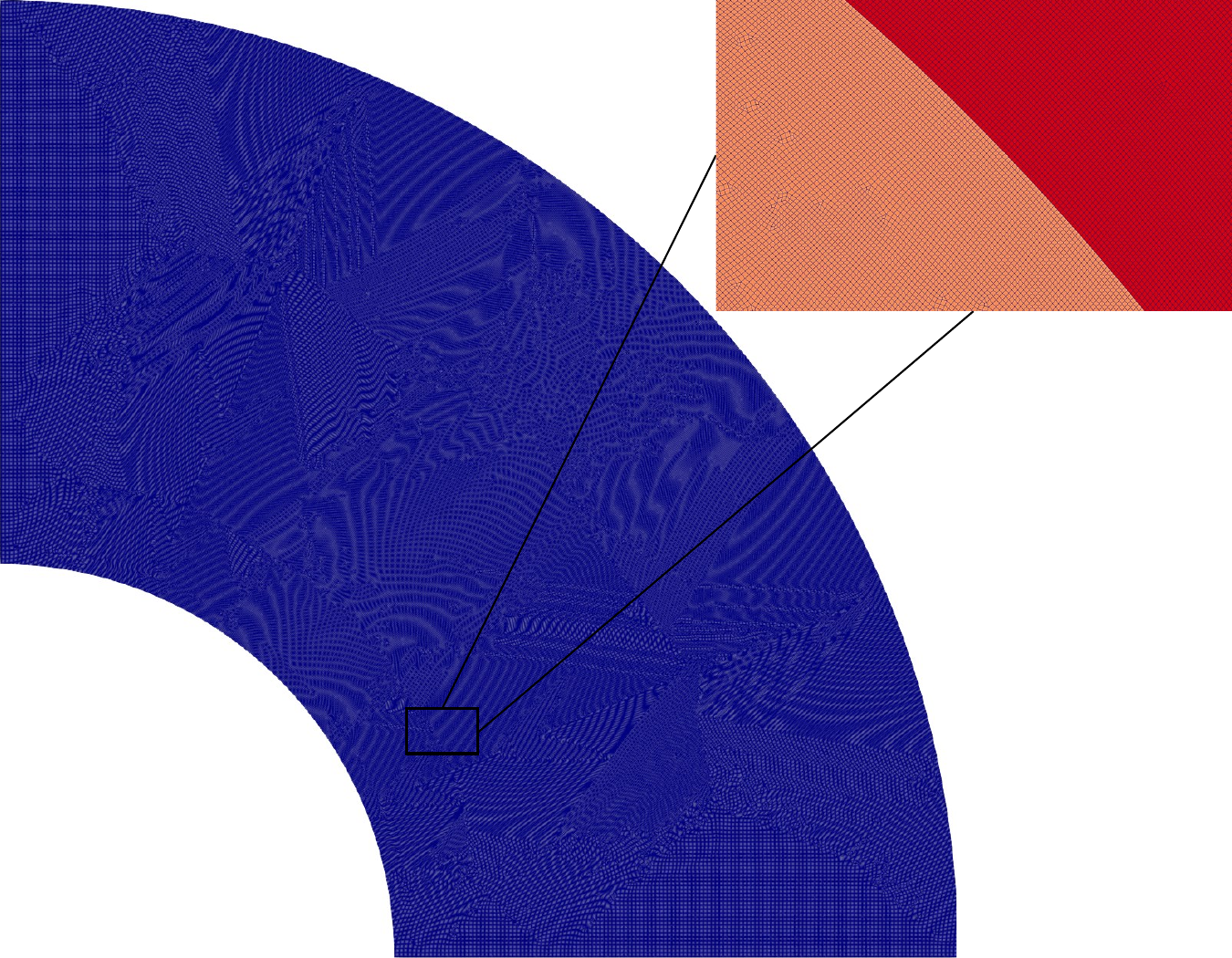}
        \hfill
      
        \caption{}
        \label{fig:wellbore_tag_mesh}         
    \end{subfigure}
    \hfill
    \begin{subfigure}{0.49\textwidth}
        \centering
        \includegraphics[width=1\linewidth]{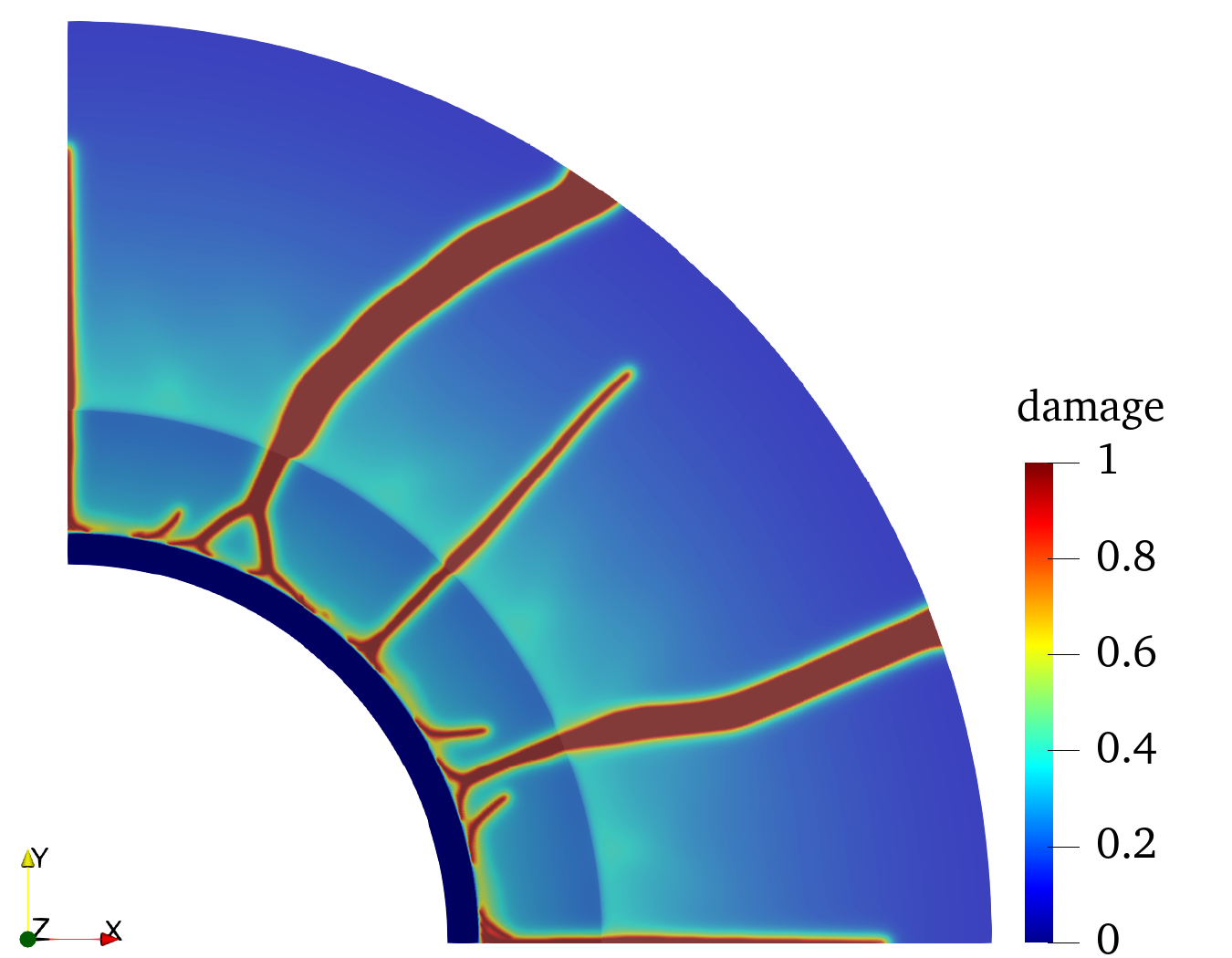}
        \caption{Present work.}
        \label{fig:wellbore_final}
    \end{subfigure}
    \begin{subfigure}{0.49\textwidth}
        \centering
        \includegraphics[width=1\linewidth]{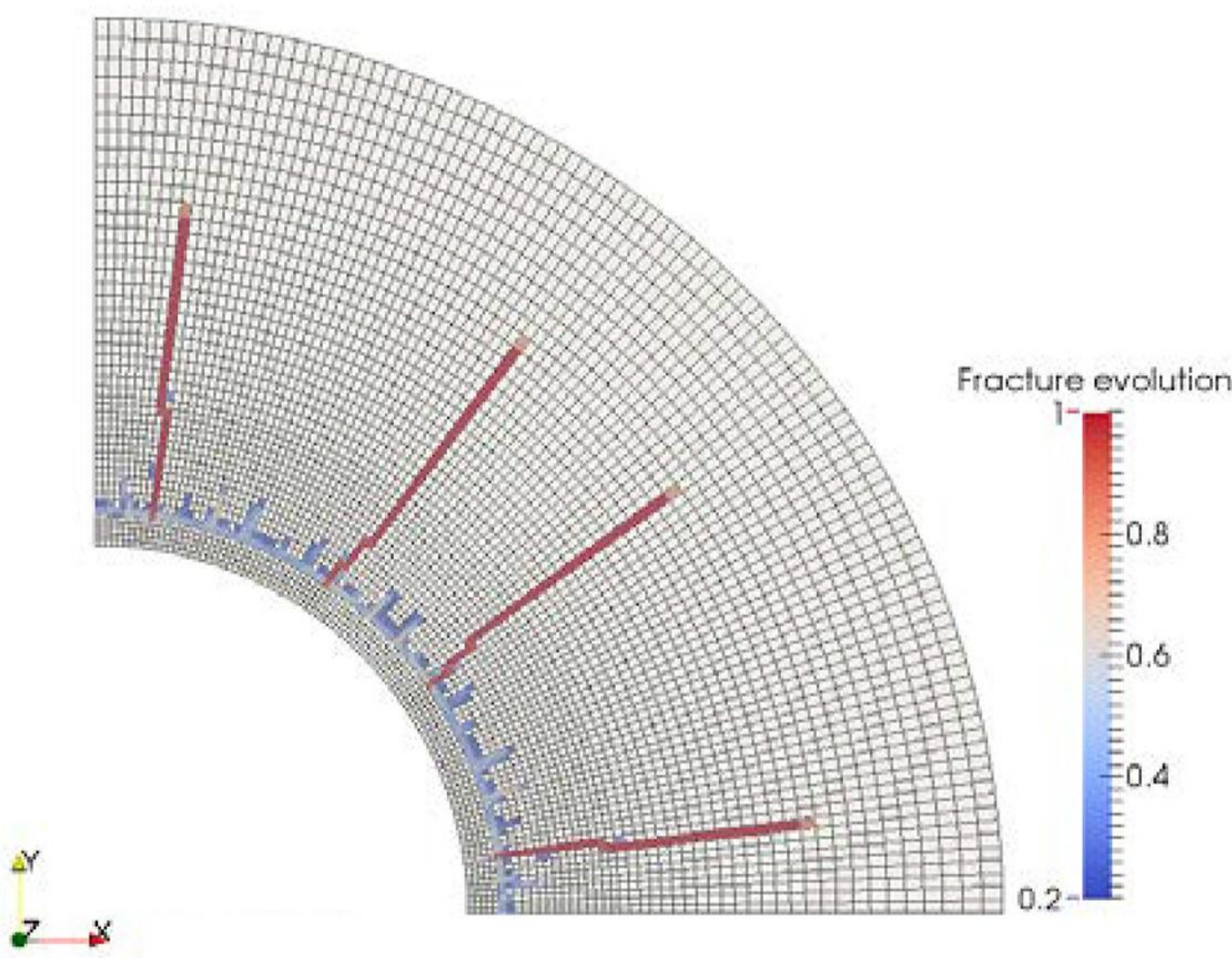}
        \caption{Result from \citet{taghipour2022novel}.}
        \label{fig:wellbore_tag_result}
    \end{subfigure}
    \caption{(a) Geometry and boundary conditions of one quadrant of the wellbore consisting of a steel casing, cement sheath, and the rock formation. The straight edges of the quadrant are on rollers. The wellbore is subjected to internal wellbore pressure $P_i$ and confining pressure $P_o$. (b) A mesh of unstructured bilinear quadrilateral elements with an average size of 0.09 mm was used for this simulation. (c) and (d): Comparison of damage profiles for the wellbore fracture problem: (c) present work and (d) \citet{taghipour2022novel}. Radial cracks develop in the cement sheath and propagate into the surrounding rock formation, while microcracking and interfacial debonding are observed predominantly near the casing-cement interface. The figure from \citet{taghipour2022novel} is reproduced with appropriate permissions from the publisher.}
    
\end{figure}

\renewcommand{\arraystretch}{1.4}
\begin{table}[h!]
\small
\centering
\caption{Material properties for the wellbore problem discussed in Section~\ref{nv:subsec3}.}
\label{table:prop_wellbore_1}
\begin{tabular}{ c c c c c}
\hline
\textbf{Properties} & \textbf{ Rock Formation} & \textbf{Cement sheath} & \textbf{Steel Casing} & \textbf{Source} \\
\hline
E (N/m$^2$) & 2$\times$10$^9$ & 8$\times$10$^9$ & 2$\times$10$^{11}$ & \citep{taghipour2022novel}\\
\hline
$\nu$ & 0.25 & 0.15 & 0.3 &\citep{taghipour2022novel}\\
\hline
$G_c$ (N/m) & 4 & 46.5 & $10^5$ & \citep{Cusini_2287725,cui2024computational}\\
\hline
$\ell$ (m) & 27$\times$10$^{-5}$ & 27$\times$10$^{-5}$ & 27$\times$10$^{-5}$ \\
\hline
\end{tabular}
\end{table}

As a final validation example, we consider a benchmark wellbore fracture problem reported in \citet{taghipour2022novel} to assess the capability of the proposed phase-field framework to capture fracture initiation, propagation, and interfacial debonding in realistic wellbore geometries. The model represents a wellbore system comprising a steel casing, a cement sheath, and the surrounding Castlegate sandstone formation. The mechanical properties of the steel casing, cement sheath, and sandstone formation, including the phase-field length scale, are summarized in Table~\ref{table:prop_wellbore_1}, with the fracture toughness values adopted from \citet{Cusini_2287725} for rock and from \citet{cui2024computational} for steel casing, respectively. The phase-field length scale was considered as $\ell = 0.27$ mm which is less than one-hundredth of the domain size as recommended in literature~\cite{Wu2020}.

Only one quarter of the wellbore system is modeled due to symmetry, as shown in Figure~\ref{fig:wellbore_tag}. The domain was discretized using non-uniform bilinear quadrilateral elements consisting of 700~thousand elements with average size 0.09~mm (one-third of the length scale) as shown in Figure~\ref{fig:wellbore_tag_mesh}. The symmetry boundary conditions on the left and bottom edges of the model result in normal displacement constraints enforced using roller boundary conditions. The confining pressure, $P_o$, is maintained constant at 1~kPa, while the inner wellbore pressure, $P_i$, is increased incrementally by 50~kPa at each load step. Here, $P_i$ represents the differential pressure between the internal wellbore pressure and the in-situ geostatic stress. This formulation enables efficient simulation of crack initiation and growth while preserving the essential mechanical response of the system.

The damage profiles obtained from the present work are shown in Figure~\ref{fig:wellbore_final} and are compared with the results reported in \citet{taghipour2022novel}, as shown in Figure~\ref{fig:wellbore_tag_result}. Very good agreement is observed between the present results and those of~\citet{taghipour2022novel}. Both approaches predict radial cracking in the surrounding rock formation, along with microcracking and interfacial debonding near the casing-cement interface. As the wellbore pressure increases, tensile circumferential stresses intensify, driving radial crack propagation through the cement sheath and into the surrounding rock formation. Once the crack reaches the steel casing, its significantly higher fracture toughness inhibits penetration, resulting in crack deflection along the casing-cement interface and eventual interfacial debonding. These results confirm that the proposed formulation accurately captures fracture behavior in multi-material wellbore systems with strong material contrasts and weak interfaces.

Overall, the results from all three validation examples demonstrate the robustness and accuracy of the proposed framework in capturing elastic response, interfacial fracture, and crack propagation in complex multi-material systems.

\section{Numerical experiments}
\label{sec:ne}
We now conduct a series of numerical experiments to investigate the effects of casing eccentricity and weak interfaces on wellbore integrity. The section concludes with a three-dimensional simulation of crack propagation in a wellbore system. The selected conditions represent a controlled, downscaled segment of the near-wellbore region rather than a full-length field well. The radial dimensions reproduce the laboratory-scale casing-cement-rock configuration of Taghipour et al.~\citep{taghipour2022novel}, while the 100 mm depth used in the three-dimensional model represents a short axial interval over which local crack nucleation, interaction, and coalescence can be resolved. The straight, circular geometry corresponds to the conventional thick-cylinder idealization of a nominally gauge, locally straight well interval~\cite{Bourgoyne1986}. Starting from this baseline, casing eccentricity represents azimuthal cement-thickness variation caused by inadequate centralization, well inclination, or borehole rugosity, whereas interface weakening represents poor cement-rock bonding associated with mud residue, voids, shrinkage, or non-uniform curing. The imposed internal pressure is interpreted as a differential pressure excursion relative to the in-situ stress state, as may occur during pressure testing, stimulation, injection, or production. Actual wells may additionally exhibit borehole ovality, washouts,  spatially variable cement properties, rock formation heterogeneity, and longer-scale axial variations; therefore, the results are intended to identify fracture mechanisms and comparative trends rather than provide direct field-scale failure-pressure predictions. It is also worth highlighting that interactions between fractures originating from adjacent wells are beyond the scope of the present study. The assumptions stated in the preamble of Section~\ref{sec:nv} are retained for the simulations presented here and are not repeated for brevity.

\begin{figure}[hbtp]
    \centering
    \includegraphics[trim= 0 66 0 63, clip, width=0.4\linewidth]{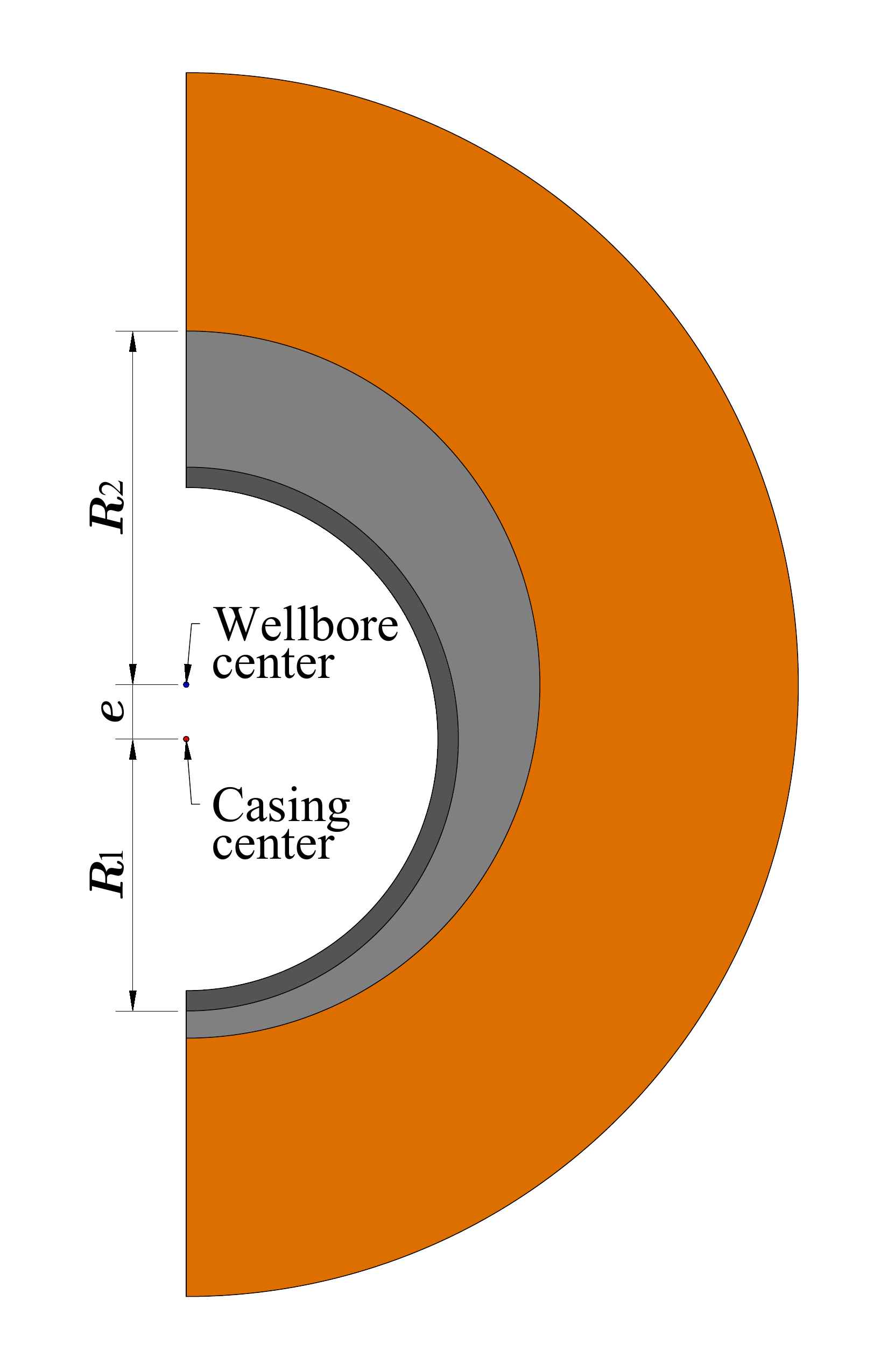}
    \caption{Geometry of the wellbore system with an eccentric casing (right half is shown). The center of the casing is offset by a distance $e$ from the center of the wellbore.}
    \label{fig:ecc_layout}
\end{figure}

\subsection{Effects of casing eccentricity}
\label{ne:subsec1}

\begin{figure}[hbtp]
    \centering
    \begin{subfigure}[b]{0.8\textwidth}
        \centering
        \includegraphics[width=0.55\linewidth]{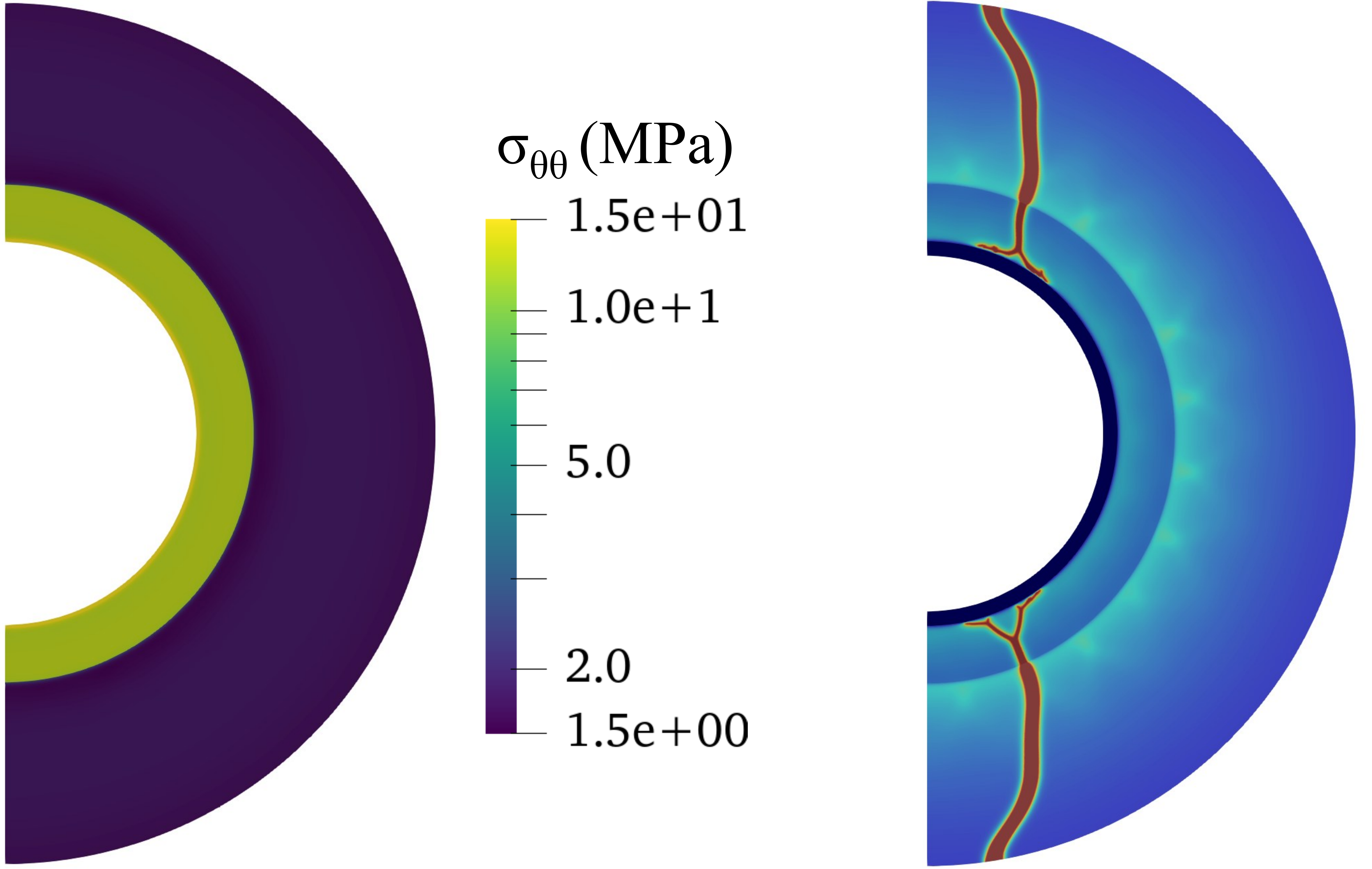}
        \caption{$e=0\%$}
        \label{fig:e=0_nuc}
    \end{subfigure}\\
    
    \begin{subfigure}[b]{0.8\textwidth}
        \centering
        \includegraphics[width=1\linewidth]{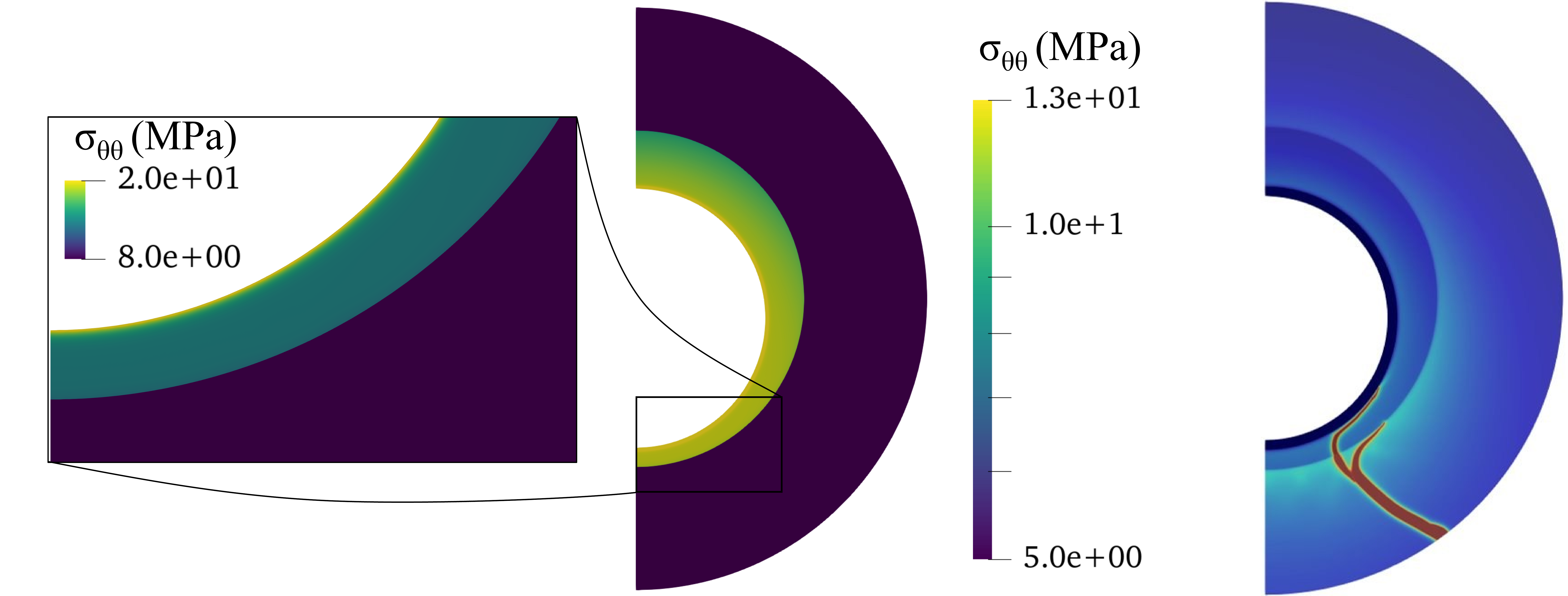}
        \caption{$e=50\%$}
        \label{fig:e=3_nuc}
    \end{subfigure} \\
    
    \begin{subfigure}[b]{0.8\textwidth}
        \centering
        \includegraphics[width=1\linewidth]{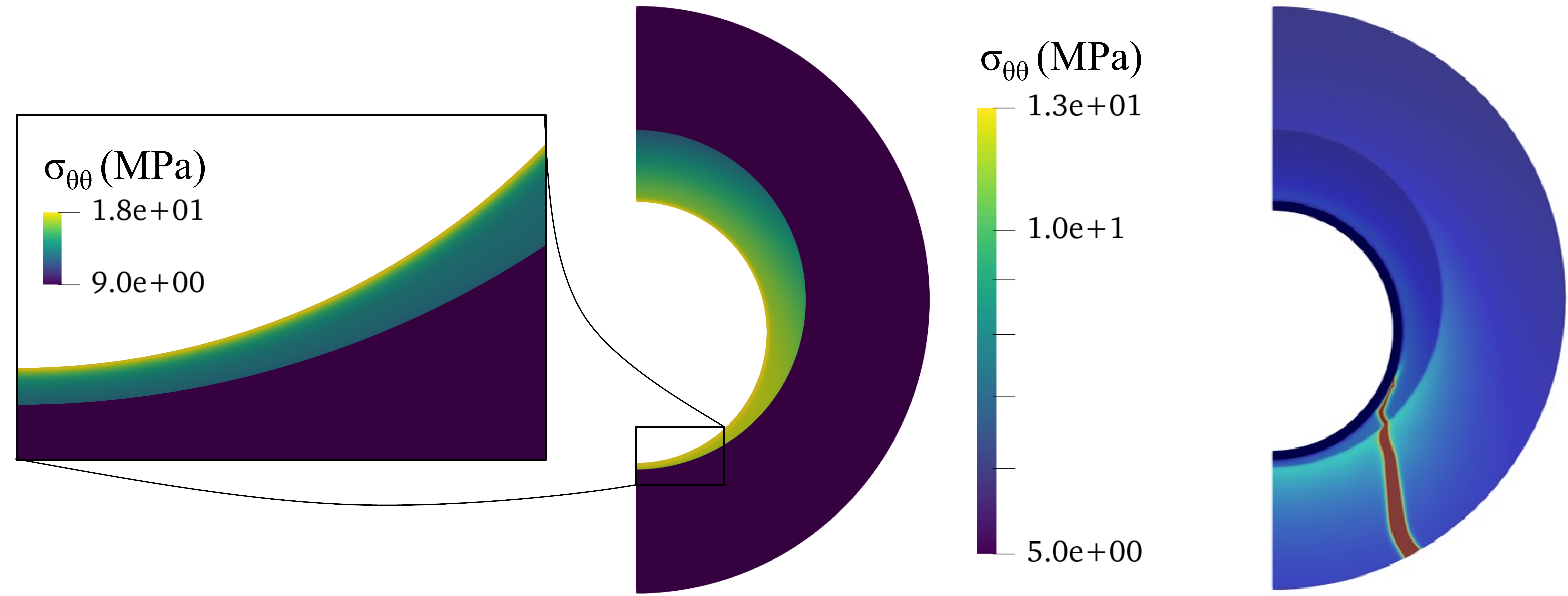}
        \caption{$e=83.33\%$}
        \label{fig:e=5_nuc}
    \end{subfigure}
    \caption{For three eccentricity cases ($e=0, 50\%, 83.33\%$), the subfigures show (i) tangential stress in the wellbore prior to crack nucleation plotted on logarithmic scale, and (ii) the corresponding damage profile at crack nucleation.}
    \label{fig:nuc}
\end{figure}

Casing eccentricity arises when the steel casing is not concentrically positioned within the wellbore, resulting in a non-uniform cement sheath in the annular gap between the casing and the borehole. As previously observed in Section~\ref{nv:subsec1}, this geometric asymmetry leads to uneven stress distributions under internal pressure, often creating regions of high tensile stresses in the cement sheath and promoting early cracking. In the following numerical example, we examine the effect of casing eccentricity on fracture behavior in a heterogeneous wellbore system consisting of a steel casing, cement sheath, and the surrounding rock formation.

A schematic of the problem is shown in Figure~\ref{fig:ecc_layout}. Due to the eccentric placement of the casing, the problem is symmetric only about the vertical plane; therefore, only half of the domain is modeled. The analysis is performed for several values of casing eccentricity, namely $e = 0, e = 0.1667, e = 0.3333, e = 0.50, e = 0.6667, e = 0.75,$ and $e=0.8333$. As discussed earlier, the eccentricity $e$ is calculated using Equation~\eqref{eq:eccentricity}. The results of the simulation are presented below with an emphasis on the mechanical response after the onset of damage.

As the internal wellbore pressure increases, cracks initiate in the cement sheath and propagate radially outward. Crack nucleation is governed primarily by the tensile circumferential (tangential) stress induced by the applied internal pressure. Figure~\ref{fig:nuc} illustrates the distribution of tangential stress in the cement sheath and surrounding rock immediately before crack initiation for each level of casing eccentricity. When $e=0$, i.e., for the concentric casing configuration, the tensile tangential stresses are uniformly distributed in both the cement sheath and the rock formation with larger magnitudes in the cement sheath and the steel casing as shown in Figure~\ref{fig:e=0_nuc}. Crack initiation is, therefore, expected to occur in the cement sheath, given its lower fracture toughness. However, because the stresses are uniform throughout the cement sheath, the precise location of crack initiation is dictated solely by numerical round-off, which effectively represents microscopic material imperfections. In contrast, increasing casing eccentricity results in the development of stress concentrations in the thinner region of the cement sheath and the adjacent rock. Consequently, damage concentrates in these high-stress regions and crack nucleation occurs there for eccentric wellbore configurations as shown in Figures~\ref{fig:e=3_nuc} and \ref{fig:e=5_nuc}.

\begin{figure}[hbtp]
    \centering
    \begin{subfigure}[b]{0.24\textwidth}
        \centering
        \includegraphics[width=0.75\linewidth]{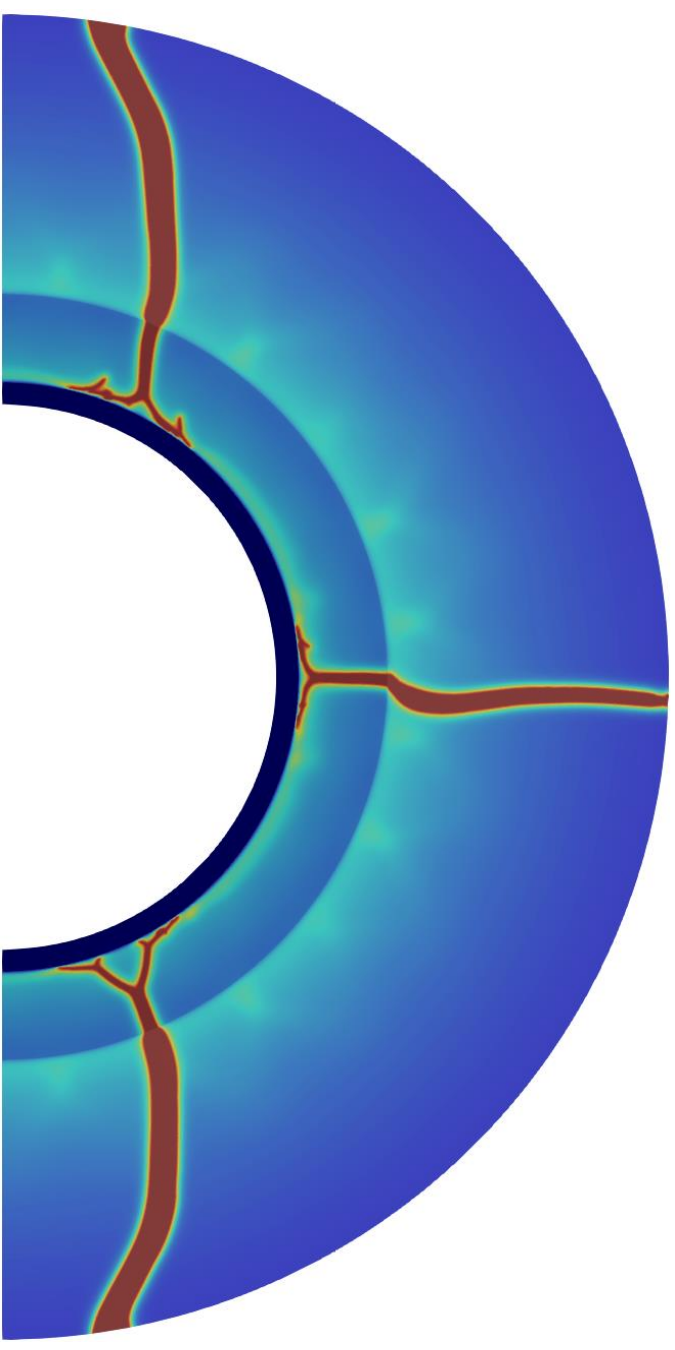}
        \caption{$e=0\%$}
        \label{fig:e=0_194}
    \end{subfigure}
    \hfill
    \begin{subfigure}[b]{0.24\textwidth}
        \centering
        \includegraphics[width=0.75\linewidth]{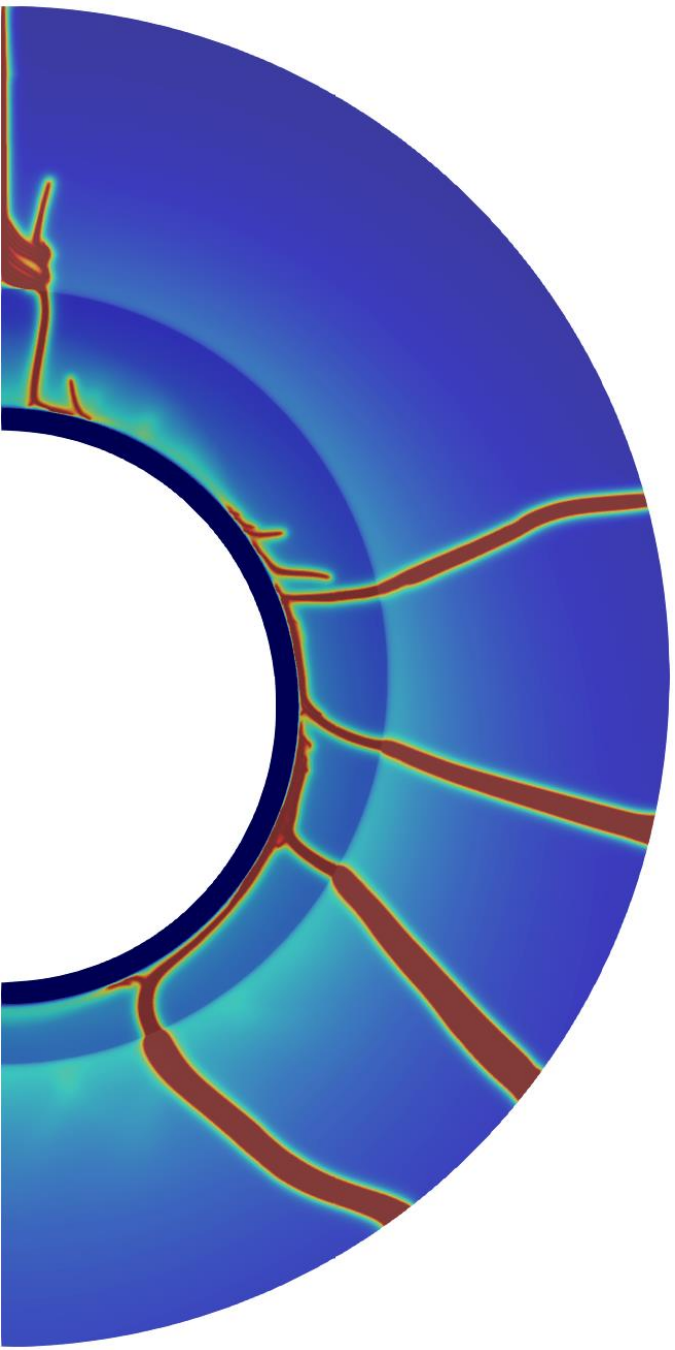}
        \caption{$e=33.33\%$}
        \label{fig:e=2_194}
    \end{subfigure}
    \hfill
    \begin{subfigure}[b]{0.24\textwidth}
        \centering
        \includegraphics[width=0.75\linewidth]{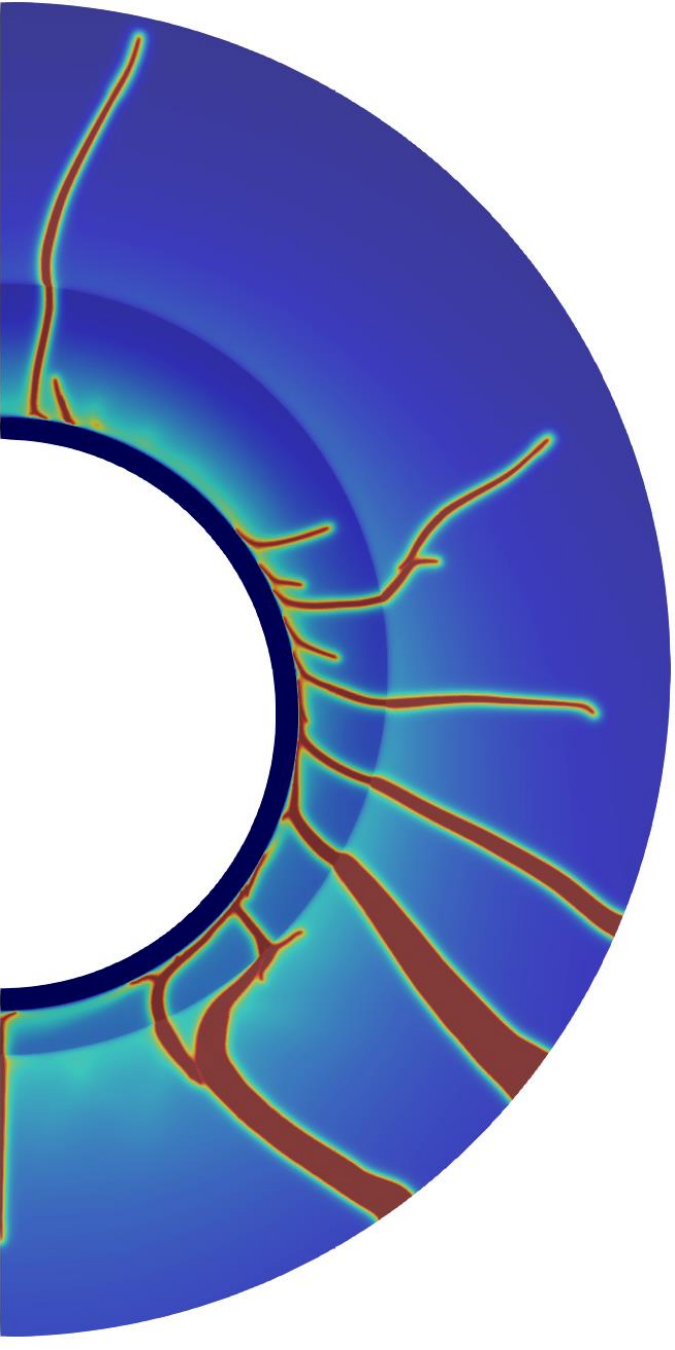}
        \caption{$e=50\%$}
        \label{fig:e=3_194}
    \end{subfigure}
    \hfill
    \begin{subfigure}[b]{0.24\textwidth}
        \centering
        \includegraphics[width=0.75\linewidth]{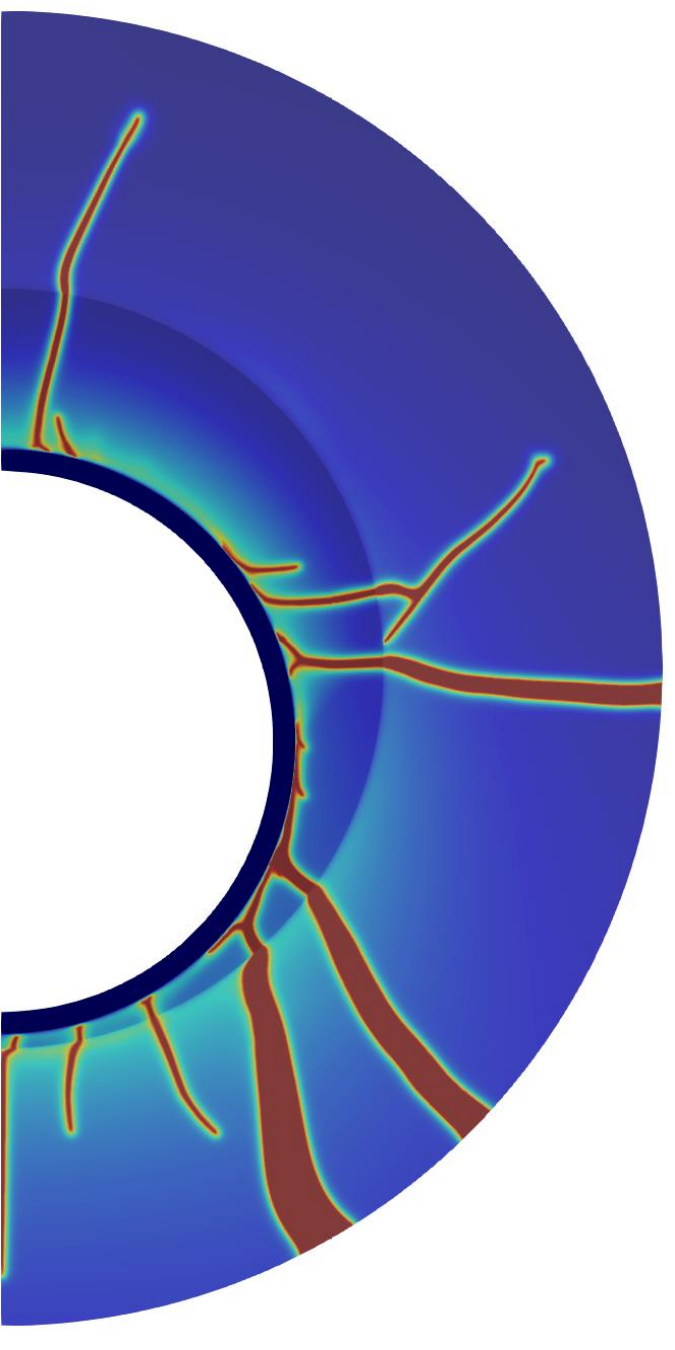}
        \caption{$e=83.33\%$}
        \label{fig:e=5_194}
    \end{subfigure}
    \centering
    \begin{subfigure}[b]{0.9\textwidth}
        \centering
        \includegraphics[width=0.4\linewidth]{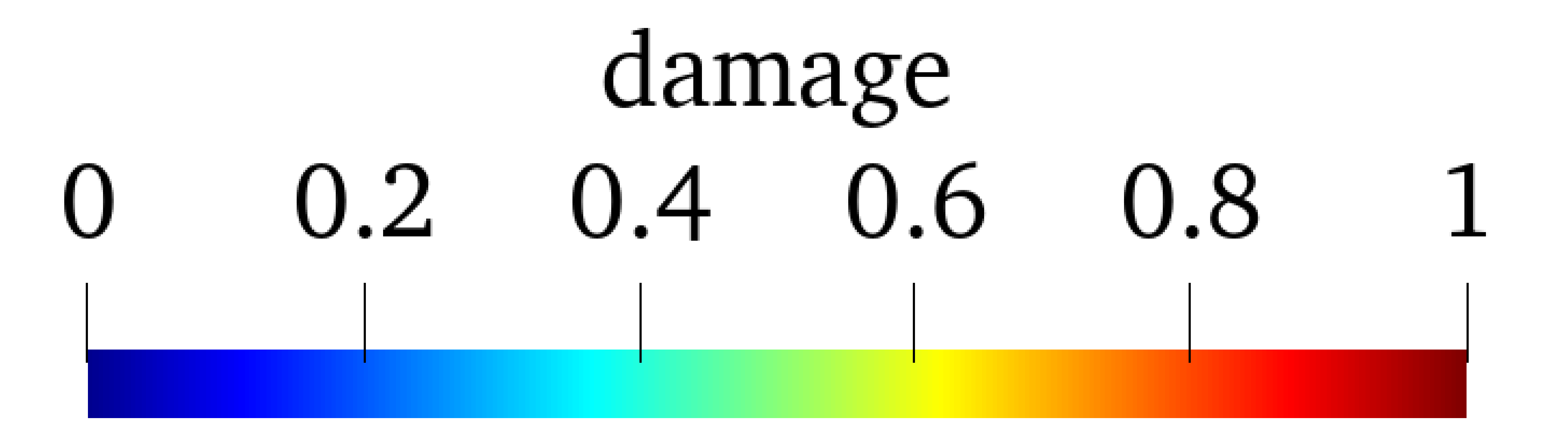}
    \end{subfigure}
    \centering
    \begin{subfigure}[b]{1\textwidth}
        \centering
        \includegraphics[width=0.6\linewidth]{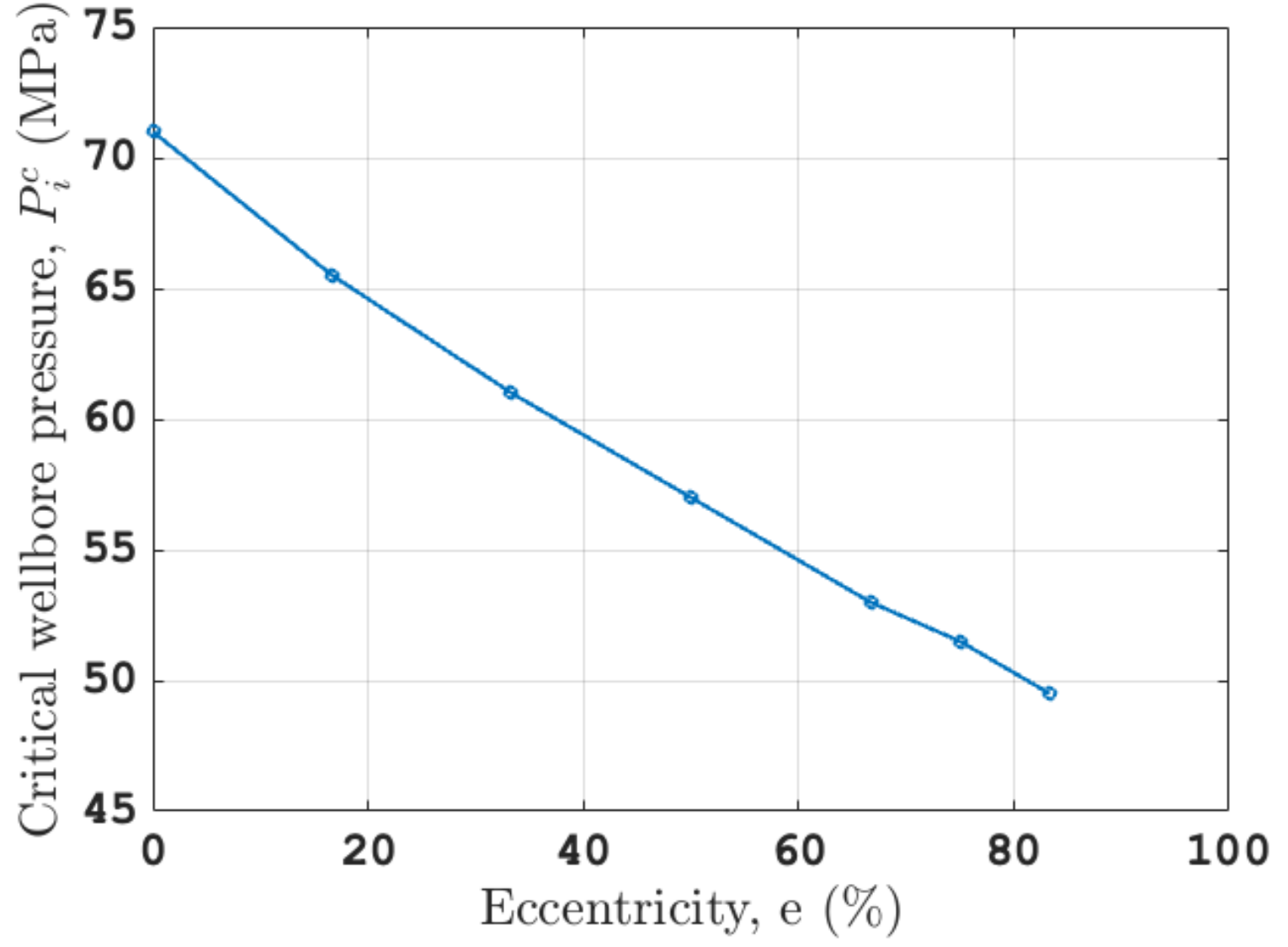}
        \caption{$P_i^c~\mathrm{versus}~e$}
        \label{fig:eccentricity_vs_pressure}
    \end{subfigure}
    \caption{(a) to (d): Damage profiles of wellbores with varying eccentricity values at a wellbore pressure of 97 MPa. The loss in symmetry of the damage pattern with increasing eccentricity can be observed. (e):  Plot of the casing eccentricity versus the critical wellbore pressure at which the crack initiates. With increasing casing eccentricity, stress concentration occurs at lower wellbore pressure, leading to early crack initiation in wellbores with eccentric casing.}
    \label{fig:194}
\end{figure}

The threshold value of the internal wellbore pressure ($P_i^c$) corresponding to crack initiation for the different eccentricities considered is reported in Table~\ref{table:nuc_pressure}. Interpreting this pressure as a measure of wellbore strength, the results clearly demonstrate an inverse relationship between casing eccentricity and wellbore strength as shown in Figure~\ref{fig:eccentricity_vs_pressure}.

\renewcommand{\arraystretch}{1.4}
\begin{table}[h!]
\small
\centering
\caption{Threshold value of wellbore pressure ($P_i^c$) at crack initiation for different values of casing eccentricity ($e$).}
\label{table:nuc_pressure}
\begin{tabular}{ c c }
\hline
\textbf{Eccentricity, $e (\%)$} & \textbf{Wellbore pressure, $P_i^c$ (MPa)}  \\
\hline
0 & 71 \\
\hline
16.67 & 65.5 \\
\hline
33.33 & 61 \\
\hline
50 & 57 \\
\hline
66.67 & 53 \\
\hline
75 & 51.5 \\
\hline
83.33 & 49.5 \\
\hline
\end{tabular}
\end{table}

\begin{figure}[hbtp]
    \begin{subfigure}[b]{0.46\textwidth}
        \centering
        \includegraphics[width=1\linewidth]{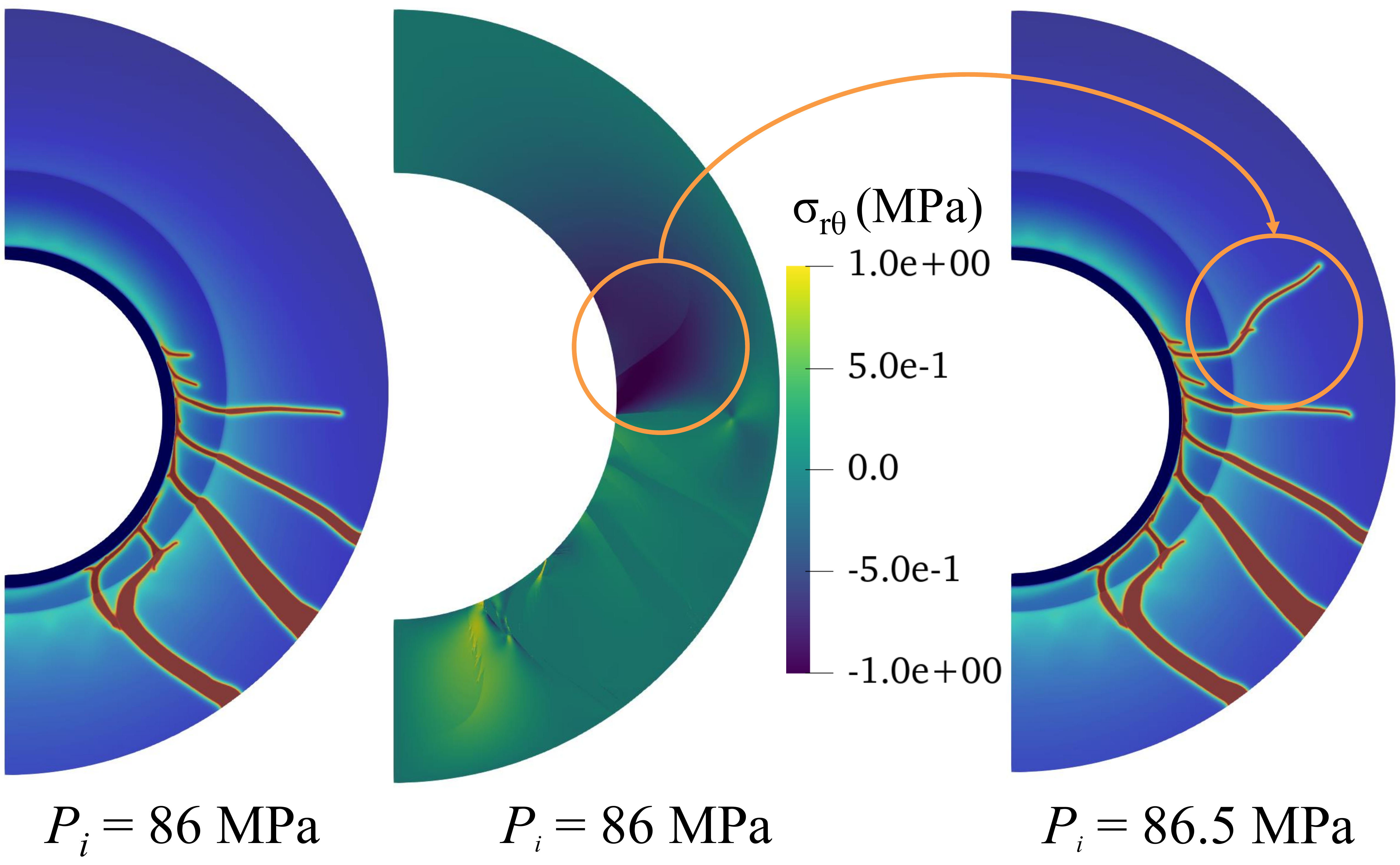}
        \caption{$e=50\%$}
        \label{fig:e=3_shear}
    \end{subfigure}
    \hfill
    \begin{subfigure}[b]{0.46\textwidth}
        \centering
        \includegraphics[width=1\linewidth]{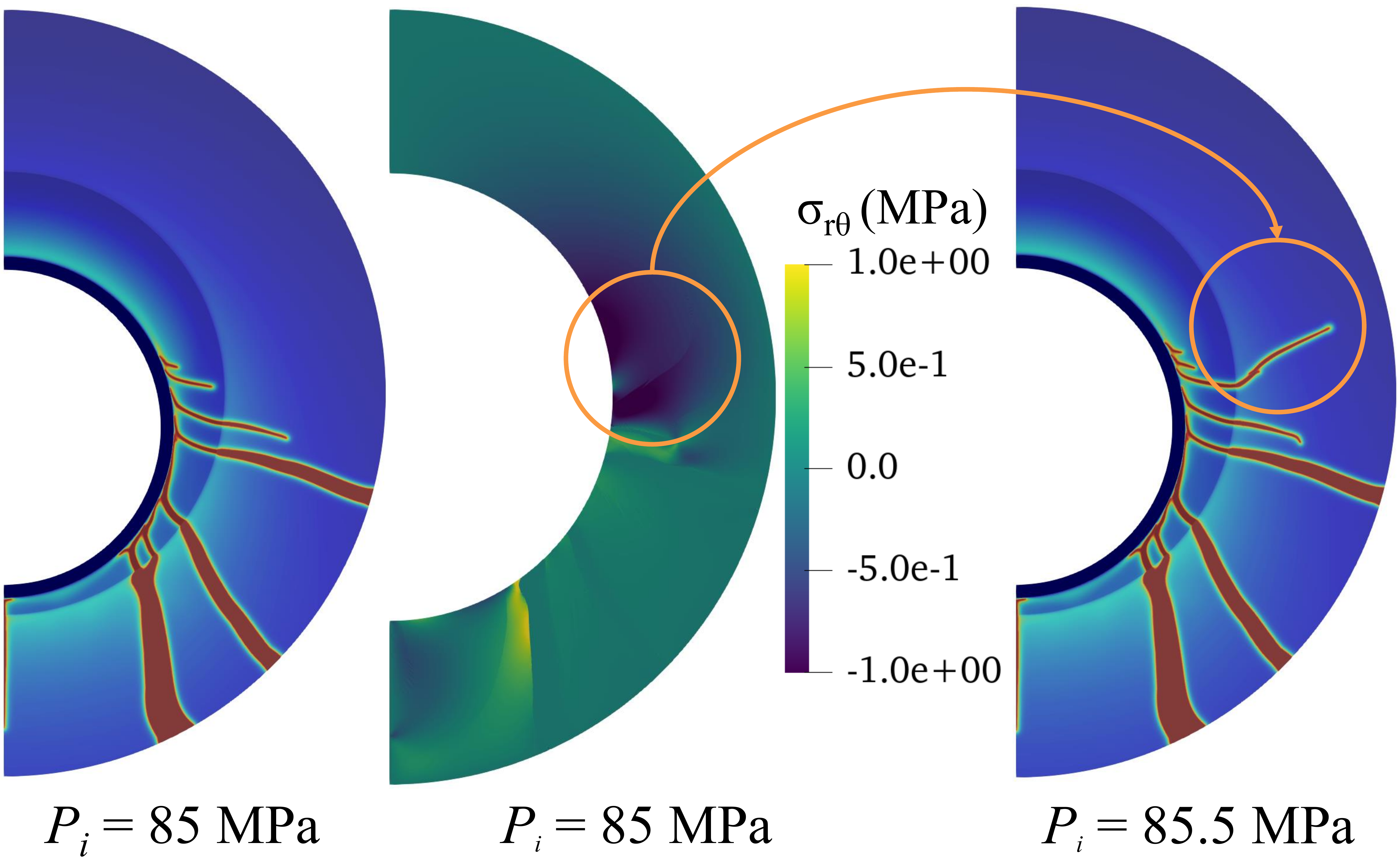}
        \caption{$e=66.67\%$}
        \label{fig:e=4_shear}
    \end{subfigure}
    \hfill
    \begin{subfigure}[b]{0.46\textwidth}
        \centering
        \includegraphics[width=1\linewidth]{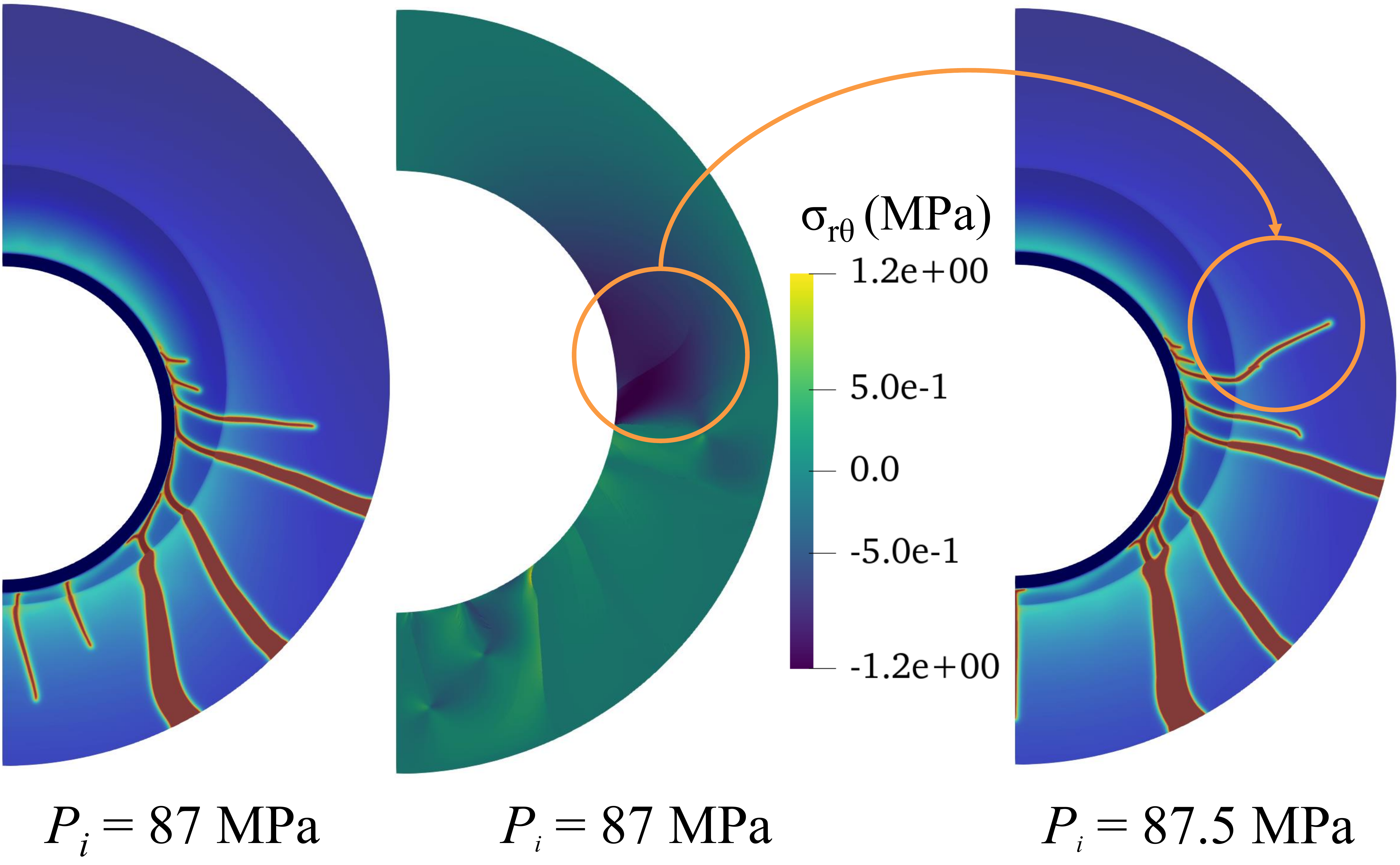}
        \caption{$e=75\%$}
        \label{fig:e=4.5_shear}
    \end{subfigure}
    \hfill
    \begin{subfigure}[b]{0.46\textwidth}
        \centering
        \includegraphics[width=1\linewidth]{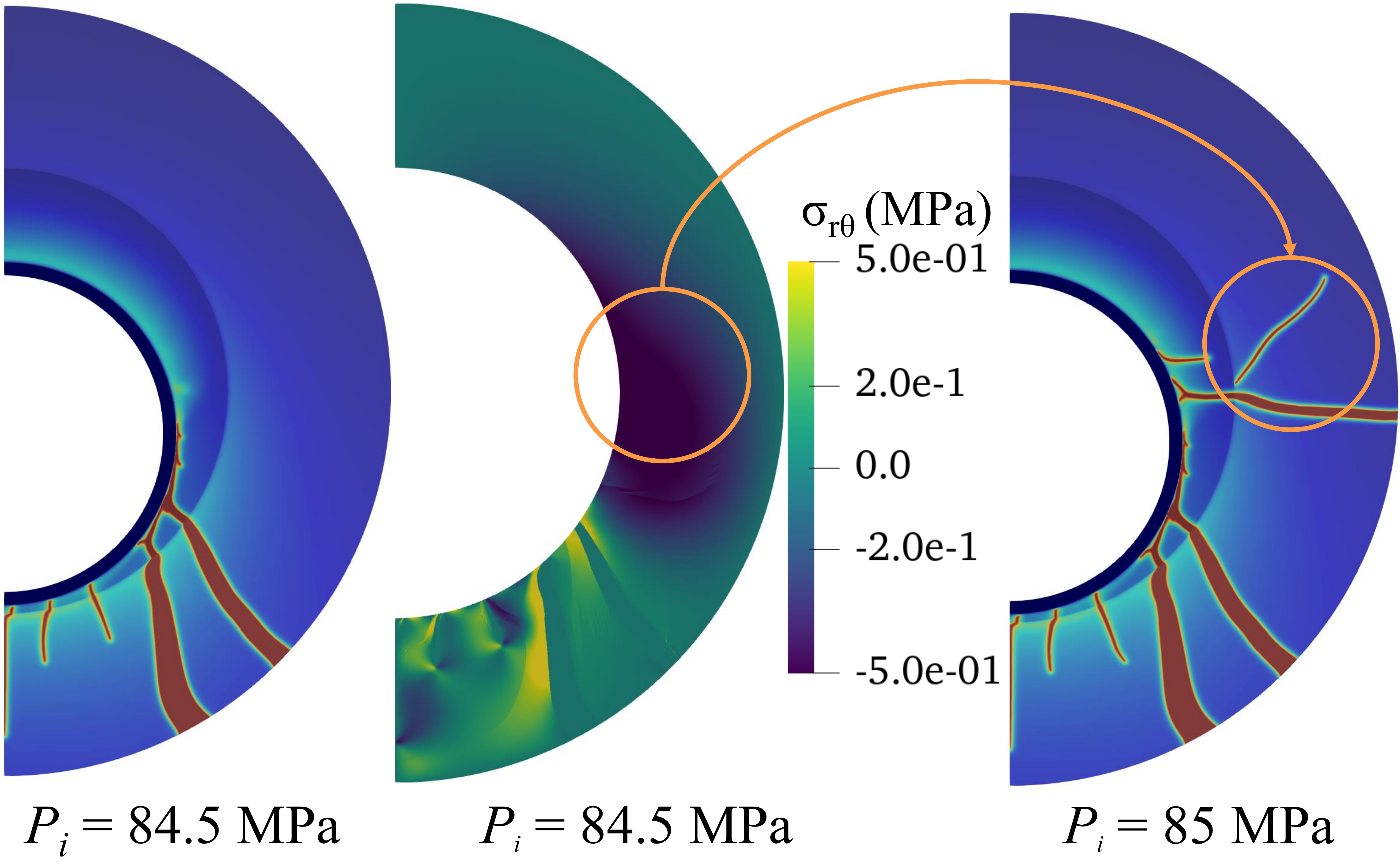}
        \caption{$e=83.33\%$}
        \label{fig:e=5_shear}
    \end{subfigure}
    \caption{For each eccentricity $e$, the subfigures show (i) damage profile before shear crack nucleation, (ii) shear stress ($\sigma_{r\theta}$) distribution, and (iii) damage profile after shear crack nucleation. Shear cracks nucleate at the  rock formation, where the magnitude of shear stress is high (indicated by the deep indigo regions in the contour). The sign represents the direction of shearing.}
    \label{fig:shear_crack}
\end{figure}

The damage profiles obtained by further increasing the wellbore pressure up to 97~MPa are shown in Figure~\ref{fig:194}. While the damage pattern remains symmetric for the concentric casing configuration, increasing casing eccentricity leads to progressively asymmetric damage profiles. In eccentric configurations, crack initiation consistently occurs in the thinner region of the cement sheath, and subsequent cracks nucleate in the thicker regions in an anticlockwise sequence. Another notable observation is that beyond a critical level of casing eccentricity ($e\geq50\%$), a non-radial crack develops in the rock at an azimuth of approximately $10^\circ$. Unlike the radial cracks, this crack propagates at an angle and not along the radial direction. As illustrated in Figure~\ref{fig:shear_crack}, this crack nucleates in a region characterized by large magnitudes of shear stresses (indicated by the deep indigo regions in the stress contour map). This threshold of approximately 50\% is an empirical observation based on the simulated cases rather than a theoretically derived critical value. Physically, this transition is consistent with the stress redistribution caused by increasing eccentricity. As the casing becomes more eccentric, the non-uniform cement sheath thickness leads to progressively higher shear stress concentrations, promoting mixed-mode fracture and the initiation of inclined cracks.

\subsection{Effects of interface weakening}
\label{ne:subsec2}

\renewcommand{\arraystretch}{1.4}
\begin{table}[h!]
\small
\centering
\caption{Properties of the wellbore materials used in the simulations discussed in Section~\ref{ne:subsec2}.}
\label{table:prop_wellbore_2}
\begin{tabular}{c c c c}
\hline
\textbf{Property} & \begin{tabular}[c]{@{}c@{}} \textbf{Rock Formation}\end{tabular} & \textbf{Cement sheath} & \textbf{Steel Casing} \\
\hline
E (N/m$^2$) & $8.28\times10^9$ & $25\times10^9$  & $200\times10^9$  \\
\hline
$\nu$ &  0.25 & 0.2  & 0.3  \\
\hline
$G_c$ (N/m) & 100 & 80 & $10^5$  \\
\hline
Source & \citep{xu2022phase} & \citep{clavijo2022coupled} & \citep{cui2024computational} \\
\hline
\end{tabular}
\end{table}

In this example, we investigate the effect of interface weakening on wellbore fracture behavior. Interface weakening in wellbore systems may arise from several factors, including cement shrinkage, voids formed due to improper cementing operations, and thermal variations~\citep{Bourgoyne1986}. 

To isolate and study the influence of interface weakening, we consider the cement sheath and rock formation to have comparable fracture energies, as reported in \citet{clavijo2022coupled} and \citet{xu2022phase}. The mechanical properties used in this study are summarized in Table~\ref{table:prop_wellbore_2}, while the geometry and boundary conditions remain the same as those shown in Figure~\ref{fig:wellbore_tag}. The phase-field length-scale parameter~$\ell = 27\times 10^{-5}$ m is chosen in accordance with the recommendation made in \cite{Wu2020}. Owing to the significantly higher fracture toughness of the steel casing, crack nucleation within the casing is unlikely, and cracks originating in the cement sheath cannot penetrate the casing. Consequently, only micro-annuli are expected at the casing-cement interface. Accordingly, the present study focuses exclusively on the cement-rock interface by varying the interface weakening parameter, $\kappa$, from $0$ to $-0.9$.

\begin{figure}[hbtp]
    \begin{subfigure}[b]{1\textwidth}
        \centering
        \includegraphics[width=0.75\linewidth]{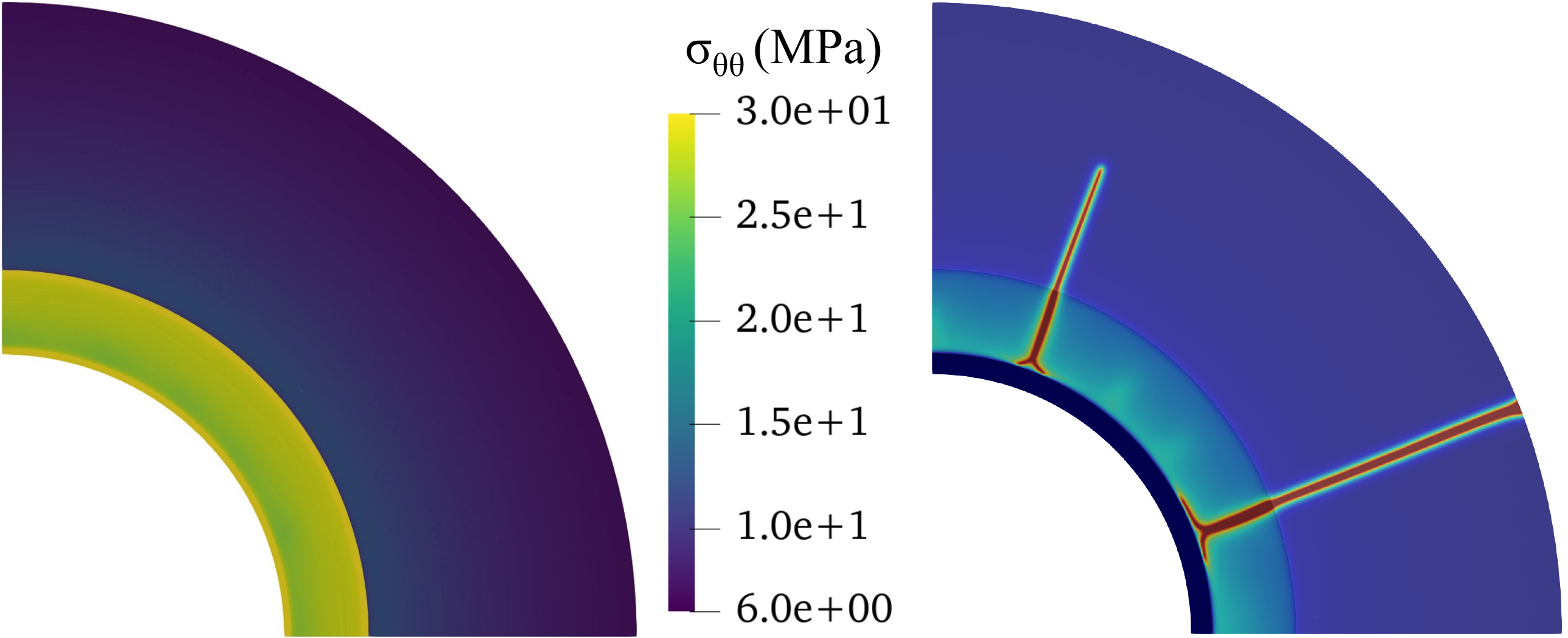}
        \caption{$\kappa=0$, crack nucleation at $P_i$ = 74 MPa}
        \label{fig:k=0_nuc}
    \end{subfigure}
    \hfill    
    \begin{subfigure}[b]{1\textwidth}
        \centering
        \includegraphics[width=0.75\linewidth]{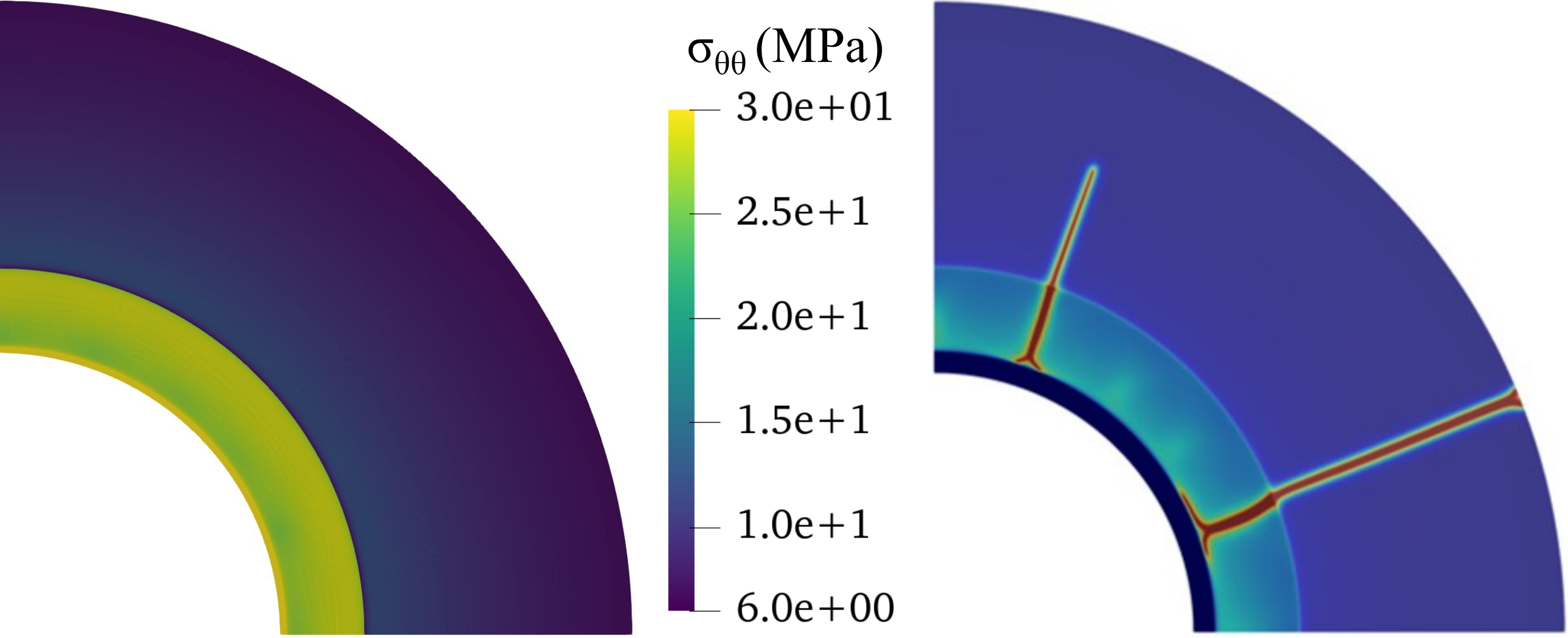}
        \caption{$\kappa=-0.5$, crack nucleation at $P_i$ = 74 MPa}
        \label{fig:k=-0.5_nuc}
    \end{subfigure}
    \hfill
    \begin{subfigure}[b]{1\textwidth}
        \centering
        \includegraphics[width=0.75\linewidth]{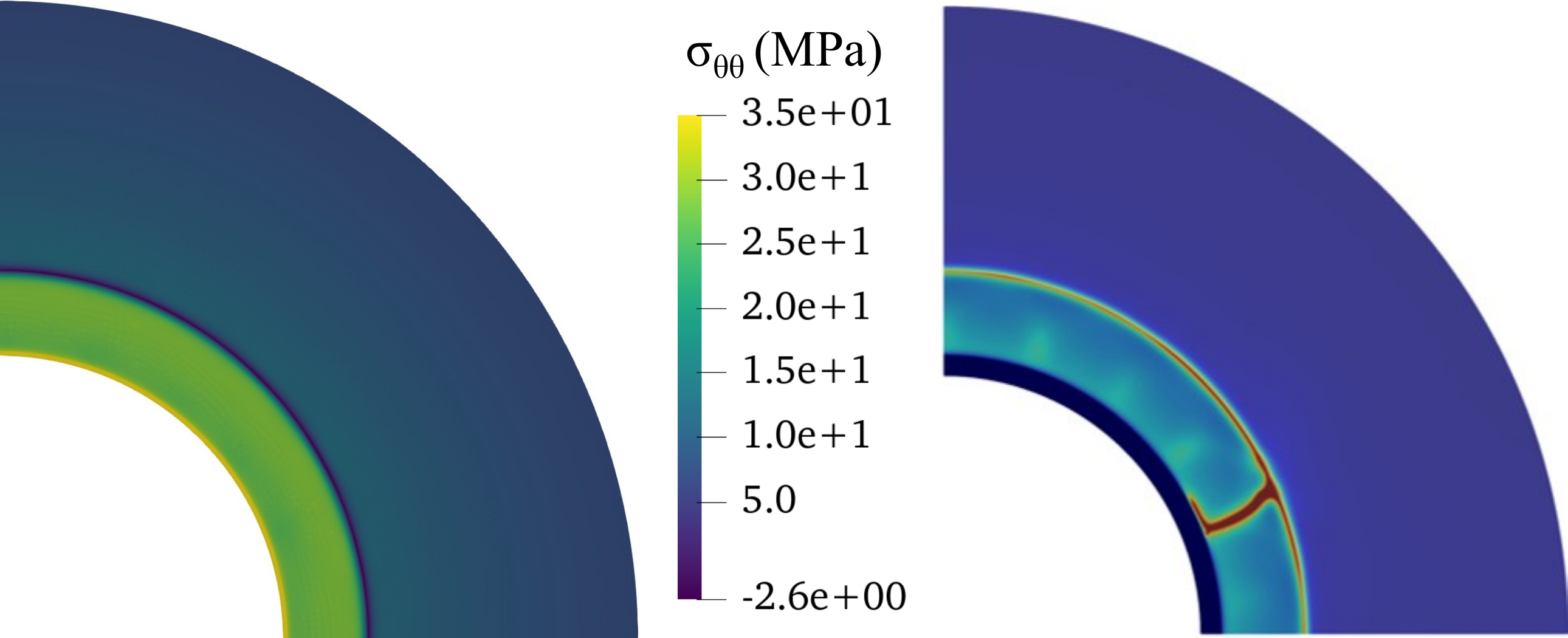}
        \caption{$\kappa=-0.9$, crack nucleation at $P_i$ = 73.5 MPa}
        \label{fig:k=-0.9}
    \end{subfigure}
    \caption{For each value of $\kappa$, the panels show: (i) tangential stress ($\sigma_{\theta\theta}$) in the cement sheath, and the rock before crack nucleation, and (ii) the corresponding damage profile at crack nucleation.}
    \label{fig:nuc_weak_int}
\end{figure}

Figure~\ref{fig:nuc_weak_int} illustrates the effect of the interfacial weakening parameter (\(\kappa\)) on crack initiation and propagation by showing the damage profile at crack onset along with the tangential stress contours in the cement sheath and rock for \(\kappa = 0, -0.5\) and \(-0.9\). For the geometry and material properties considered here, the influence of interface weakening becomes significant for \(\kappa \leq -0.7\); for higher values, the interface remains sufficiently strong for the crack to penetrate through it. Interestingly, the tangential stress along the interface decreases with decreasing \(\kappa\), becoming compressive for \(\kappa = -0.9\), as shown in Figure~\ref{fig:k=-0.9}. This indicates that the penetration of a radial crack in the rock formation is increasingly difficult as $\kappa$ decreases.

\begin{figure}[hbtp]
    \centering
    \begin{subfigure}[b]{1\textwidth}
        \centering
        \includegraphics[width=0.7\linewidth]{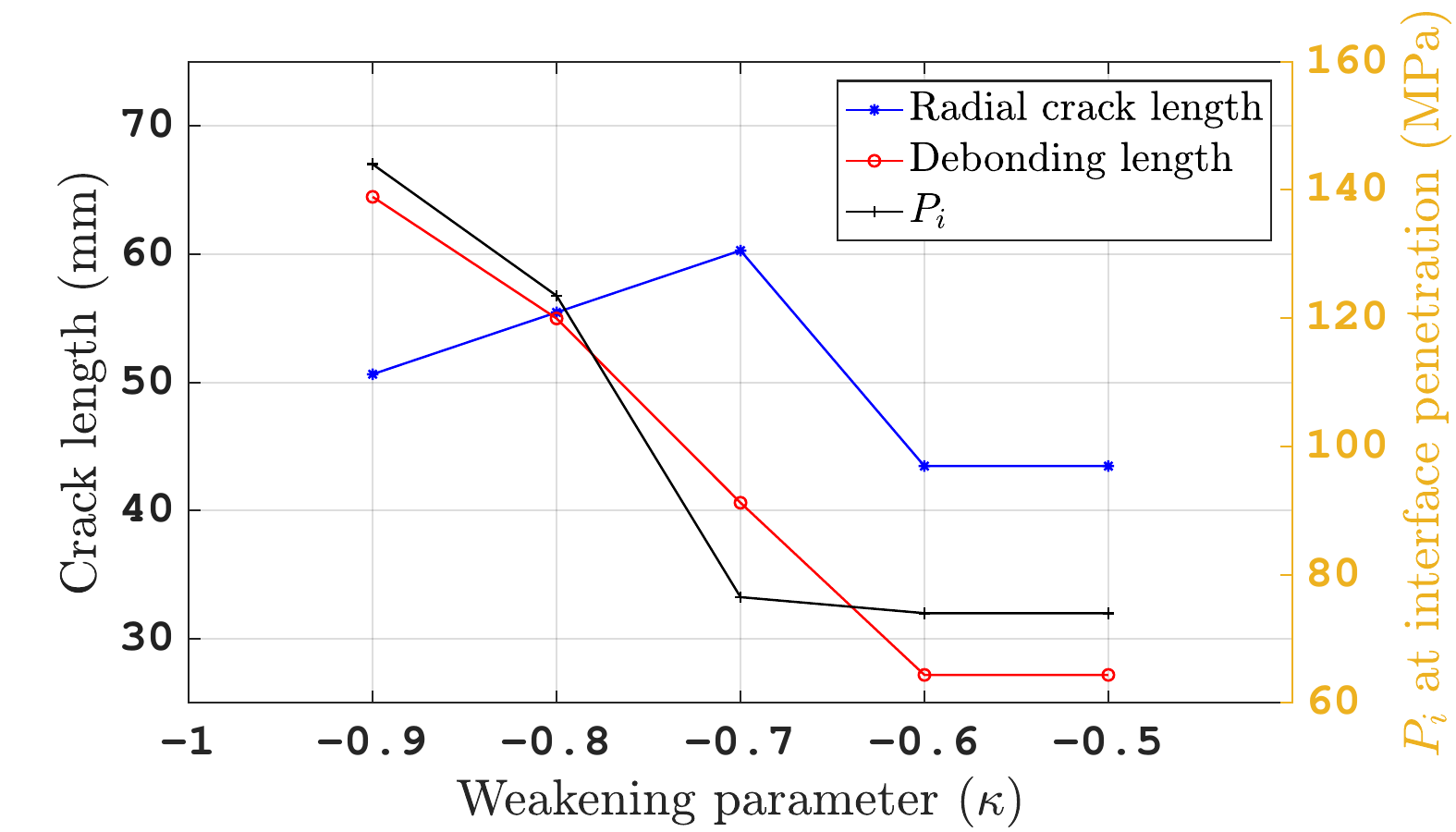}
        \caption{Variation of crack length and wellbore pressure at interface penetration with the weakening parameter}
        \label{fig:crack_length_plot}
    \end{subfigure}
    \begin{subfigure}[b]{0.3\textwidth}
        \centering
        \includegraphics[width=1\linewidth]{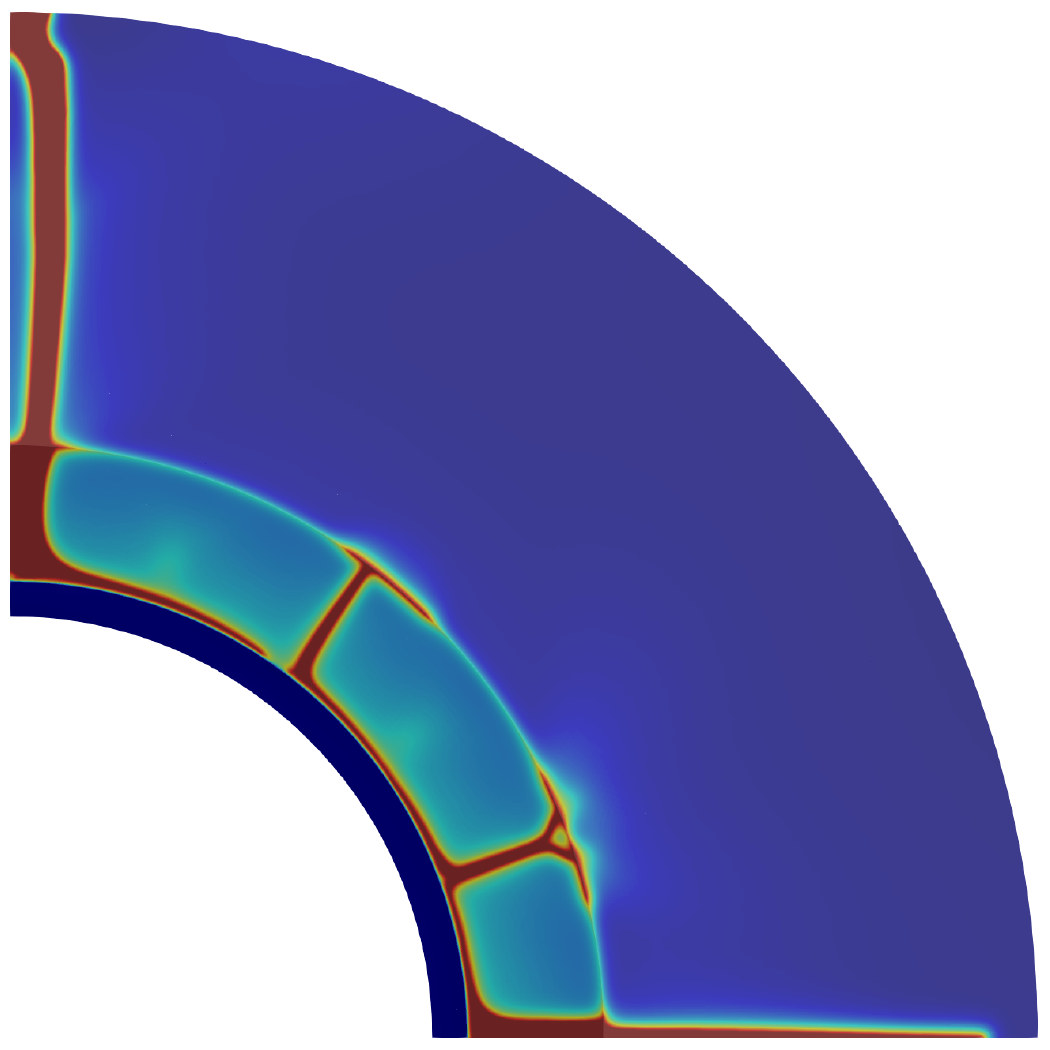}
        \caption{$\kappa=-0.7$}
        \label{fig:k=-0.7_325}
    \end{subfigure}
    \hfill    
    \begin{subfigure}[b]{0.3\textwidth}
        \centering
        \includegraphics[width=1\linewidth]{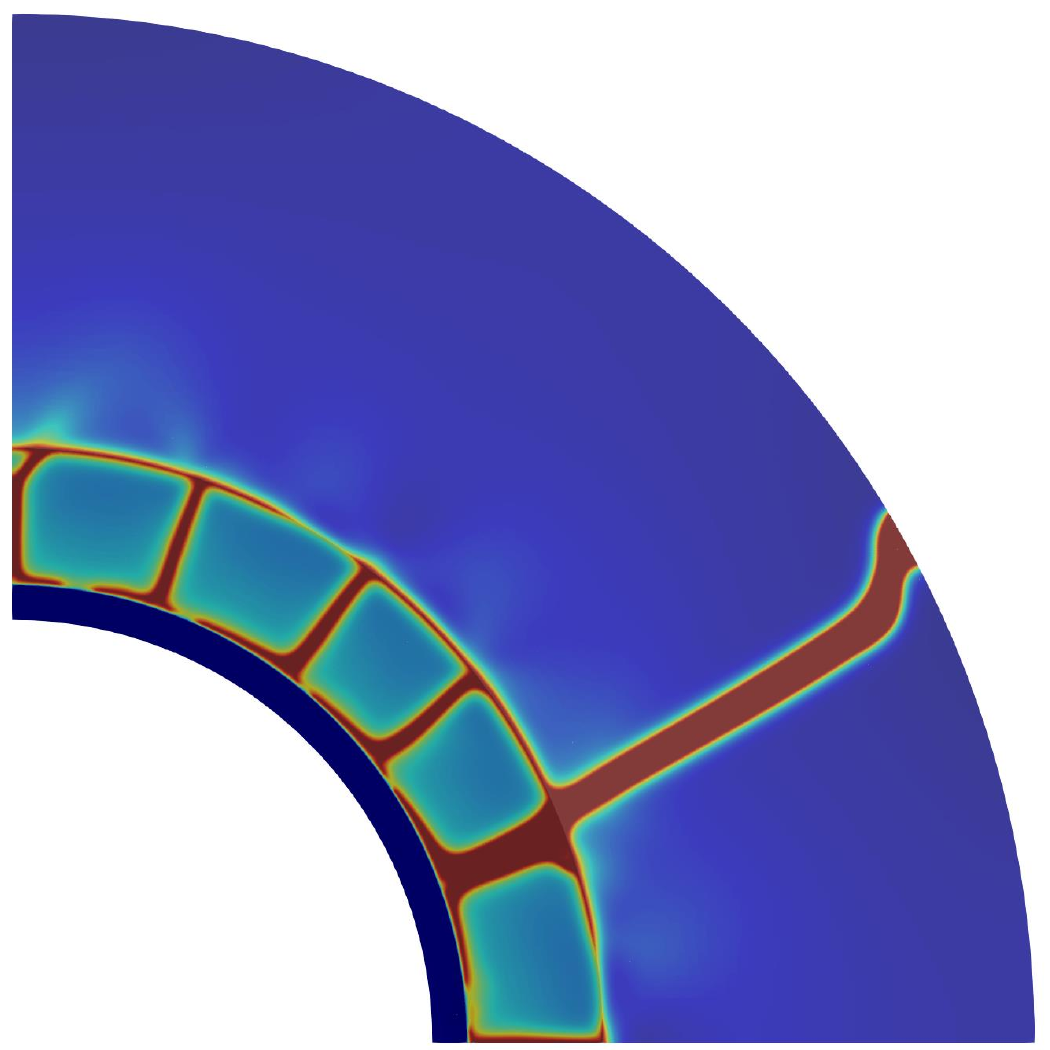}
        \caption{$\kappa=-0.8$}
        \label{fig:k=-0.8_325}
    \end{subfigure}
    \hfill
    \begin{subfigure}[b]{0.3\textwidth}
        \centering
        \includegraphics[width=1\linewidth]{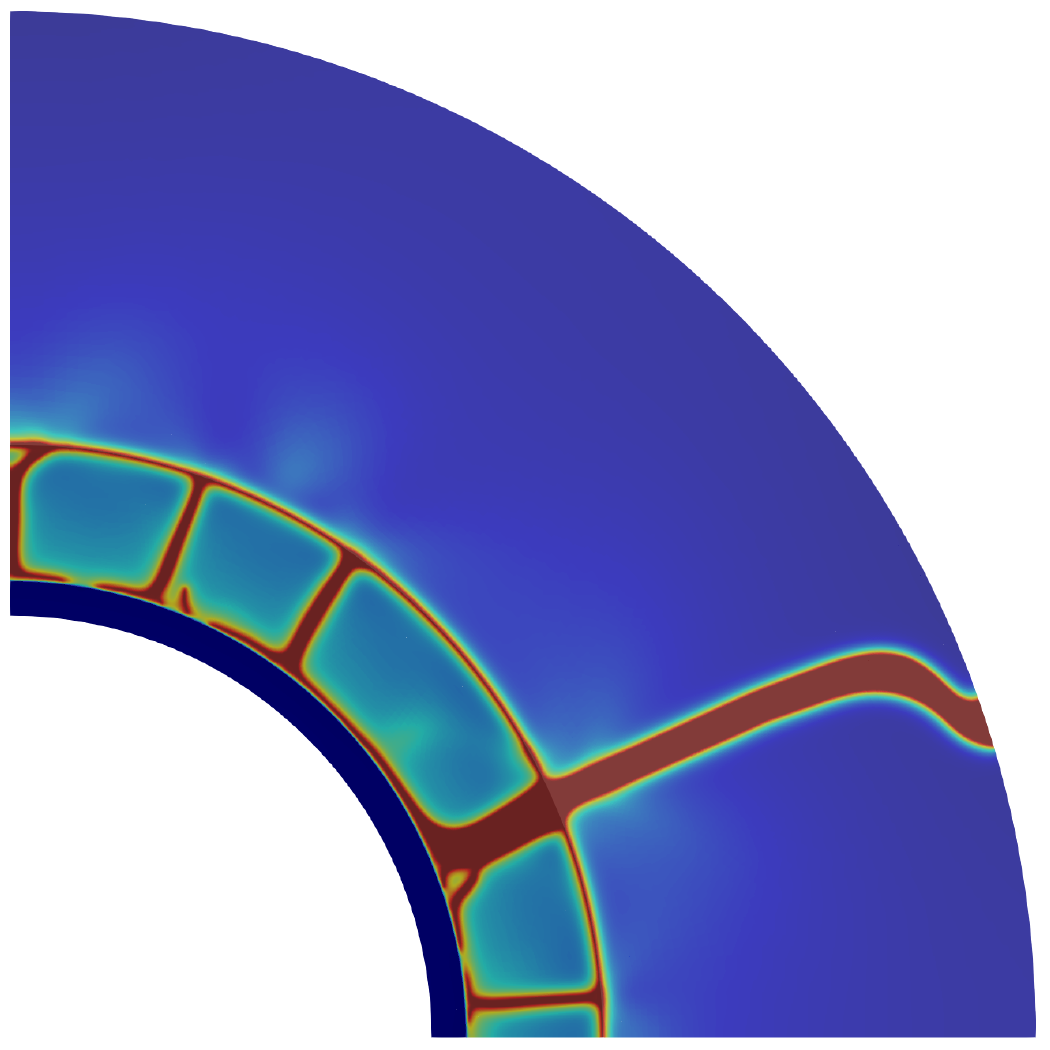}
        \caption{$\kappa=-0.9$}
        \label{fig:k=-0.9_325}
    \end{subfigure}
    \centering
    \begin{subfigure}[b]{0.9\textwidth}
        \centering
        \includegraphics[width=0.4\linewidth]{damage_legend.pdf}
    \end{subfigure}
        \hfill
    \caption{
(a): Variation of radial crack length, debonding length, and wellbore pressure at interface penetration with respect to the weakening parameter ($\kappa$). The interface penetration gets delayed as the interface weakens. (b) to (d): Damage profiles at a wellbore pressure of 162.5~MPa for three different values of interface weakening.}
    \label{fig:damage_weak_int}
\end{figure}

Figure~\ref{fig:damage_weak_int} presents the damage contours at a wellbore pressure of 162.5~MPa for different values of $\kappa$. In heterogeneous wellbore systems, crack propagation is governed by the competition between penetration across material interfaces and deflection along weak interfaces. As $\kappa$ decreases, interfacial deflection becomes increasingly favorable, while penetration into the adjacent material becomes less favorable. For the geometry and material properties considered here, interfacial debonding along the cement-rock interface becomes prominent for $\kappa \leq -0.7$. When a crack reaches the weakened interface, it is temporarily arrested because penetration into the adjacent layer is energetically less favorable. This promotes the nucleation of additional radial cracks within the cement sheath. In contrast, when the crack readily penetrates the rock formation for $\kappa>-0.7$, localized stress relaxation in the cement sheath suppresses further crack initiation.

Figure~\ref{fig:crack_length_plot} shows a quantitative measure, where the total radial crack length, total debonding length, and the wellbore pressure at interface penetration are plotted for different values of the weakening parameter. The plot highlights the change in the dominant failure mode of the wellbore system from radial cracking to debonding as \(\kappa\) value decreases. The debonding length and the wellbore pressure at interface penetration increase monotonically as the weakening parameter decreases (i.e., the interface is weakened further). This is because, as the interface strength decreases, the debonding increases, and with increased debonding, the discontinuity between the materials increases, making the rock formation behave as a separate system in which cracks must nucleate independently. This requires higher wellbore pressure, thereby delaying crack penetration into the rock formation. In Figures~\ref{fig:k=-0.7_325}–\ref{fig:k=-0.9_325}, it is observed that the number of radial cracks in the cement sheath is higher for $\kappa = -0.8$ than for $\kappa = -0.9$. This occurs because, for $\kappa = -0.9$, the interface strength is reduced by 90\%, resulting in complete debonding of the cement-rock interface at an earlier stage compared to the case with $\kappa = -0.8$. Early debonding induces localized stress relaxation in the cement sheath, thereby suppressing the nucleation of additional radial cracks. Once the cement sheath is fully debonded from the rock formation, the same effect described above applies: the rock formation behaves as an independent system requiring higher pressure for crack nucleation. Consequently, crack penetration into the rock formation occurs at a higher wellbore pressure of 144~MPa for $\kappa = -0.9$ versus 123.5~MPa for $\kappa = -0.8$, consistent with the trend of increasing $P_i$ with greater interface weakening noted earlier.

\begin{figure}[hbtp]
    \begin{subfigure}[b]{0.5\textwidth}
        \centering
        \includegraphics[width=1\linewidth]{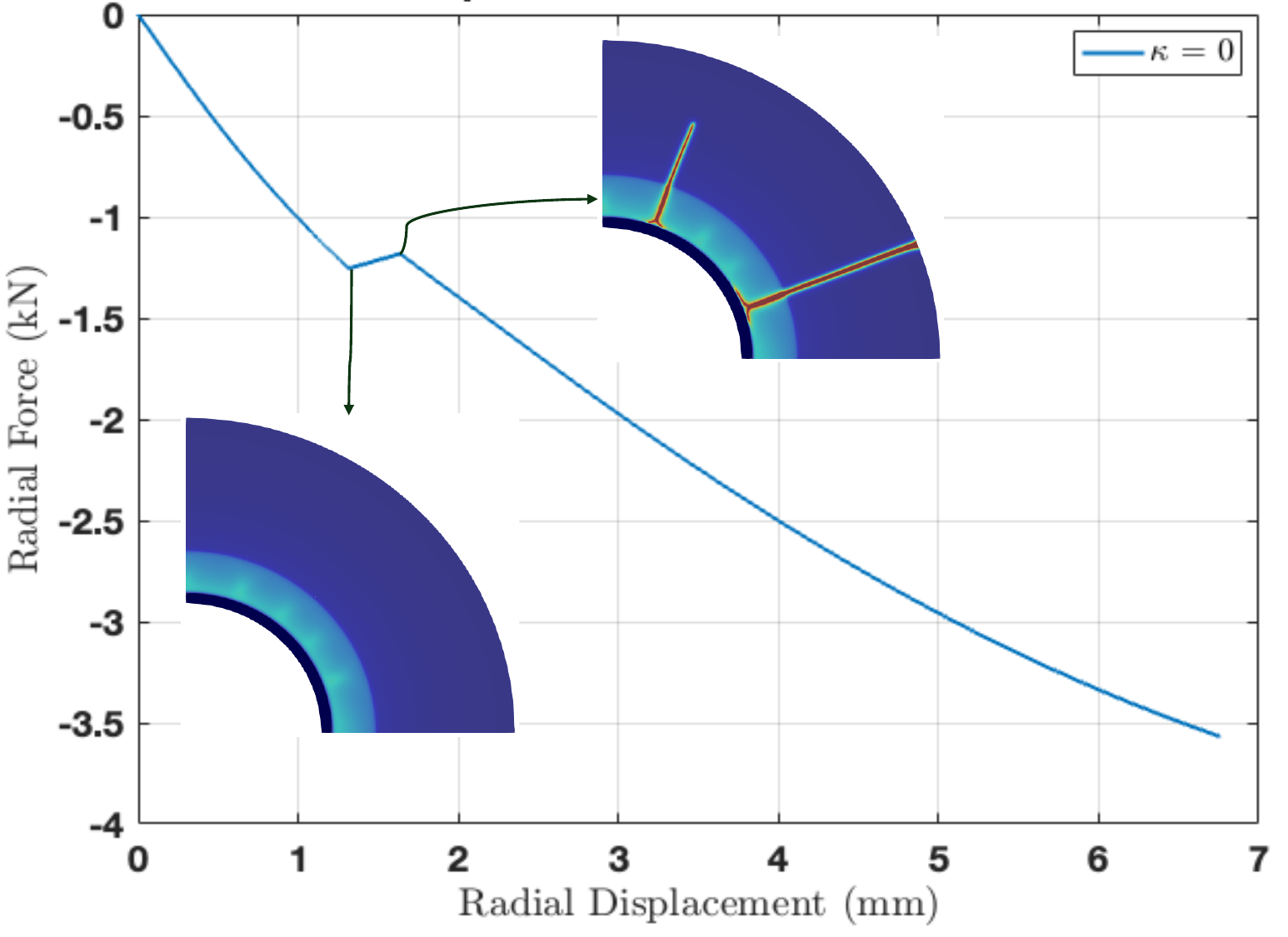}
        \caption{Radial load--displacement plot for $\kappa=0$}
        \label{fig:k=0_radial_plot}
    \end{subfigure}
    \hfill    
    \begin{subfigure}[b]{0.5\textwidth}
        \centering
        \includegraphics[width=1\linewidth]{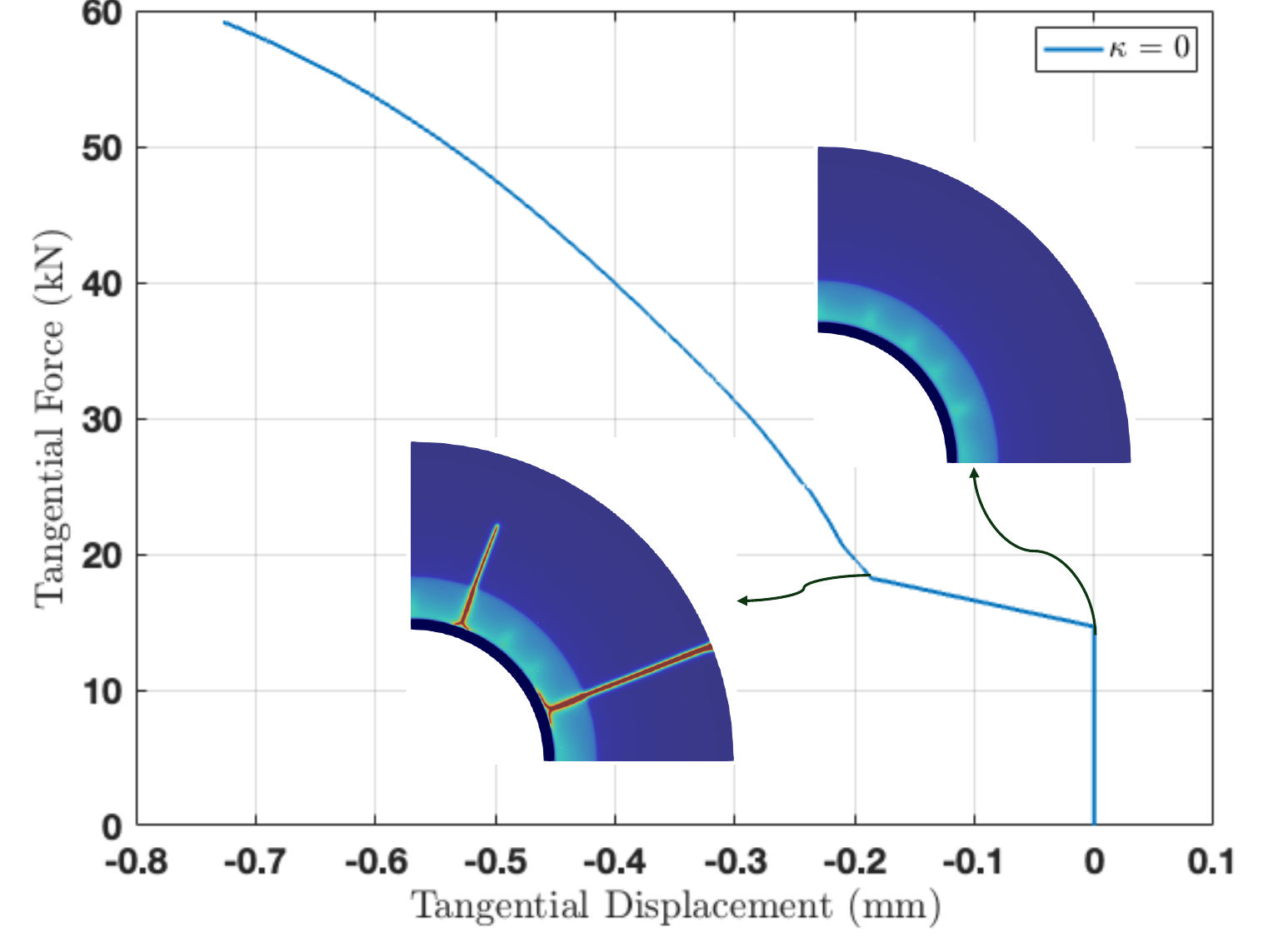}
        \caption{Tangential load--displacement plot for $\kappa=0$}
        \label{fig:k=0_tan_plot}
    \end{subfigure}
    \hfill
    \begin{subfigure}[b]{0.5\textwidth}
        \centering
        \includegraphics[width=1\linewidth]{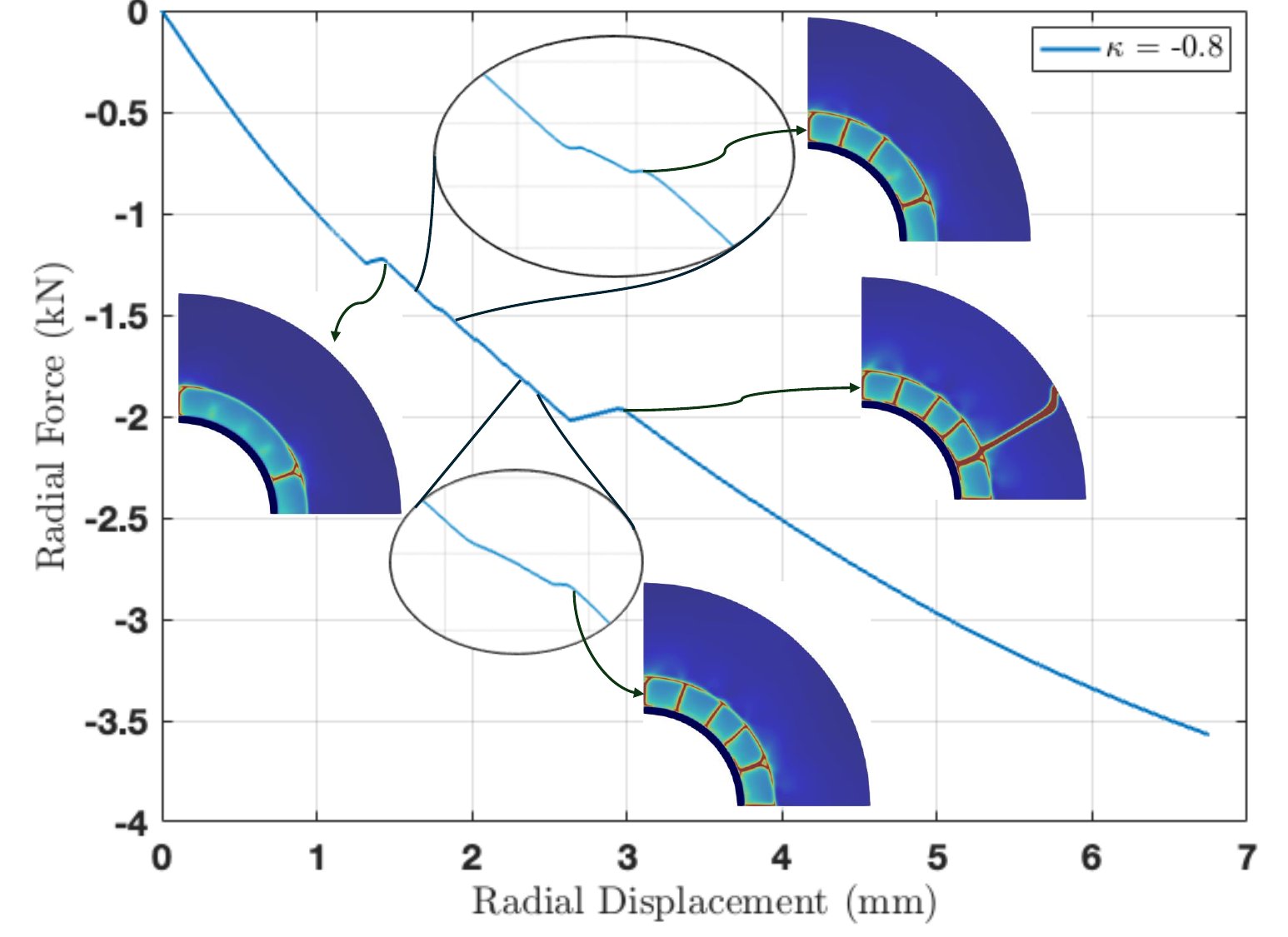}
       \caption{Radial load-displacement plot for $\kappa=-0.8$}
        \label{fig:k=-0.8_radial_plot}
    \end{subfigure}
    \hfill
    \begin{subfigure}[b]{0.5\textwidth}
        \centering
        \includegraphics[width=1\linewidth]{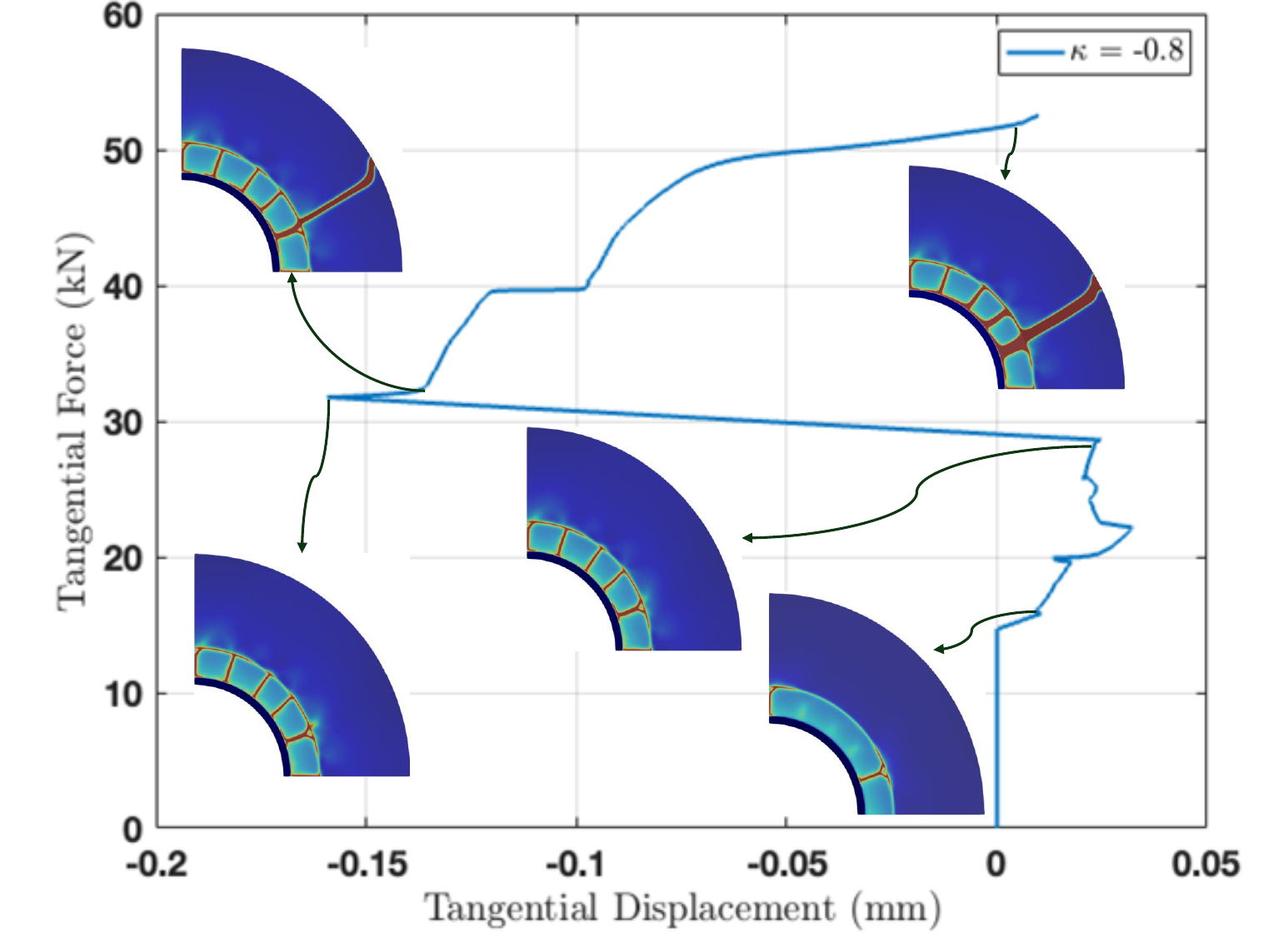}
        \caption{Tangential load-displacement plot for $\kappa=-0.8$}
        \label{fig:k=-0.8_tan_plot}
    \end{subfigure}
    \caption{Load--displacement curves obtained by summing the values over the entire inner surface of the wellbore for the neutrally bonded cement-rock interface $(\kappa=0)$ (a and b) and when the strength of the cement-rock interface is degraded by 80\% $(\kappa=-0.8)$ (c and d).}
    \label{fig:load_disp_k=0}
\end{figure}

Figure~\ref{fig:load_disp_k=0} shows the load--displacement response at the inner surface of the wellbore for a neutrally bonded cement-rock interface and for an interface weakened by 80\%, respectively. The load and displacement values in the plot are obtained by summing the values over the entire inner surface of the wellbore. Since the model is subjected to traction boundary conditions, crack nucleation and propagation are reflected as sudden displacement jumps at nearly constant load. For the neutrally bonded case $(\kappa=0)$, the radial displacement increases with the applied radial force until the onset of cracking, where a distinct kink appears in the response. This kink corresponds to the simultaneous nucleation of two radial cracks. A similar trend is observed in the tangential load--displacement response. Prior to crack initiation, the tangential displacement remains nearly zero due to confinement. After crack nucleation, tangential displacement increases due to deformation induced by cracking and lateral expansion, producing a kink at crack onset. For the weakened-interface case $(\kappa=-0.8)$, the load--displacement curves show multiple kinks corresponding to the sequential nucleation of individual radial cracks, followed by a larger response change when the crack penetrates the cement-rock interface.

Such crack proliferation may create potential leakage pathways for formation fluids (oil, water, or gas), thereby increasing the risk of loss of zonal isolation. Despite these variations in damage evolution, the ultimate failure mode in all cases occurs through crack penetration into the rock formation.

\subsection{Three-dimensional analysis of a wellbore}
\label{ne:subsec3}
\begin{figure} [hbtp]
    \centering
    \begin{subfigure}[b]{0.3\textwidth}
        \centering
        \includegraphics[width=1\linewidth]{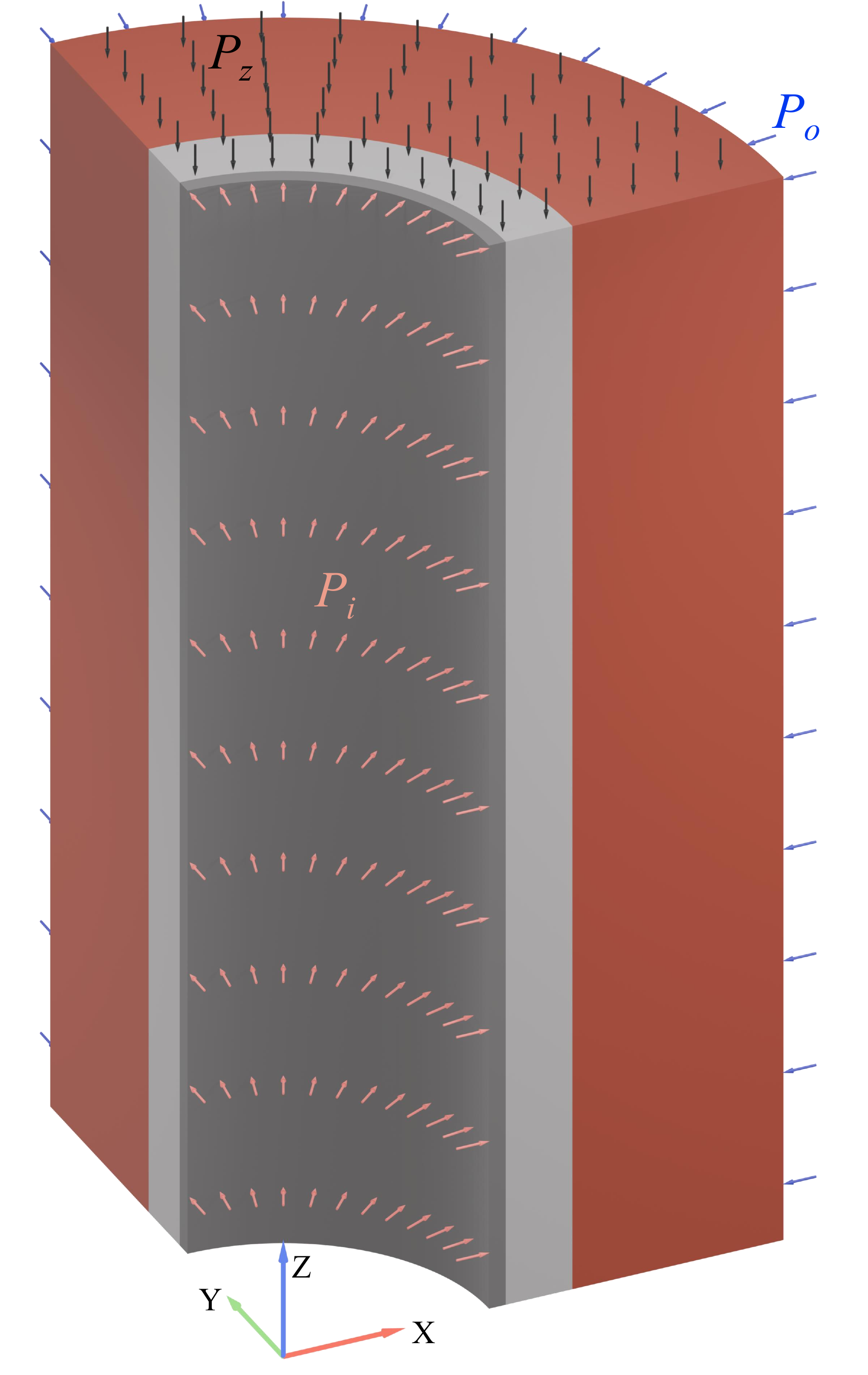}
        \caption{}
        \label{fig:wellbore_3d_layout}
    \end{subfigure}
    \hfill
        \begin{subfigure}[b]{0.32\textwidth}
        \centering
        \includegraphics[width=1\linewidth]{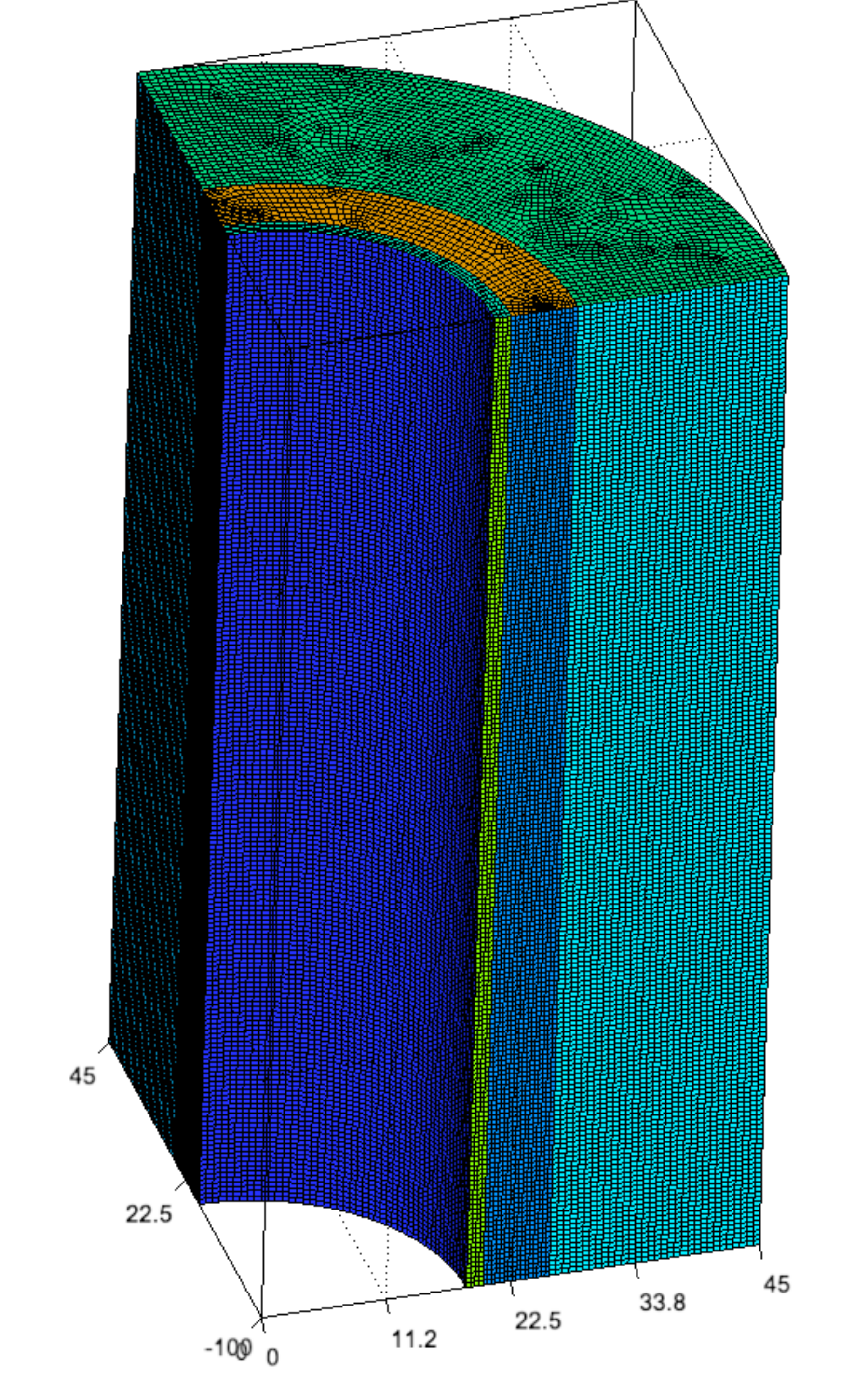}
        \caption{}
        \label{fig:wellbore_3d_initial_mesh}
    \end{subfigure}
    \hfill
        \begin{subfigure}[b]{0.3\textwidth}
        \centering
        \includegraphics[width=1\linewidth]{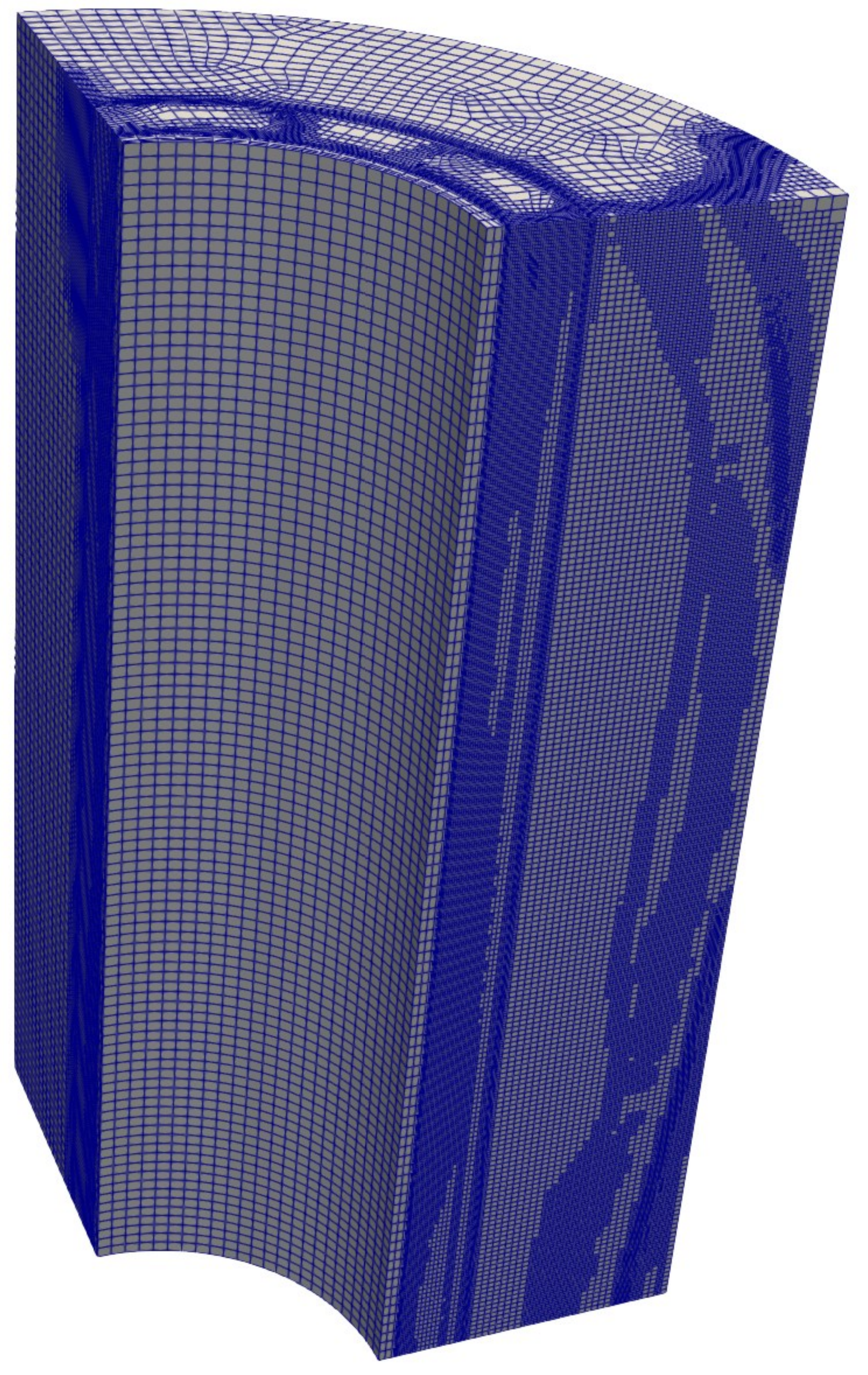}
        \caption{}
        \label{fig:wellbore_3d_final_mesh}
    \end{subfigure}

    \caption{(a) Three-dimensional geometry and boundary conditions of one quadrant of the wellbore consisting of a steel casing, cement sheath, and the rock formation. The wellbore is fixed normally in the boundary planes parallel to the $y-z, x-z,$ and $x-y$ planes to prevent rigid-body motion. The wellbore is subjected to internal wellbore pressure $P_i$, confining pressure $P_o$, and an overburden pressure $P_z$. (b) The initial mesh used for the simulation consisted of 724000 trilinear hexahedral elements. (c) Adaptively refined final mesh at the end of the simulation.}
    \label{fig:wellbore_3d}
\end{figure}

As a final example, we present a 3D simulation of a wellbore in which the cement-rock interface is weakened by 90\% ($\kappa = -0.9$). The mechanical properties used in this study remain identical to the previous example (see Table~\ref{table:prop_wellbore_2}). The wellbore depth is specified as 100~mm. The wellbore is constrained in the normal direction on the boundary planes parallel to the $x$–$y$, $y$–$z$, and $x$–$z$ planes to eliminate rigid-body motion. The model is subjected to an overburden pressure $(P_z)$ and a confining pressure $(P_o)$ of 1~kPa, while the internal pressure $(P_i)$ is increased incrementally by 50~kPa per load step. Like previous sections, $P_i$ is interpreted as the differential pressure between the actual wellbore pressure and the in-situ geostatic stresses. The thicknesses of the casing, cement sheath, and rock formation are also identical to the two-dimensional simulations presented earlier (see Figure~\ref{fig:wellbore_tag}). A schematic of the geometry and boundary conditions is shown in Figure~\ref{fig:wellbore_3d_layout}. This example is intended to demonstrate the ability of the framework to capture depth-dependent crack nucleation, non-uniform crack growth, and crack coalescence mechanisms that cannot be represented under plane-strain assumptions.

\begin{figure}[hbtp]
    \begin{subfigure}[b]{0.3\textwidth}
        \centering
        \includegraphics[width=0.8\linewidth]{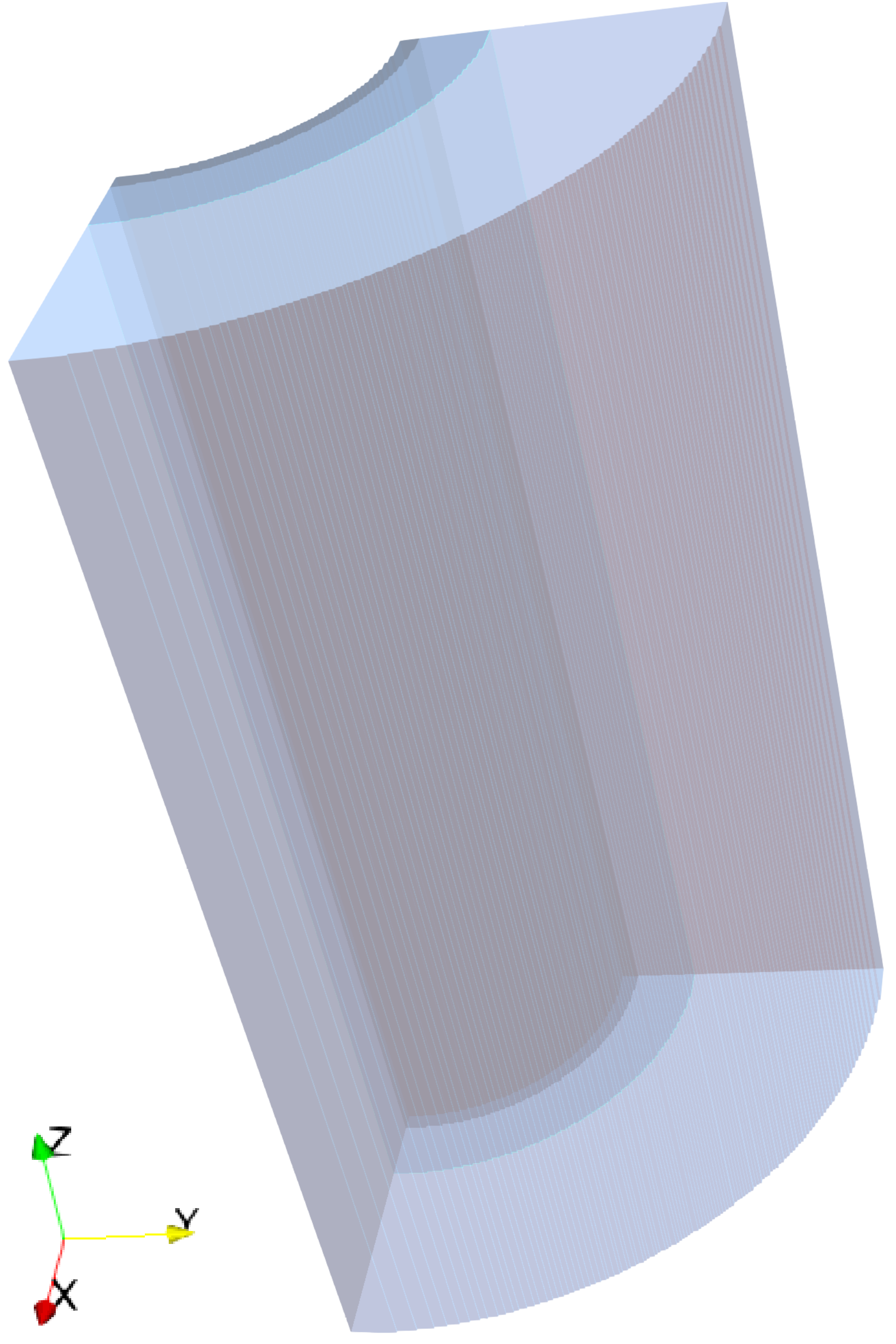}
        \caption{$P_i=$ 0~MPa}
        \label{fig:wellbore_3d_0}
    \end{subfigure}
    \hfill    
    \begin{subfigure}[b]{0.3\textwidth}
        \centering
        \includegraphics[width=0.8\linewidth]{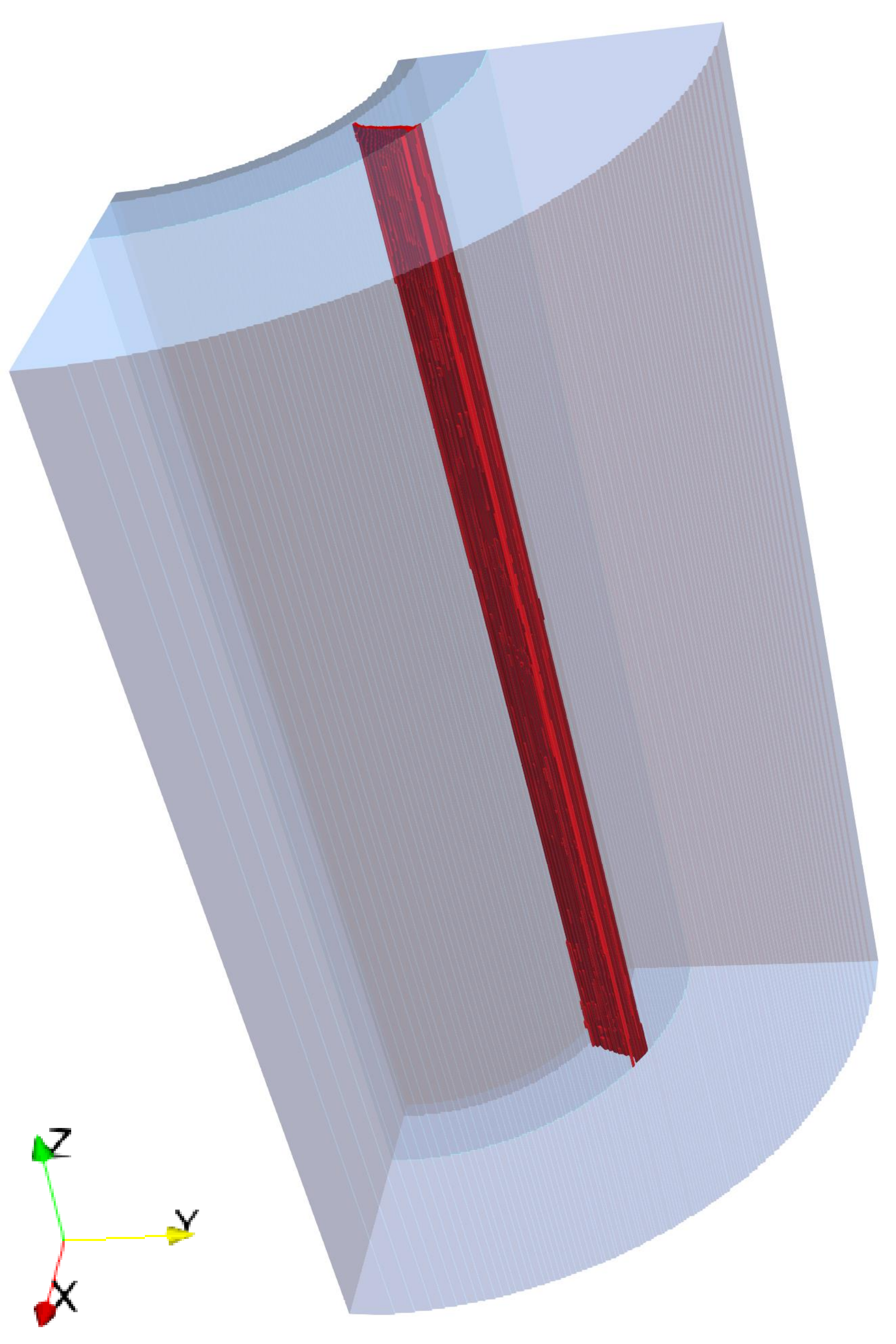}
        \caption{$P_i=$ 70.5~MPa}
        \label{fig:wellbore_3d_1}
    \end{subfigure}
    \hfill    
    \begin{subfigure}[b]{0.3\textwidth}
        \centering
        \includegraphics[width=0.8\linewidth]{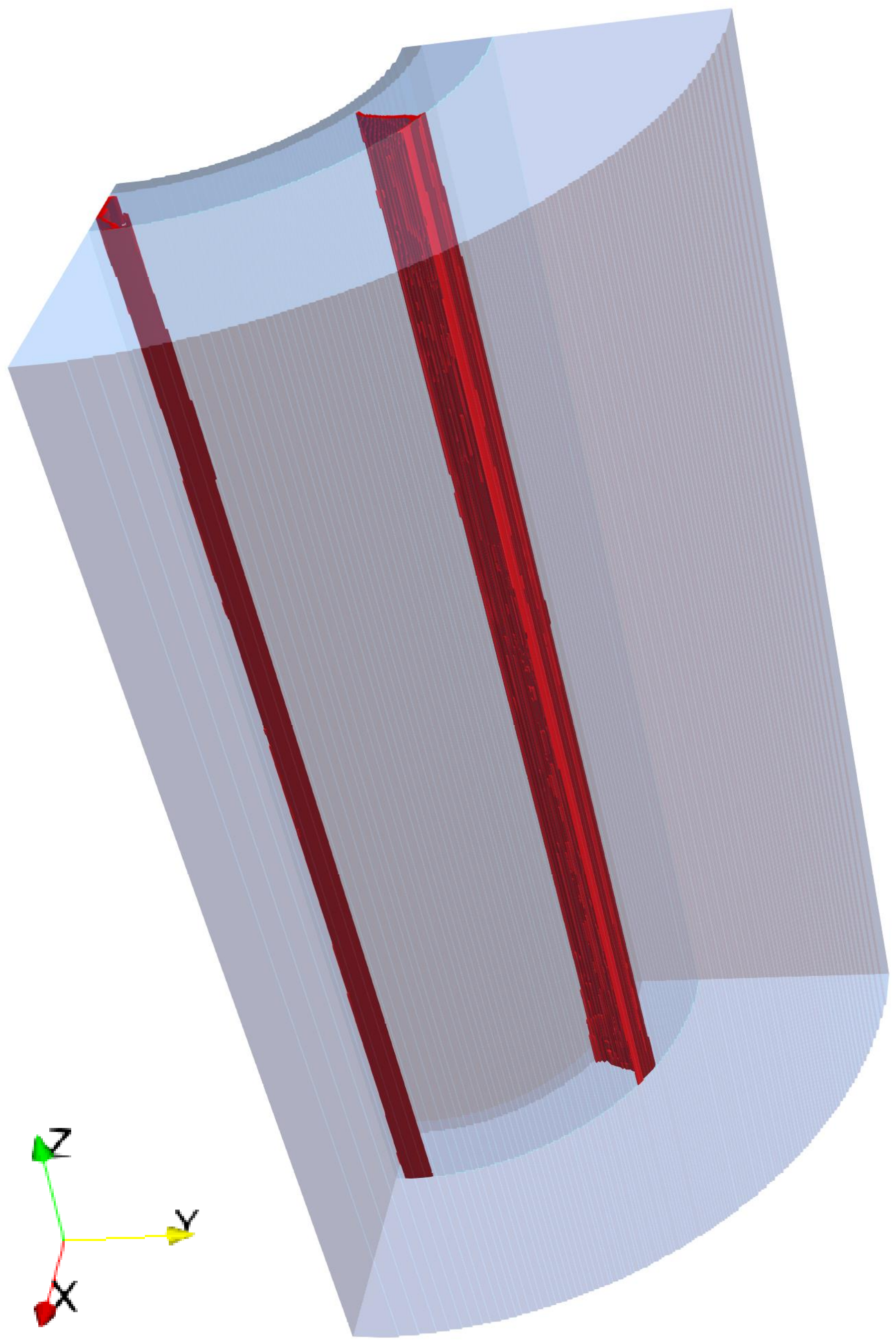}
        \caption{$P_i=$ 75~MPa}
        \label{fig:wellbore_3d_2}
    \end{subfigure}
    \hfill    
    \begin{subfigure}[b]{0.3\textwidth}
        \centering
        \includegraphics[width=0.8\linewidth]{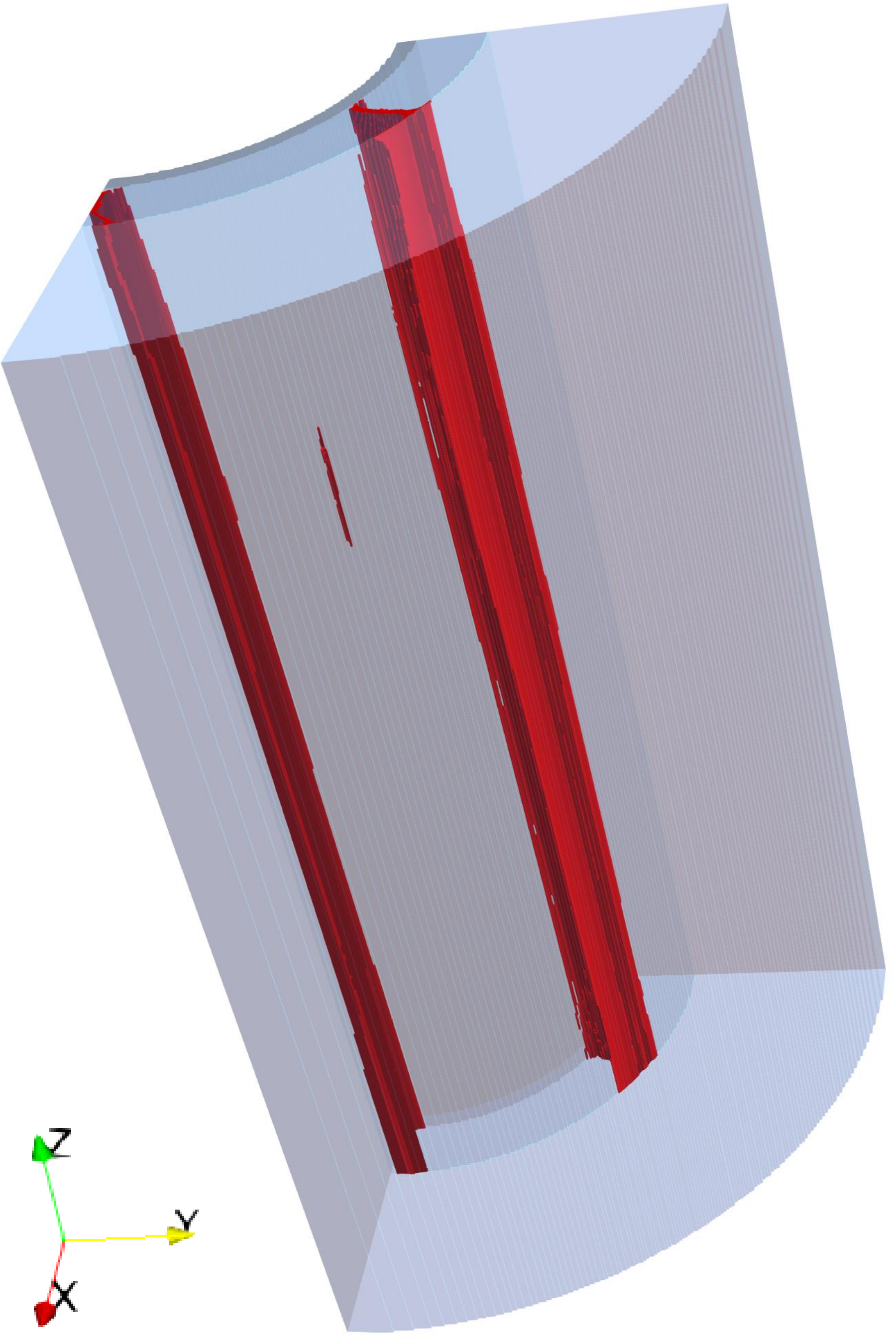}
        \caption{$P_i=$ 101.5~MPa}
        \label{fig:wellbore_3d_3}
    \end{subfigure}
    \hfill    
    \begin{subfigure}[b]{0.3\textwidth}
        \centering
        \includegraphics[width=0.8\linewidth]{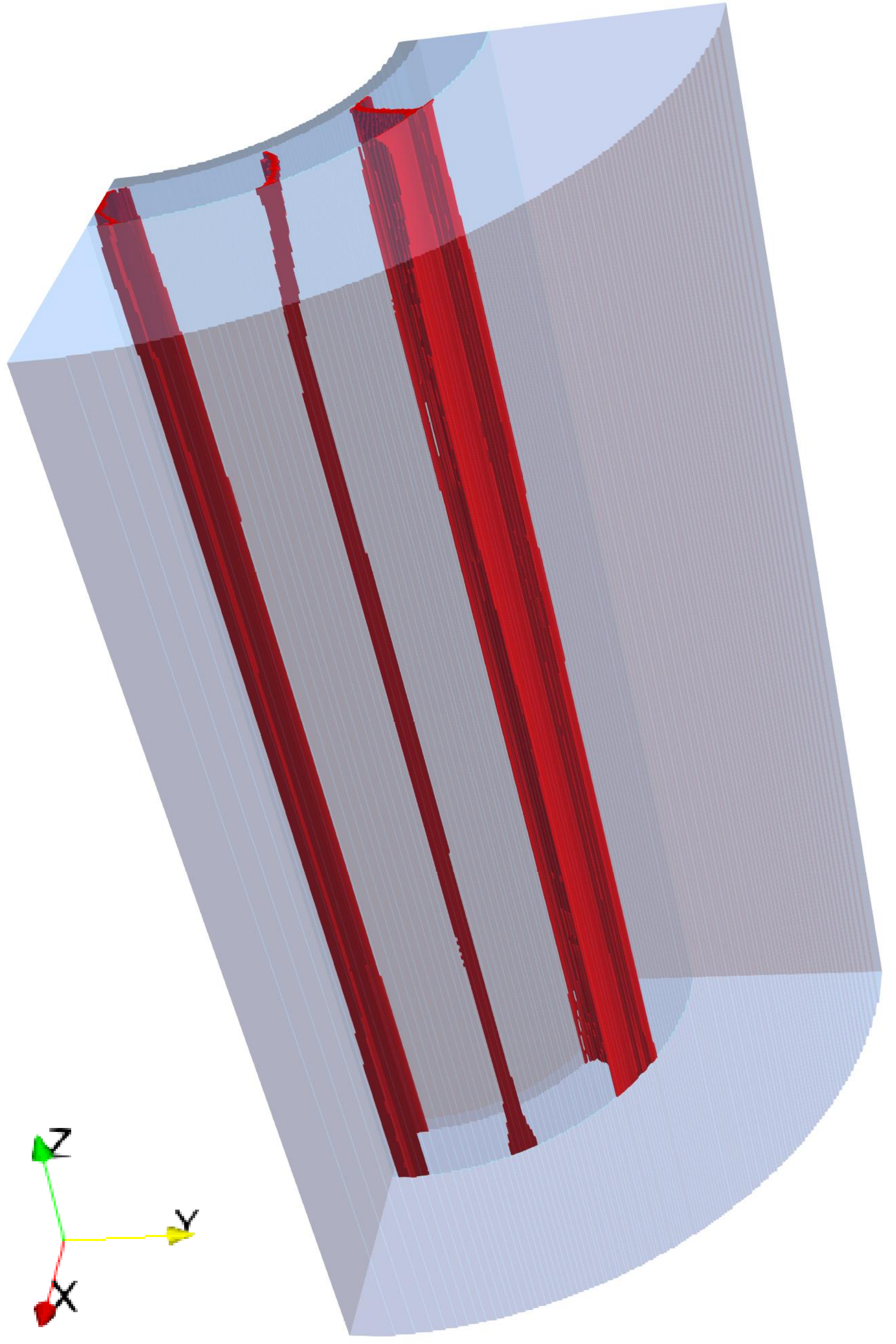}
        \caption{$P_i=$ 105~MPa}
        \label{fig:wellbore_3d_4}
    \end{subfigure}
    \hfill    
    \begin{subfigure}[b]{0.3\textwidth}
        \centering
        \includegraphics[width=0.8\linewidth]{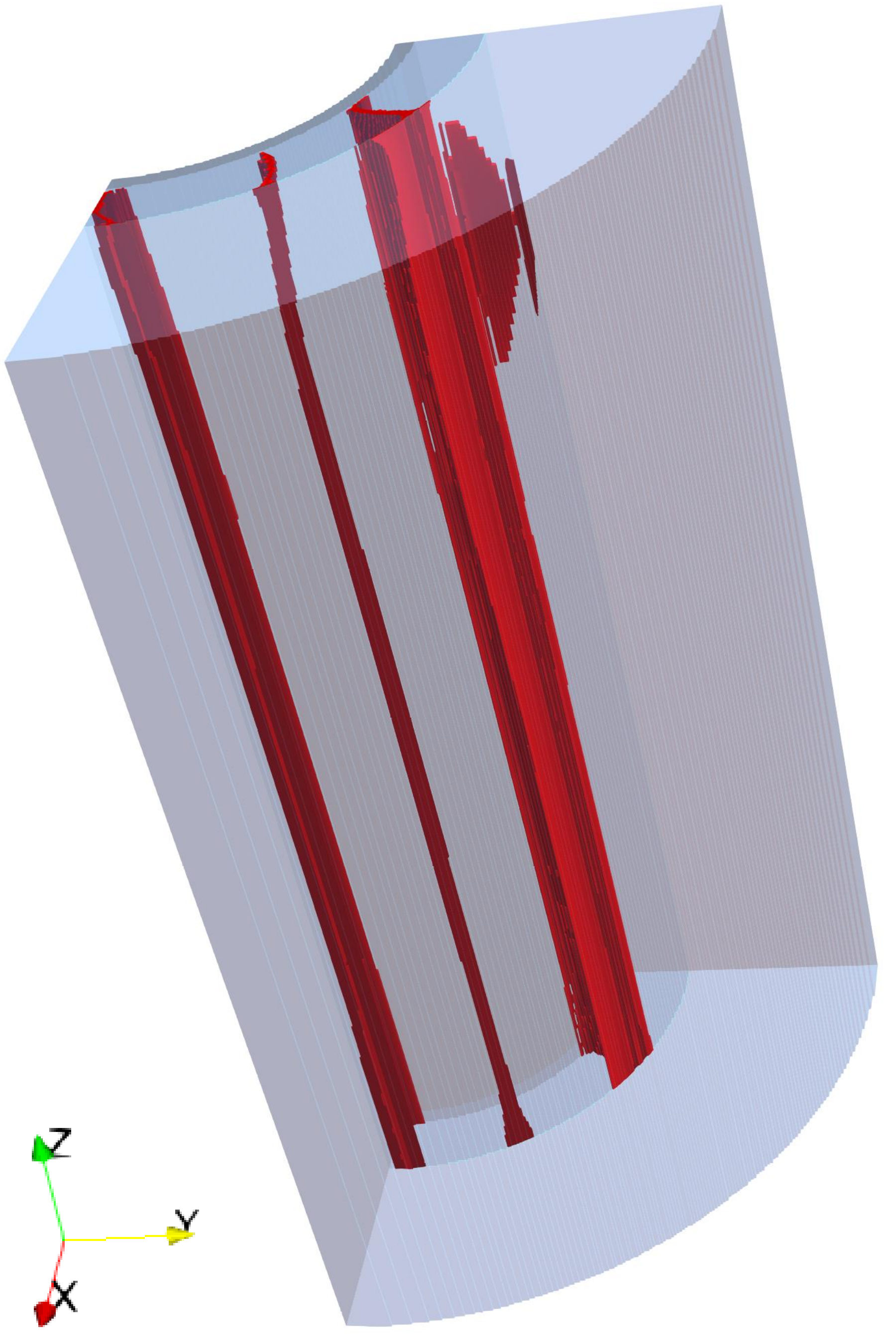}
        \caption{$P_i=$ 106.5~MPa}
        \label{fig:wellbore_3d_5}
    \end{subfigure}
    \hfill    
    \begin{subfigure}[b]{0.3\textwidth}
        \centering
        \includegraphics[width=0.8\linewidth]{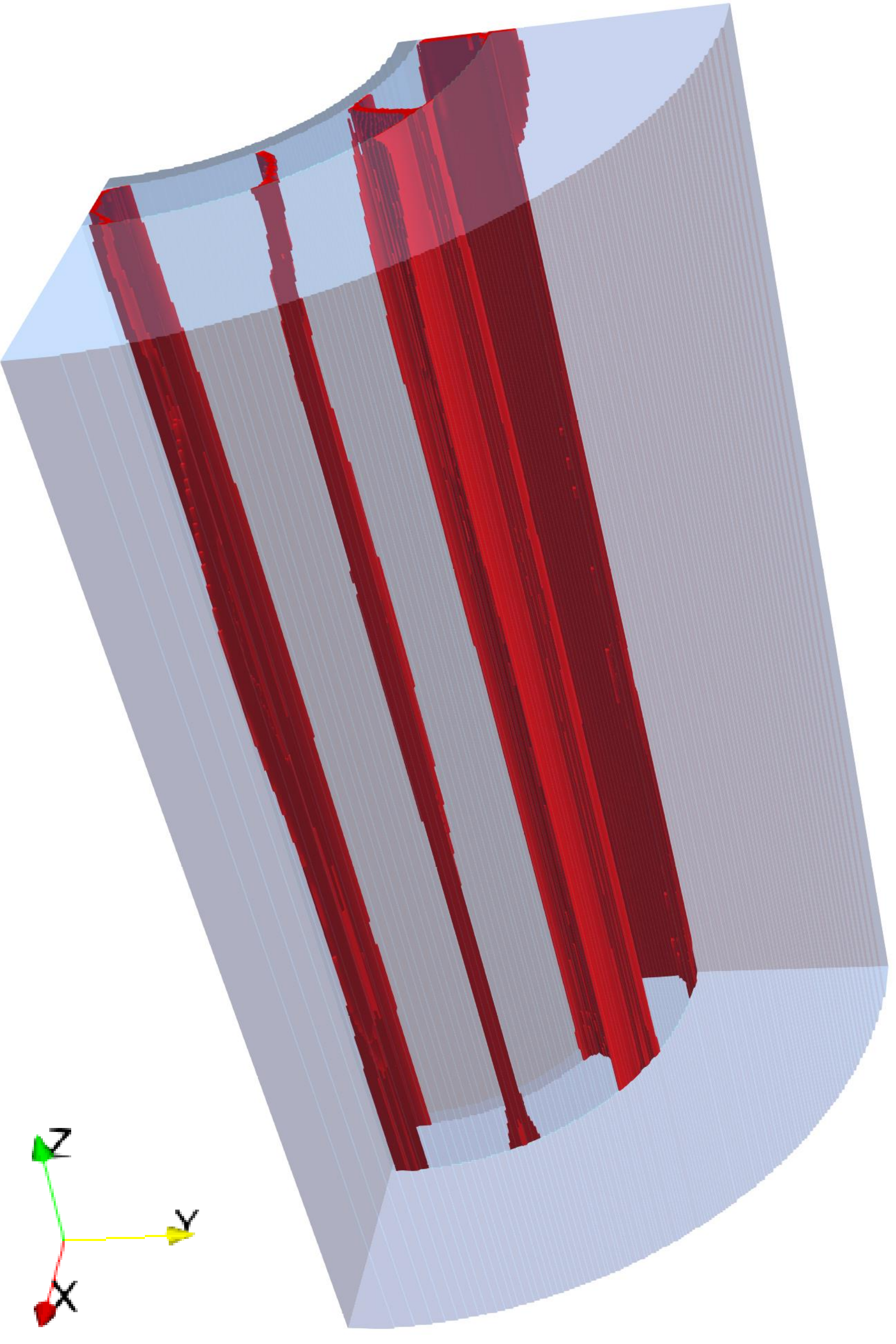}
        \caption{$P_i=$ 128~MPa}
        \label{fig:wellbore_3d_7}
    \end{subfigure}
    \hfill    
    \begin{subfigure}[b]{0.3\textwidth}
        \centering
        \includegraphics[width=0.8\linewidth]{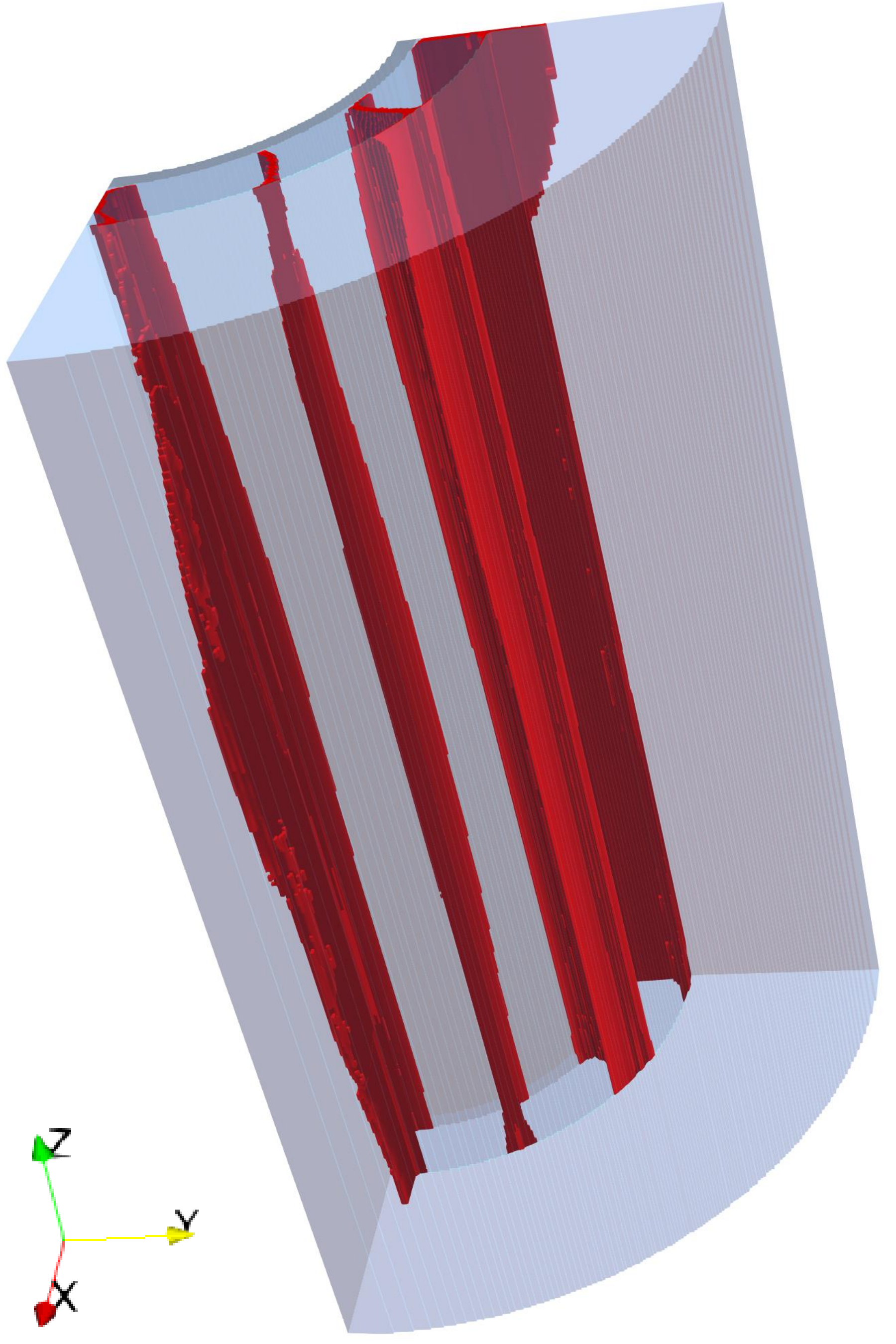}
        \caption{$P_i=$ 128.5~MPa}
        \label{fig:wellbore_3d_8}
    \end{subfigure}
    \hfill    
    \begin{subfigure}[b]{0.3\textwidth}
        \centering
        \includegraphics[width=0.8\linewidth]{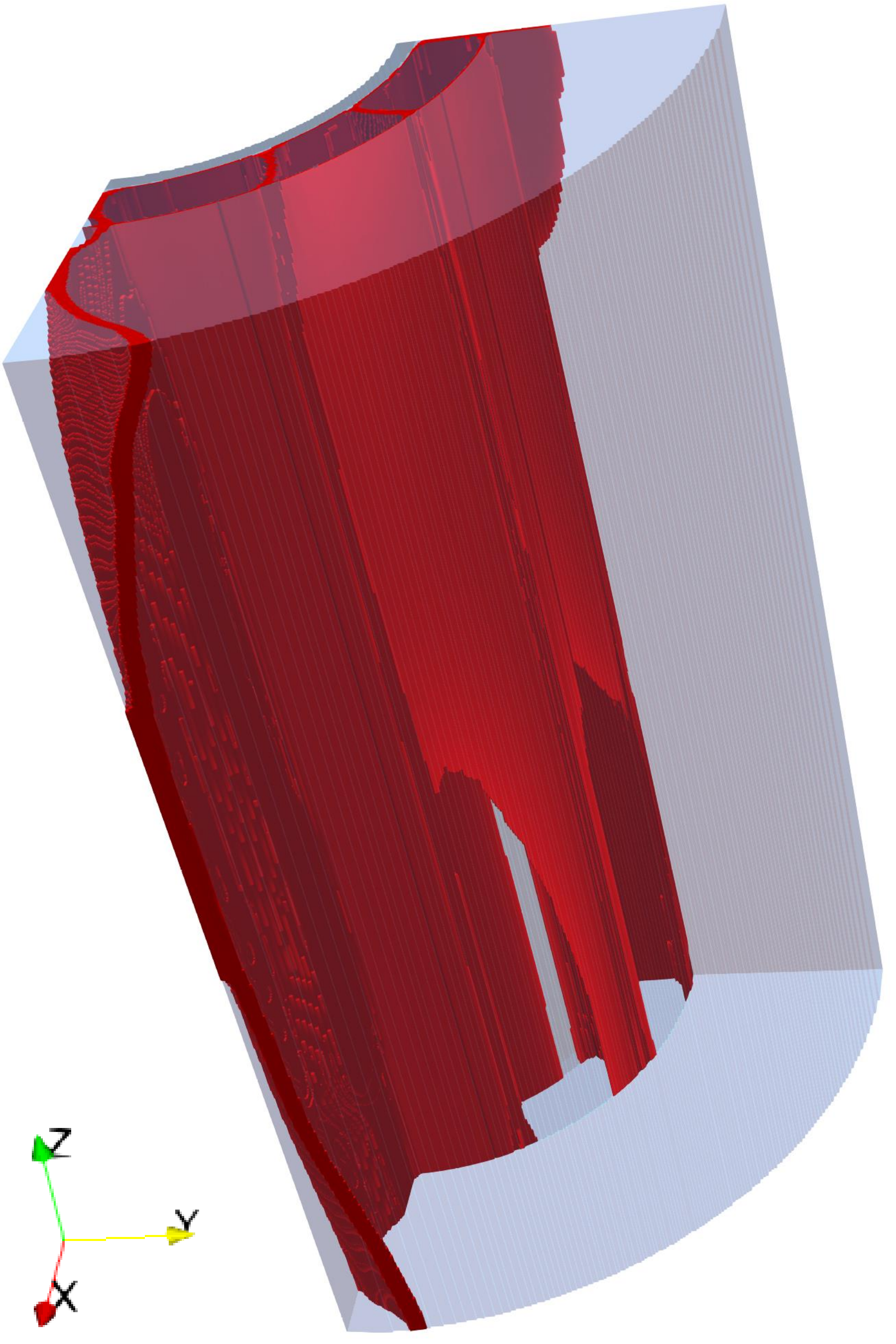}
        \caption{$P_i=$ 129~MPa}
        \label{fig:wellbore_3d_9}
    \end{subfigure}
    \caption{Damage profile at different internal pressures of the wellbore. The finite elements with a damage value greater than 0.9, which represent the cracked region, are shown in red.}
    \label{fig:wellbore_3d_damage}
\end{figure}

The evolution of fracture in the three-dimensional wellbore proceeds through a sequence of nucleation, arrest, interaction, and eventual coalescence, as described below and shown in Figure~\ref{fig:wellbore_3d_damage}. Figure~\ref{fig:wellbore_3d_1} shows the development of a through-the-thickness initial radial crack in the cement sheath at an internal pressure of 70.5~MPa. This crack is arrested at the weakened cement-rock interface, where slight debonding occurs. Similar to the two-dimensional case, continued loading while the crack remains arrested at the interface leads to the nucleation of a new through-the-thickness radial crack in the cement sheath, followed by further interfacial debonding, as shown in Figure~\ref{fig:wellbore_3d_2}. With further increase in the internal casing pressure, we observe nucleation of a radial crack in the upper one-third of the wellbore depth as shown in Figure~\ref{fig:wellbore_3d_3}. Unlike the previous cracks, which nucleated through the entire thickness of the model, the third crack nucleates only over a limited depth interval. The earlier cracks generate a stress shadow in this region, preventing this crack from nucleating through the full depth of the model. Figure~\ref{fig:wellbore_3d_4} shows that as this radial crack propagates towards the top and the bottom of the model, its orientation deviates from the z-axis, highlighting the limitations of the plane-strain assumption and motivating the need for fully three-dimensional analysis. Figure~\ref{fig:wellbore_3d_5} shows that, at an internal pressure of 106.5~MPa, another radial crack nucleates in the same way. At a wellbore pressure of 128 MPa, two radial cracks begin to penetrate the interface at different depths: the left crack penetrates near the mid-depth of the domain, while the right crack penetrates near the top boundary -- see Figure~\ref{fig:wellbore_3d_7}. Finally, as seen in Figures~\ref{fig:wellbore_3d_8} and \ref{fig:wellbore_3d_9}, the cracks begin to coalesce along the cement-rock interface with increasing wellbore pressure, and simultaneously, one of the penetrated cracks extends throughout the depth of the wellbore resulting in complete failure. This numerical simulation demonstrates the potential of the phase-field fracture model for detailed simulation of complex wellbore failure behavior and its utility in guiding engineering operations such as designing cementing jobs and interpreting cement bond logs, which are collected before and after cement curing to evaluate structural integrity and hydraulic isolation properties of the well~\citep{Bourgoyne1986}.

Although this is not the key focus of this paper, we provide a brief overview of the computational expense of this three-dimensional simulation. The simulation began with approximately 0.4 million degrees of freedom, which increased to 113 million as damage evolved and adaptive mesh refinement was activated. The simulation required approximately 20,000 core-hours on an in-house high-performance computing cluster equipped with AMD 64-core processors (2.45~GHz). In contrast, running the same problem on a uniformly refined mesh without adaptive mesh refinement would result in approximately 12 billion degrees of freedom, rendering the problem computationally intractable in practice.

Taken together, the numerical experiments demonstrate that geometric imperfections and interface weakening strongly alter both crack-initiation pressure and post-initiation crack evolution. These effects are not limited to changes in failure load; they also modify crack topology, interfacial debonding, and the transition from radial to non-radial fracture modes.

\section{Limitations}
\label{sec:limitations}
Although the present results provide useful insights into the effects of casing eccentricity and interface weakening on crack propagation and debonding in the cement sheath, we acknowledge that the model has several limitations that must be addressed in future studies. 

Firstly, the numerical framework relies on idealized representations of the wellbore system, which may not capture the full complexity of interfacial damage and failure under realistic wellbore conditions. Secondly, the rock formation and cement sheath are assumed to be isotropic and non-porous, whereas natural rock formations and cementitious materials can exhibit significant heterogeneity, anisotropy, and porosity. Thirdly, although the model provides a means to model weak interfaces, it expects interface strength as an input parameter and precisely characterizing the interface strength under operating conditions at the field-scale may be non-trivial. Fourthly, the phase-field fracture model adopted here is not the complete model, i.e., the crack nucleation is governed by purely energetic considerations and lacks the notion of material strength unlike strength-based phase-field models described in~\cite{kumar2020,lopezpamies2025}. 

Furthermore, the phase-field fracture model adopted here considers only crack opening at the interface and does not account for interfacial sliding, which may influence the evolution of shear cracks and debonding -- a limitation that is particularly relevant to the shear-dominated regions of the eccentric casing configuration examined in Figure~\ref{fig:shear_crack}, where interface friction is expected to play a non-negligible role. Discrete element simulations of frictional discontinuities in rocks have shown that permitting interfacial sliding, rather than restricting failure to tensile separation, can substantially alter both the magnitude and spatial distribution of tangential stress, and consequently the location and mode of instability \citep{Duan2018}. The interface sliding effect might act as an energy dissipation mechanism, thereby reducing the shear stress and delaying the nucleation of shear cracks \citep{xu2026}. Extending the present framework toward mixed-mode fracture behavior -- for instance, through dual tensile–shear phase-field formulations \citep{xu2026} coupled with frictional damage models on closed crack faces \citep{Xu2024} -- represents a natural direction for capturing these effects. Finally, the model accounts only for mechanical stresses and does not consider the effect of thermal stress encountered at greater subsurface depths. The simulations are an extension of the lab-scale experiments by \citep{taghipour2022novel} and consider only the local response of an individual wellbore system. 

Consequently, the effects of field-scale heterogeneity, wellbore interactions, and spatial variations in in-situ conditions are not captured. Field-scale simulations that account for the coupled thermo-mechanical loading of a poroelastic material using a complete phase-field fracture model, while incorporating interface sliding, are interesting future directions to enable more realistic and predictive simulations for failure of heterogeneous wellbore systems.

\section{Conclusion}
\label{sec:conclusion}
This paper investigates crack growth mechanisms in heterogeneous wellbore systems using a hybrid phase-field method with weak interfaces. A series of numerical experiments was performed to isolate the effects of interface strength and casing eccentricity on failure mechanisms in wellbore systems.

The results demonstrate that casing eccentricity strongly influences both the wellbore pressure at crack initiation and the resulting crack paths. Eccentric configurations generate stress concentrations in the thinner region of the cement sheath, leading to preferential crack initiation and propagation from this location. An inverse relationship between wellbore strength -- measured as the internal pressure at crack nucleation -- and casing eccentricity is observed. For the cases considered, inclined cracks deviating from the radial direction emerge in the rock formation beyond a critical eccentricity threshold ($e \geq 50\%$), driven by localized regions of high shear stress.

The study also examines the influence of interface strength on wellbore failure, focusing on the cement-rock interface. This interface is critical because it can undergo either crack penetration or interfacial delamination, unlike the casing-cement interface, which primarily debonds. The results show that interface strength governs both the crack initiation pressure and the resulting crack patterns. Weaker interfaces promote crack nucleation at lower pressures. When the interface is sufficiently weak, radially propagating cracks in the cement sheath are deflected along the interface rather than penetrating into the rock formation. This deflection delays stress relaxation within the sheath and promotes the nucleation of additional radial cracks, increasing the likelihood of sustained casing pressure (due to the leakage of deep gas toward the surface) and potential wellbore failure.

In addition to these mechanistic insights, the study presents preliminary three-dimensional simulations demonstrating the applicability of the phase-field framework to crack propagation in heterogeneous wellbore systems. Overall, the proposed framework provides a robust and physically consistent approach for modeling fracture behavior in multi-material wellbore systems, with potential applications in wellbore integrity assessment and design. Future work will focus on extending the framework to more realistic field scenarios, including inclined wellbores with varying inclination and azimuth, depth-dependent in-situ stresses, interactions between fractures originating from multiple wells, poroelastic effects in the cement and rock formation, and fully coupled hydro-thermo-mechanical processes.

\appendix
\section{Closed-form analytical expressions for elastic stresses in a pipe with an eccentric bore}
\label{sec:app_closed_form}

\begin{figure}[hbtp]
    \centering
    \includegraphics[
    width=0.6\linewidth]{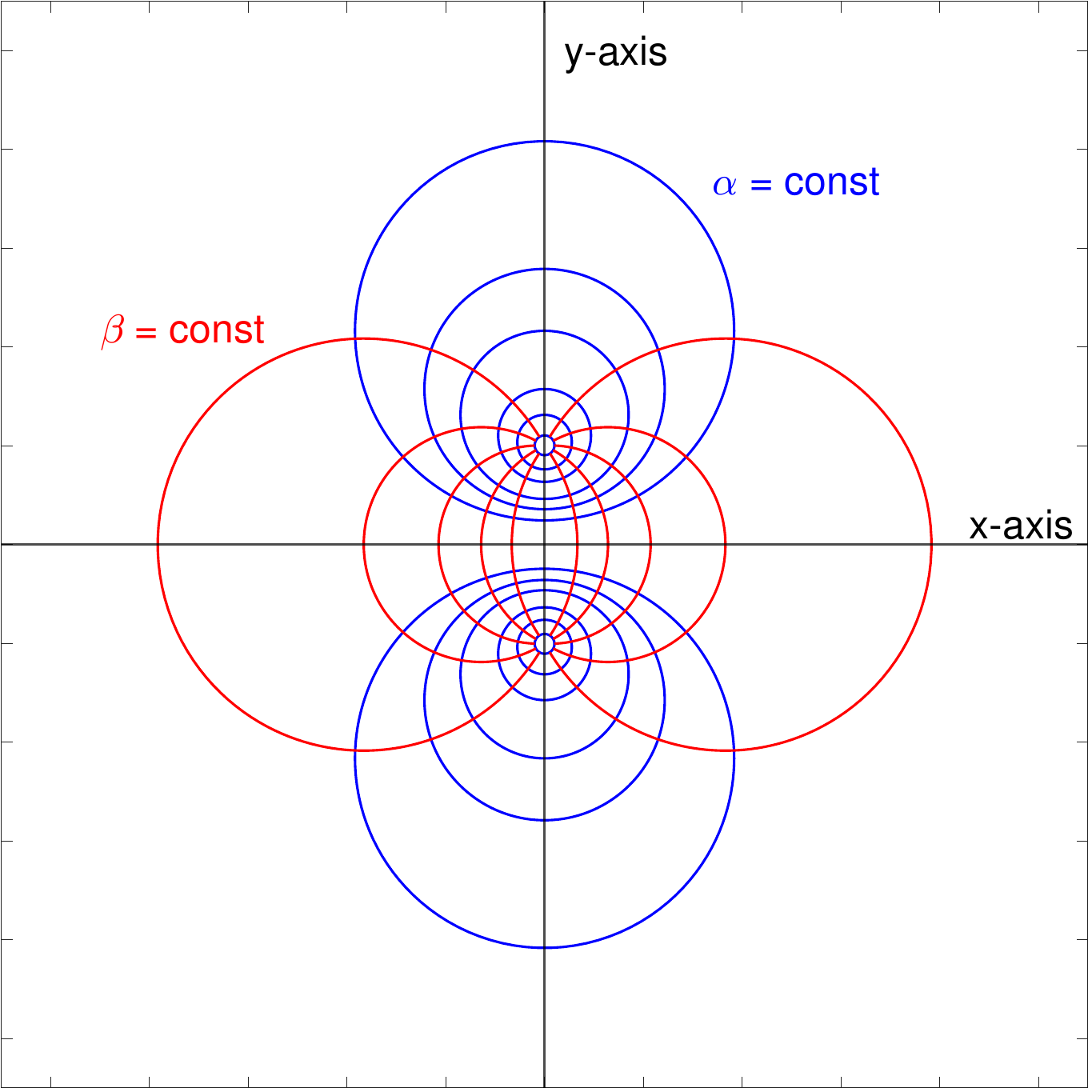}
    \caption{Bipolar coordinate system in terms of $\alpha$ and $\beta$ with respect to the Cartesian coordinate system: The blue circles represent contours with a constant $\alpha$ value and the red circles represent contours with a constant $\beta$ value. The boundaries of the pipe with eccentric bore are considered to be constant $\alpha$ circles.}
    \label{fig:bipolar_coordinate_system}
\end{figure}

Figure~\ref{fig:bipolar_coordinate_system} shows the bipolar coordinate system with respect to the Cartesian coordinate system. For the problem of a pipe with an eccentric bore, the boundaries of the pipes are considered as constant $\alpha$ circles. Following this assumption, the closed-form analytical expressions for elastic stresses in a pipe with an eccentric bore, as given by \citet{jeffery1921ix} in a bipolar coordinate system, are as follows,

\text{Tangential stress at the inner surface of the pipe:} 
\begin{equation} 
\begin{split}
    \sigma_{\beta\beta} &= -P_1 + 4(P_1-P_2)M (\cosh \alpha_1 - \cos \beta) \\
    &\qquad\quad\{\sinh (\alpha_1 - \alpha_2) \cos \beta + \sinh \alpha_1 \cosh (\alpha_1 - \alpha_2)\}
\end{split}
\label{eq:circum_stress_jeff_a1}
\end{equation}

\text{Tangential stress at the outer surface of the pipe:}
\begin{equation}
\begin{split}
    \sigma_{\beta\beta} &= -P_2 - 4(P_1-P_2)M (\cosh \alpha_2 - \cos \beta) \\
    &\qquad\quad\{\sinh (\alpha_1 - \alpha_2) \cos \beta + \sinh \alpha_2 \cosh (\alpha_1 - \alpha_2)\}
\end{split}
\label{eq:circum_stress_jeff_a2}
\end{equation}
where
\begin{align*} 
\sigma_{\beta\beta} &= \text{Tangential stress}, \\
P_1 &= \text{Traction on the inner surface of the pipe}, \\
P_2 &= \text{Traction on the outer surface of the pipe}, \\
\alpha_1 &= \text{Inner surface expressed in bipolar coordinates}, \\
\alpha_2 &= \text{Outer surface expressed in bipolar coordinates}, \\
M &=\frac{1}{2} \text{csch}(\alpha_1-\alpha_2)\{\sinh^2 \alpha_1 + \sinh^2 \alpha_2\}^{-1}, \\
\beta &= \text{Azimuthal coordinate (bipolar or hyperbolic angle) along the surface.} \\
\end{align*}

This expression requires us to know the bipolar coordinates of the boundaries of the pipe, i.e., $\alpha_1$ and $\alpha_2$. This section describes the steps involved in obtaining the bipolar coordinates of the pipe boundaries for a given eccentricity (e), radius of inner circle ($R_1)$ and radius of outer circle ($R_2)$ as shown in Figure~\ref{fig:jeff_model}. 

Consider $\alpha_1$ and $\alpha_2$ as the coordinates of the inner and outer circle of the pipe, respectively, and the coordinate $\beta$ varies from 0 to $2\pi$ for a given $\alpha$. Due to symmetry, we consider that $\beta$ varies from 0 to $\pi$ where $0^\circ$ corresponds to the thick part of the pipe and $180^\circ\,(\text{or}\,\pi\text{~radians})$ corresponds to the thinnest part of the pipe. The values of $\alpha_1$ and $\alpha_2$ required for the analytical solution are computed as follows:

The equation of a constant $\alpha$ circle in the Cartesian coordinate system is given by,
\begin{equation}
x^2+(y-y_c)^2=R^2
\end{equation}
\qquad where,
\begin{align*} 
y_c &= a \coth \alpha = \text{Coordinate of the circle's center}, \\
R &= a \operatorname{csch} \alpha =\text{Radius of the circle}, \\
a &= \text{Focal distance in the bipolar axis}.
\end{align*}

The eccentricity between the two circles is given as, 
\begin{equation}
    \begin{split}
        e &= y_{c2} -y_{c1} \\
          &= a \coth \alpha_2 -a \coth \alpha_1 \\
    \end{split}
\end{equation}
\qquad substituting, 
\begin{equation}
    \alpha = \operatorname{cosech}^{-1} \bigg(\frac{R}{a}\bigg), 
    \label{eq:ecc_alpha}
\end{equation}
\qquad we get 
\begin{equation}
              e = a \bigg[\coth \bigg( \operatorname{cosech}^{-1} \bigg(\frac{R_2}{a}\bigg)\bigg) - \coth \bigg( \operatorname{cosech}^{-1} \bigg(\frac{R_1}{a}\bigg)\bigg)\bigg]
              \label{eq:eccentricity_eqn}
\end{equation}
        
On solving Equation~\eqref{eq:eccentricity_eqn} for a known value of eccentricity and radii of inner and outer circles, we get the value of $a$, and using this value of $a$ and the corresponding radius in Equation~\eqref{eq:ecc_alpha}, we get the value of the corresponding $\alpha$ of the circles. For the geometry of the pipe considered in Section~\ref{nv:subsec1}, we get the values of $a$, $\alpha_1$, and $\alpha_2$ as 187.5 mm, 1.719, and 1.437, respectively. Further, the relation between the bipolar angular coordinate ($\beta$) in the bipolar coordinate system and the physical azimuth angle ($\theta$) in the polar coordinate system is given as,

\begin{equation}
\theta(\beta) = \arg \left[ \left( \cosh(\alpha_2) \cos\beta - \cosh(\alpha_1 - \alpha_2) \right) + i \sinh(\alpha_2) \sin\beta \right]
\label{eq:beta_to_theta}
\end{equation}

\section{Mesh convergence study for the pipe with an eccentric bore problem}
\label{app:mesh_convergence}

\begin{figure}[hbtp]
    \centering
    \includegraphics[
    width=0.6\linewidth]{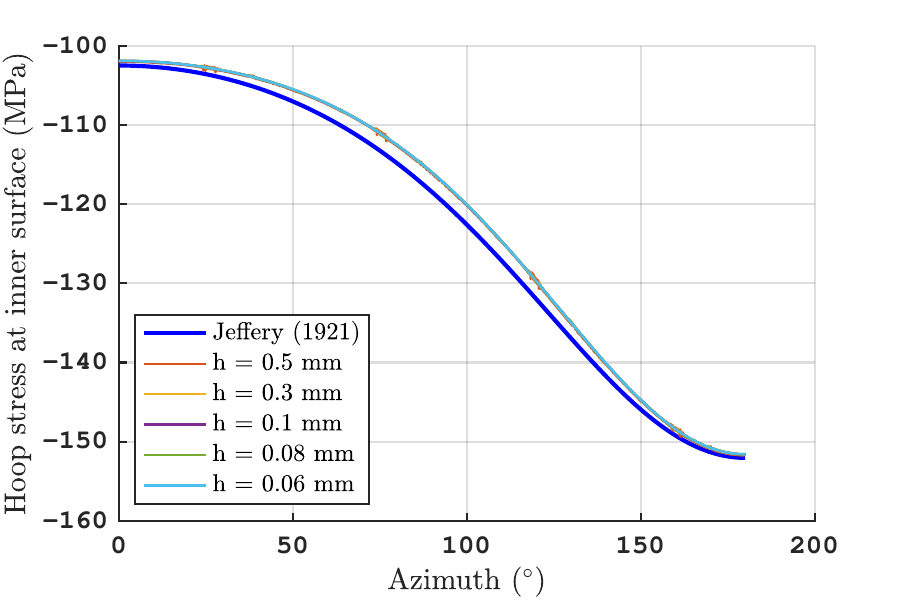}
    \caption{Result for the mesh convergence study for the tangential stress along the inner surface of the pipe with eccentric bore, compared against the closed-form solution of \citet{jeffery1921ix}.}
    \label{fig:jeff_circum_stress_mesh_convergence}
\end{figure}

A mesh-convergence study consisting of bilinear quadrilateral elements with element sizes of 0.5~mm, 0.3~mm, 0.1~mm, 0.08~mm, and 0.06~mm was performed to check the convergence of the FEM solution as shown in Figure~\ref{fig:jeff_circum_stress_mesh_convergence}. The finest mesh resolves the minimum ligament thickness of 15~mm with approximately 250 elements across it. The recovered hoop stress along the inner surface is essentially unchanged across this range, indicating the FEM solution itself has converged. The maximum relative error between the FEM solution and the closed-form result is 2.1\%, occurring at an azimuth of $\theta \approx 94^\circ$. Given that stress is a post-processed quantity in a displacement-based finite element formulation, we regard the 2.1\% error as acceptable. Although NURBS-enhanced finite elements or higher-order elements could yield better accuracy, we continue using bilinear quadrilateral elements to balance accuracy, implementation simplicity, and computational cost for the eccentric wellbore geometries considered in this study.

\newpage
\bibliographystyle{elsarticle-num-names}
\bibliography{sample}

@article{francfort1998revisiting,
  title={Revisiting brittle fracture as an energy minimization problem},
  author={Francfort, Gilles A and Marigo, J-J},
  journal={Journal of the Mechanics and Physics of Solids},
  volume={46},
  number={8},
  pages={1319--1342},
  year={1998},
  publisher={Elsevier}
}

@article{bourdin2000numerical,
  title={Numerical experiments in revisited brittle fracture},
  author={Bourdin, Blaise and Francfort, Gilles A and Marigo, Jean-Jacques},
  journal={Journal of the Mechanics and Physics of Solids},
  volume={48},
  number={4},
  pages={797--826},
  year={2000},
  publisher={Elsevier}
}

@article{munshi2025phase,
  title={Phase-field modeling of crack propagation in layered materials with weak interfaces},
  author={Munshi, WN and Annavarapu, C and Mulay, SS and Rodriguez-Ferran, A},
  journal={International Journal for Computational Methods in Engineering Science and Mechanics},
  pages={1--12},
  year={2025},
  publisher={Taylor \& Francis}
}

@article{munshi2025modeling,
title = {Modeling three-dimensional crack interaction in layered materials with weakly bonded interfaces using a phase-field model},
journal = {Engineering Fracture Mechanics},
volume = {329},
pages = {111624},
year = {2025},
issn = {0013-7944},
doi = {https://doi.org/10.1016/j.engfracmech.2025.111624},
url = {https://www.sciencedirect.com/science/article/pii/S0013794425008252},
author = {Wasim Niyaz Munshi and Chandrasekhar Annavarapu and Shantanu Shashikant Mulay and Antonio Rodríguez-Ferran}
}

@article{ambati2015review,
  title={A review on phase-field models of brittle fracture and a new fast hybrid formulation},
  author={Ambati, Marreddy and Gerasimov, Tymofiy and De Lorenzis, Laura},
  journal={Computational Mechanics},
  volume={55},
  number={2},
  pages={383--405},
  year={2015},
  publisher={Springer}
}

@article{miehe2010thermodynamically,
  title={Thermodynamically consistent phase-field models of fracture: Variational principles and multi-field FE implementations},
  author={Miehe, Christian and Welschinger, Fabian and Hofacker, Martina},
  journal={International journal for numerical methods in engineering},
  volume={83},
  number={10},
  pages={1273--1311},
  year={2010},
  publisher={Wiley Online Library}
}

@article{amor2009regularized,
  title={Regularized formulation of the variational brittle fracture with unilateral contact: Numerical experiments},
  author={Amor, Hanen and Marigo, Jean-Jacques and Maurini, Corrado},
  journal={Journal of the Mechanics and Physics of Solids},
  volume={57},
  number={8},
  pages={1209--1229},
  year={2009},
  publisher={Elsevier}
}

@article{moes1999finite,
  title={A finite element method for crack growth without remeshing},
  author={Mo{\"e}s, Nicolas and Dolbow, John and Belytschko, Ted},
  journal={International journal for numerical methods in engineering},
  volume={46},
  number={1},
  pages={131--150},
  year={1999},
  publisher={Wiley Online Library}
}

@article{gustafsson2022phase,
  title={Phase field models of interface failure for bone application-evaluation of open-source implementations},
  author={Gustafsson, Anna and Isaksson, Hanna},
  journal={Theoretical and Applied Fracture Mechanics},
  volume={121},
  pages={103432},
  year={2022},
  publisher={Elsevier}
}

@article{taghipour2022novel,
  title={Novel laboratory setup for realistic wellbore cement and formation integrity studies},
  author={Taghipour, Ali and Ghaderi, Amir and Cerasi, Pierre and Gheibi, Sohrab},
  journal={Journal of Petroleum Science and Engineering},
  volume={208},
  pages={109664},
  year={2022},
  publisher={Elsevier}
}

@article{Karma2001,
  author  = {Karma, Alain and Kessler, David A. and Levine, Herbert},
  title   = {Phase-field model of Mode III dynamic fracture},
  journal = {Physical Review Letters},
  volume  = {87},
  number  = {4},
  pages   = {045501},
  year    = {2001},
  doi     = {10.1103/PhysRevLett.87.045501}
}

@techreport{Cusini_2287725,
  author       = {Cusini, Matteo and Fei, Fan},
  title        = {Phase Field Modeling of Near-Wellbore Hydraulic Fracture Nucleation and Propagation},
  institution  = {Lawrence Livermore National Laboratory (LLNL), Livermore, CA (United States)},
  doi          = {10.2172/2287725},
  url          = {https://www.osti.gov/biblio/2287725},
  place        = {United States},
  year         = {2024},
  month        = {02}}

@article{jeffery1921ix,
  title={IX. Plane stress and plane strain in bipolar co-ordinates},
  author={Jeffery, George Barker},
  journal={Philosophical Transactions of the Royal Society of London. Series A, Containing Papers of a Mathematical or Physical Character},
  volume={221},
  number={582-593},
  pages={265--293},
  year={1921},
  publisher={The Royal Society London}
}

@article{KHAN2025110672,
title = {Adaptive phase-field modeling of fracture propagation in layered media: Effects of mechanical property mismatches, layer thickness, and interface strength},
journal = {Engineering Fracture Mechanics},
volume = {314},
pages = {110672},
year = {2025},
issn = {0013-7944},
doi = {https://doi.org/10.1016/j.engfracmech.2024.110672},
url = {https://www.sciencedirect.com/science/article/pii/S001379442400835X},
author = {Salman Khan and Ishank Singh and Chandrasekhar Annavarapu and Antonio Rodríguez-Ferran}
}

@article{bangerth2007deal,
  title={deal. II—a general-purpose object-oriented finite element library},
  author={Bangerth, Wolfgang and Hartmann, Ralf and Kanschat, Guido},
  journal={ACM Transactions on Mathematical Software (TOMS)},
  volume={33},
  number={4},
  pages={24--es},
  year={2007},
  publisher={ACM New York, NY, USA}
}

@article{MUNSHI2026109901,
title = {A detailed guide to an open-source implementation of the hybrid phase field method for 3D fracture modeling in deal.II},
journal = {Computer Physics Communications},
volume = {319},
pages = {109901},
year = {2026},
issn = {0010-4655},
doi = {https://doi.org/10.1016/j.cpc.2025.109901},
url = {https://www.sciencedirect.com/science/article/pii/S0010465525004023},
author = {Wasim Niyaz Munshi and Marc Fehling and Wolfgang Bangerth and Chandrasekhar Annavarapu}
}

@article{xu2022phase,
  title={Phase-field modeling of crack growth and interaction in rock},
  author={Xu, Bin and Xu, Tao and Xue, Yanchao and Heap, Michael J and Ranjith, PG and Wasantha, PLP and Li, Zhiguo},
  journal={Geomechanics and Geophysics for Geo-Energy and Geo-Resources},
  volume={8},
  number={6},
  pages={180},
  year={2022},
  publisher={Springer}
}

@article{clavijo2022coupled,
  title={A coupled phase-field and reactive-transport framework for fracture propagation in poroelastic media},
  author={Clavijo, Santiago Pena and Addassi, Mouadh and Finkbeiner, Thomas and Hoteit, Hussein},
  journal={Scientific Reports},
  volume={12},
  number={1},
  pages={17819},
  year={2022},
  publisher={Nature Publishing Group UK London}
}

@article{xi2020experimental,
  title={Experimental and numerical investigations of accumulated plastic deformation in cement sheath during multistage fracturing in shale gas wells},
  author={Xi, Yan and Li, Jun and Tao, Qian and Guo, Boyun and Liu, Gonghui},
  journal={Journal of Petroleum Science and Engineering},
  volume={187},
  pages={106790},
  year={2020},
  publisher={Elsevier}
}

@article{goodwin1992cement,
  title={Cement sheath stress failure},
  author={Goodwin, KJ and Crook, RJ},
  journal={SPE drilling engineering},
  volume={7},
  number={04},
  pages={291--296},
  year={1992},
  publisher={SPE}
}

@article{kuhn2019phase,
  title={Phase field modeling of interface effects on cracks in heterogeneous materials},
  author={Kuhn, Charlotte and M{\"u}ller, Ralf},
  journal={PAMM},
  volume={19},
  number={1},
  pages={e201900378},
  year={2019},
  publisher={Wiley Online Library}
}

@article{kumar2020,
title = {Revisiting nucleation in the phase-field approach to brittle fracture},
journal = {Journal of the Mechanics and Physics of Solids},
volume = {142},
pages = {104027},
year = {2020},
issn = {0022-5096},
doi = {https://doi.org/10.1016/j.jmps.2020.104027},
url = {https://www.sciencedirect.com/science/article/pii/S0022509620302623},
author = {Aditya Kumar and Blaise Bourdin and Gilles A. Francfort and Oscar Lopez-Pamies}
}

@article{lopezpamies2025,
title = {Classical variational phase-field models cannot predict fracture nucleation},
journal = {Computer Methods in Applied Mechanics and Engineering},
volume = {433},
pages = {117520},
year = {2025},
issn = {0045-7825},
doi = {https://doi.org/10.1016/j.cma.2024.117520},
url = {https://www.sciencedirect.com/science/article/pii/S0045782524007746},
author = {Oscar Lopez-Pamies and John E. Dolbow and Gilles A. Francfort and Christopher J. Larsen}}

@article{Ding2024,
title = {An adaptive phase field modeling of fatigue crack growth using variable-node elements and explicit cycle jump scheme},
journal = {Computer Methods in Applied Mechanics and Engineering},
volume = {429},
pages = {117200},
year = {2024},
issn = {0045-7825},
doi = {https://doi.org/10.1016/j.cma.2024.117200},
url = {https://www.sciencedirect.com/science/article/pii/S0045782524004560},
author = {Junlei Ding and Tiantang Yu and Weihua Fang and Sundararajan Natarajan}
}

@article{Ortiz1999,
author = {Ortiz, M. and Pandolfi, A.},
title = {Finite-deformation irreversible cohesive elements for three-dimensional crack-propagation analysis},
journal = {International Journal for Numerical Methods in Engineering},
volume = {44},
number = {9},
pages = {1267-1282},
year = {1999}
}

@article{Camacho1996,
	title={Computational modelling of impact damage in brittle materials},
	author={Camacho, G. T. and Ortiz, M.},
	journal={International Journal of Solids and Structures},
	volume={33},
	number={20-22},
	pages={2899--2938},
	year={1996}}

@article{Khan2023a,
  title={Adaptive phase-field modeling of fracture propagation in bi-layered materials},
  author={Khan, Salman and Muix{\'\i}, Alba and Annavarapu, Chandrasekhar and Rodr{\'\i}guez-Ferran, Antonio},
  journal={Engineering Fracture Mechanics},
  volume={292},
  pages={109650},
  year={2023},
  publisher={Elsevier}
}

@article{Khan2023b,
    author = {Salman Khan and Alba Muixí and Chandrasekhar Annavarapu and Antonio Rodríguez-Ferran},
    title = {Investigation on the Effect of Material Mismatch Between Two Dissimilar Materials Using an Adaptive Phase-field Method},
    journal = {International Journal of Advances in Engineering Sciences and Applied Mathematics},
    year = {2023}
}

@article{Jain2023a,
title = {Adaptive phase-field modeling of fracture in orthotropic composites},
journal = {Engineering Fracture Mechanics},
volume = {292},
pages = {109673},
year = {2023},
issn = {0013-7944},
doi = {https://doi.org/10.1016/j.engfracmech.2023.109673},
url = {https://www.sciencedirect.com/science/article/pii/S0013794423006318},
author = {Ishank Jain and Alba Muixí and Chandrasekhar Annavarapu and Shantanu S. Mulay and Antonio Rodríguez-Ferran}
}

@article{Jain2023b,
title = {Numerical modeling of fracture propagation in orthotropic composite materials using an adaptive phase-field method},
journal = {International Journal of Advances in Engineering Sciences and Applied Mathematics},
volume = {15},
pages = {144–154},
year = {2023},
author = {Ishank Jain and Chandrasekhar Annavarapu and Shantanu S. Mulay and Antonio Rodríguez-Ferran}
}

@article{Ghosh2019,
title = {A stabilized finite element method for enforcing stiff anisotropic cohesive laws using interface elements},
journal = {Computer Methods in Applied Mechanics and Engineering},
volume = {348},
pages = {1013-1038},
year = {2019},
issn = {0045-7825},
doi = {https://doi.org/10.1016/j.cma.2019.02.007},
url = {https://www.sciencedirect.com/science/article/pii/S0045782519300787},
author = {Gourab Ghosh and Ravindra Duddu and Chandrasekhar Annavarapu}
}

@article{Ghosh2021,
author = {Gourab Ghosh and Ravindra Duddu and Chandrasekhar Annavarapu},
title = {On the robustness of the stabilized finite element method for delamination analysis of composites using cohesive elements},
journal = {International Journal for Computational Methods in Engineering Science and Mechanics},
volume = {22},
number = {6},
pages = {538--558},
year = {2021},
publisher = {Taylor \& Francis},
doi = {10.1080/15502287.2021.1896607},
URL = {https://doi.org/10.1080/15502287.2021.1896607},
eprint = {https://doi.org/10.1080/15502287.2021.1896607}
}

@incollection{Wu2020,
title = {Chapter One - Phase-field modeling of fracture},
editor = {Stéphane P.A. Bordas and Daniel S. Balint},
booktitle   = {Advances in Applied Mechanics},
publisher = {Elsevier},
volume = {53},
pages = {1-183},
year = {2020},
issn = {0065-2156},
doi = {https://doi.org/10.1016/bs.aams.2019.08.001},
url = {https://www.sciencedirect.com/science/article/pii/S0065215619300134},
author = {Jian-Ying Wu and Vinh Phu Nguyen and Chi Thanh Nguyen and Danas Sutula and Sina Sinaie and Stéphane P.A. Bordas}
}

@article{Annavarapu2016,
  author = {Annavarapu, C. and Settgast, R. R. and Vitali, E. and Morris, J. P.},
  title = {A local crack-tracking strategy to model three-dimensional crack propagation with embedded methods},
  journal = {Computer Methods in Applied Mechanics and Engineering},
  year = {2016},
  volume = {311},
  pages = {815--837}
}

@article{zhang2019phase,
  title={Phase field modeling of fracture in fiber reinforced composite laminate},
  author={Zhang, Peng and Hu, Xiaofei and Bui, Tinh Quoc and Yao, Weian},
  journal={International Journal of Mechanical Sciences},
  volume={161},
  pages={105008},
  year={2019},
  publisher={Elsevier}
}

@article{seiler2020efficient,
  title={An efficient phase-field model for fatigue fracture in ductile materials},
  author={Seiler, Martha and Linse, Thomas and Hantschke, Peter and K{\"a}stner, Markus},
  journal={Engineering Fracture Mechanics},
  volume={224},
  pages={106807},
  year={2020},
  publisher={Elsevier}
}

@article{denli2020phase,
  title={A phase-field model for fracture of unidirectional fiber-reinforced polymer matrix composites},
  author={Denli, Funda Aksu and G{\"u}ltekin, Osman and Holzapfel, Gerhard A and Dal, H{\"u}sn{\"u}},
  journal={Computational Mechanics},
  volume={65},
  number={4},
  pages={1149--1166},
  year={2020},
  publisher={Springer}
}

@article{messaoudi2025fracture,
  title={Fracture modeling of CNT/epoxy nanocomposites based on phase-field method using multiscale strategy},
  author={Messaoudi, I and Mallek, H and Mellouli, H and Wali, M and Dammak, F},
  journal={Proceedings of the Institution of Mechanical Engineers, Part L: Journal of Materials: Design and Applications},
  volume={239},
  number={2},
  pages={319--334},
  year={2025},
  publisher={SAGE Publications Sage UK: London, England}
}

@article{wang2014three,
  title={Three-dimensional analysis of cement sheath integrity around Wellbores},
  author={Wang, Wꎬ and Taleghani, A Dahi},
  journal={Journal of Petroleum Science and Engineering},
  volume={121},
  pages={38--51},
  year={2014},
  publisher={Elsevier}
}

@article{gu2022numerical,
  title={Numerical investigation of cement interface debonding in deviated shale gas wells considering casing eccentricity and residual drilling fluid},
  author={Gu, Chenwang and Li, Xiaorong and Feng, Yongcun and Deng, Jingen and Gray, Kenneth},
  journal={International Journal of Rock Mechanics and Mining Sciences},
  volume={158},
  pages={105197},
  year={2022},
  publisher={Elsevier}
}

@article{xi2024failure,
  title={Failure evaluation mechanism of cement sheath sealing integrity under casing eccentricity during multistage fracturing},
  author={Xi, Yan and Yao, Yu and Guo, Xue-Li and Li, Jun and Tian, Yu-Dong and Liu, Gong-Hui},
  journal={Petroleum Science},
  volume={21},
  number={5},
  pages={3428--3445},
  year={2024},
  publisher={Elsevier}
}

@article{zhao2020cohesive,
  title={A cohesive-element-based model to evaluate interfacial behavior of casing--cement sheath for high-pressure, high-temperature wellbore integrity considering casing eccentricity},
  author={Zhao, Xinbo and Yang, Xiujuan and Lin, Xingyu and Mei, Yue},
  journal={Advances in Mechanical Engineering},
  volume={12},
  number={4},
  pages={1687814019898300},
  year={2020},
  publisher={SAGE Publications Sage UK: London, England}
}

@inproceedings{zheng2023influence,
  title={The influence of casing eccentricity on zonal isolation},
  author={Zheng, Danzhu and Ozbayoglu, Evren and Miska, Stefan and Silvio, Baldino and Liu, Yaxin and Wang, Junzhe},
  booktitle={ARMA US Rock Mechanics/Geomechanics Symposium},
  pages={ARMA--2023},
  year={2023},
  organization={ARMA}
}

@article{cheng2017numerical,
  title={Numerical stress analysis for the multi-casing structure inside a wellbore in the formation using the boundary element method},
  author={Cheng, Wan and Jin, Yan and Chen, Mian and Jiang, Guo-Sheng},
  journal={Petroleum Science},
  volume={14},
  number={1},
  pages={126--137},
  year={2017},
  publisher={Springer}
}

@inproceedings{dou2020numerical,
  title={Numerical modelling of the cement sheath with radial cracks under casing eccentric conditions},
  author={Dou, Haoyu and Dong, Xuelin and Gao, Deli},
  booktitle={ARMA US Rock Mechanics/Geomechanics Symposium},
  pages={ARMA--2020},
  year={2020},
  organization={ARMA}
}

@article{liu2018impact,
  title={Impact of casing eccentricity on cement sheath},
  author={Liu, Kui and Gao, Deli and Taleghani, Arash Dahi},
  journal={Energies},
  volume={11},
  number={10},
  pages={2557},
  year={2018},
  publisher={MDPI}
}

@article{zhao2023integrity,
  title={Integrity and Failure Analysis of Cement Sheath Subjected to Coalbed Methane Fracturing.},
  author={Zhao, Lingyun and Yang, Heng and Wei, Yuanlong and Bu, Yuhuan and Jing, Shaorui and Zhou, Peiming},
  journal={Fluid Dynamics \& Materials Processing},
  volume={19},
  number={2},
  year={2023}
}

@article{anya2023novel,
  title={A novel apparatus and method for lab-scale study of wellbore integrity using CT imaging and analysis},
  author={Anya, Alexander and Emadi, Hossein and Watson, Marshall},
  journal={Journal of Petroleum Science and Engineering},
  volume={220},
  pages={111209},
  year={2023},
  publisher={Elsevier}
}

@article{vraalstad2020digital,
  title={Digital cement integrity: A methodology for 3D visualization of cracks and microannuli in well cement},
  author={Vr{\aa}lstad, Torbj{\o}rn and Skorpa, Ragnhild},
  journal={Sustainability},
  volume={12},
  number={10},
  pages={4128},
  year={2020},
  publisher={MDPI}
}

@article{nassan2024experimental,
  title={Experimental investigation of wellbore integrity during geological carbon sequestration: Thermal-and pressure-cycling experiments},
  author={Nassan, Taofik H and Kirch, Martin and Freese, Carsten and Alkan, Hakan and Baganz, Dirk and Amro, Mohd},
  journal={Gas Science and Engineering},
  volume={124},
  pages={205253},
  year={2024},
  publisher={Elsevier}
}

@book{driscoll1986,
  title     = "Groundwater and Wells",
  author    = "Fletcher Driscoll",
  year      = 1986,
  publisher = "Johnson Screens",
  address   = "Minnesota, USA"
}

@book{Bourgoyne1986,
  author    = {Bourgoyne, Adam T. Jr. and Millheim, Keith K. and Chenevert, Martin E. and Young, F. S. Jr.},
  title     = {Applied Drilling Engineering},
  publisher = {Society of Petroleum Engineers},
  series    = {SPE Textbook Series},
  volume    = {2},
  address   = {Richardson, TX},
  year      = {1986},
  isbn      = {978-1555630010}
}

@article{Xu2026,
  author  = {Xu, Bin and Xu, Tao and Baktheer, Abedulgader and Lan, Xiaocong and Heap, Michael J. and Amitrano, David and Liu, Ben and Aldakheel, Fadi},
  title   = {An energy-based dual phase-field model for shear cracking patterns in heterogeneous geomaterials},
  journal = {Engineering Fracture Mechanics},
  volume  = {332},
  pages   = {111774},
  year    = {2026},
  doi     = {10.1016/j.engfracmech.2025.111774}
}

@article{Xu2024,
  author  = {Xu, Bin and Xu, Tao and Xue, Yanchao and Heap, Michael J. and Wasantha, P. L. P. and Li, Zhiguo},
  title   = {Phase field modeling of mixed-mode crack in rocks incorporating heterogeneity and frictional damage},
  journal = {Engineering Fracture Mechanics},
  volume  = {298},
  pages   = {109936},
  year    = {2024},
  doi     = {10.1016/j.engfracmech.2024.109936}
}

@article{Duan2018,
  author  = {Duan, Kang and Wu, Wei and Kwok, Chung Yee},
  title   = {Discrete element modelling of stress-induced instability of directional drilling boreholes in anisotropic rock},
  journal = {Tunnelling and Underground Space Technology},
  volume  = {81},
  pages   = {55--67},
  year    = {2018},
  doi     = {10.1016/j.tust.2018.07.001}
}

@article{cui2024computational,
  title={Computational predictions of hydrogen-assisted fatigue crack growth},
  author={Cui, Chuanjie and Bortot, Paolo and Ortolani, Matteo and Mart{\'\i}nez-Pa{\~n}eda, Emilio},
  journal={International Journal of Hydrogen Energy},
  volume={72},
  pages={315--325},
  year={2024},
  publisher={Elsevier}
}

\end{document}